%% file: MAIN.tex
\documentclass[conference, letterpaper, 10pt]{IEEEtran}
\IEEEoverridecommandlockouts

\input{preamble}

\usepackage[
  letterpaper,            
  top=0.72in,             
  bottom=1.00in,          
  left=0.63in,            
  right=0.63in,
  heightrounded           
]{geometry}

\begin{document}

\title{An Extensible Quantum Network Simulator Built on ns-3: \textit{Q2NS} Design and Evaluation}

\author{
\IEEEauthorblockN{
Adam Pearson, 
Francesco Mazza,~\IEEEmembership{Graduate~Student~Member,~IEEE}, \\
Marcello~Caleffi,~\IEEEmembership{Senior~Member,~IEEE}, Angela~Sara~Cacciapuoti,~\IEEEmembership{Senior~Member,~IEEE}}

\thanks{
A preliminary 5-page conference version of this manuscript is available at \cite{PeaMazCal-26}.
}

\thanks{
This work has been funded by the European Union under Horizon Europe ERC-CoG grant QNattyNet, n.101169850. Views and opinions expressed are however those of the author(s) only and do not necessarily reflect those of the European Union or the European Research Council Executive Agency. Neither the European Union nor the granting authority can be held responsible for them.
}
}

\maketitle

\begin{abstract}
As quantum networking hardware remains costly and not yet widely accessible, simulation tools are essential for the design and evaluation of quantum network architectures and protocols.
However, designing a scalable and computationally efficient quantum network simulator is intrinsically challenging: (i) quantum dynamics must be emulated on classical computing platforms while capturing the stateful and non-local nature of entanglement, a unique quantum resource without any classical networking analog; (ii) moreover, quantum networking is inherently hybrid, as protocol execution also fundamentally depends on classical signaling. This makes a tight and faithful co-simulation of quantum operations and classical message exchanges a core requirement. In this light, we present \textit{Q2NS}, a modular and extensible quantum network simulator, built on top of ns-3, designed to seamlessly integrate quantum-network primitives with ns-3’s established classical protocol stack. \textit{Q2NS} adopts a modular architecture that decouples protocol control logic from node- and channel-level operations, enabling rapid prototyping and adaptation across heterogeneous and evolving Quantum Internet scenarios. \textit{Q2NS} natively supports multiple quantum state representations through a unified plug-in interface, allowing interchangeable state-vector, density-matrix, and stabilizer backends. We validate \textit{Q2NS} through realistic use-case studies and comprehensive benchmarks, demonstrating superior computational efficiency over representative state-of-the-art alternatives, while preserving modeling flexibility. Finally, we provide a dedicated visualization tool that jointly captures physical and entanglement-enabled connectivity and supports entangled-state manipulations, facilitating an intuitive interpretation of entanglement dynamics and protocol behavior. Overall, \textit{Q2NS} offers a flexible, open, and scalable simulation platform for advancing Quantum Internet research.
\end{abstract}


\section{Introduction}

The development of a fully operational Quantum Internet \cite{CalCac-25, CacCal-26, Kim-08, DurLamHeu-17, VanSatBen-21, PirDur-19} promises transformative applications across secure communications \cite{BenBra-14, MinPitRob-19, CaoZhaWan-22}, distributed quantum computing \cite{CirEkeHue-99, CalAmoFer-22}, and high-precision sensing \cite{DegReiFri-17, ZhaZhu-21, GiaWinCon-25}. 
However, progress toward a functional Quantum Internet is still constrained by the high cost, limited availability, and early-stage maturity of quantum hardware \cite{IllCalMan-22, MazCalCac-25, CaldAvHan-25, CacPelIll-25}.
Consequently, realistic and accessible simulation platforms have become essential for investigating quantum network architectures, protocols, and control mechanisms under controlled and reproducible conditions. Indeed, by enabling the modeling, validation, and reproducibility of complex quantum communication scenarios, simulators support rapid prototyping, quantitative performance assessment, and the exploration of novel design principles, bridging the gap between theoretical design and practical implementation.

However, designing an effective simulator for quantum networks is far from trivial due to the unconventional principles and phenomena of quantum mechanics: qubits cannot be copied, quantum states degrade rapidly, and local operations can have non-local effects when acting on entangled states \cite{CacCalVan-20, IllCalMan-22}. Indeed, entanglement -- the key resource of any quantum network -- is inherently non-local and stateful: local actions on one node can instantaneously affect correlated states across distant nodes. Consequently, a feasible simulator must be capable of maintaining a persistent, global awareness of the overall quantum state, while enabling tight in-network coordination to capture these non-local interactions.
Moreover, since the entire entanglement life-cycle -- spanning its generation, distribution, and exploitation -- together with other quantum communication protocols, relies on classical signaling for synchronization, control, and coordination \cite{CirZolKim-97, IllCalMan-22, MazCalCac-25}, an effective quantum network simulator must seamlessly integrate both classical and quantum communication primitives within a unified architectural design.
At the same time, accurately capturing quantum phenomena is computationally challenging, as the ability to simulate quantum systems on classical computing machines is fundamentally limited by the exponential growth of the quantum state-space with the qubit number.

From all of the above, it is evident that \textit{the design of the simulator architecture is itself a determining factor in achieving both accuracy and scalability}, even before the implementation of individual software modules or libraries. A well-designed framework must capture not only event-driven dynamics, but also the global control logic governing the life-cycle of entangled states. Without such architectural coherence, a simulator cannot faithfully reproduce the operational constraints and resource dependencies that define entanglement-enabled networking.

\begin{table*}[t]
\centering
\caption{Comparison of selected quantum network simulators}
\label{tab:1}
\setlength{\tabcolsep}{3.5pt}
\renewcommand{\arraystretch}{1.15}
\begin{tabular}{@{}l l c c c c c c@{}}
\hline
\textbf{Simulator} & \textbf{Language(s)} & \textbf{Open Source} & \textbf{Classical Stack} & \textbf{Timing Realism} & \textbf{Extensibility} & \textbf{General Purpose} \\
\hline
\textbf{SQUANCH}~\cite{bartlett2018squanch}         & Python          & \checkmark & Limited      & Limited   & Moderate & $\times$ \\
\textbf{QuNetSim}~\cite{diadamo2021qunetsim}        & Python          & \checkmark & \checkmark   & $\times$  & High     & \checkmark \\
\textbf{NetSquid}~\cite{coopmans2021netsquid}       & C++/Python      & $\times$   & $\times$     & \checkmark& Limited & \checkmark  \\
\textbf{SeQUeNCe}~\cite{wu2021sequence}             & C++/Python      & \checkmark & $\times$     & \checkmark& Moderate & \checkmark  \\
\textbf{NetQASM SDK}~\cite{dahlberg2022netqasm}     & C++/Python      & \checkmark & Limited      & Limited   & High     & $\times$ \\
\textbf{QNE-ADK}~\cite{qneadk2022}                  & C++/Python      & $\times$   & Limited      & Limited   & High     & $\times$ \\
\textbf{QuiSP}~\cite{satoh2021quisp}                & C++             & \checkmark & Limited      & \checkmark& Moderate & \checkmark  \\
\textbf{QKDNetSim(+)}~\cite{dervisevic2024qkdnetsim, soler2024qkdnetsimplus}   & C++ (ns-3)      & \checkmark & \checkmark   & \checkmark& Limited & $\times$ \\
\textbf{ns-3 Quantum Network  Module}~\cite{ns3quantum2025}   & C++ (ns-3)    & \checkmark    & \checkmark    & \checkmark    & Moderate  & $\times$ \\
\textbf{qns-3}~\cite{Lin2025CFA}    & C++ (ns-3)    & \checkmark    & \checkmark    & \checkmark    & Moderate    & \checkmark \\
\textbf{*\textit{Q2NS}*}                               & C++ (ns-3)      & \checkmark        & \checkmark   & \checkmark& High     & \checkmark \\
\hline
\end{tabular}

\vspace{0.25em}
\end{table*}

In this light, we present \textit{Q2NS}, a modular and inherently extensible quantum network simulation framework, built on top of ns-3, a de facto reference for discrete-event classical network simulation \cite{ns3}. \textit{Q2NS} is designed to natively support the inherent interplay between quantum and classical communication primitives, within a unified architecture. This, in turn, ensures accurate modeling of entanglement dynamics and protocol control logic.

As detailed in Sec.~\ref{sec:2}, the architectural design of \textit{Q2NS} enforces a clear separation of concerns between the operational functionalities of modules and entities and the overarching control logic. This architectural principle is realized through the \texttt{NetController}, which maintains quantum state awareness and enables flexible adaptation to heterogeneous and rapidly evolving quantum networking paradigms \cite{CalCac-25}. Furthermore, the modular architecture incorporates a unified plug-in interface for heterogeneous quantum state representations, currently supporting \textit{state-vector}, \textit{density-matrix}, and \textit{stabilizer} backends. This extensibility allows researchers to select representation-specific optimizations tailored to the simulation scenario, ranging from exact state-vector dynamics to computationally efficient stabilizer formalism for Clifford circuits. At the same time, \textit{Q2NS} leverages ns-3's mature classical networking stack, by supporting faithful co-simulation of quantum operations and classical communication, while facilitating the integration of new protocols and quantum noise models with minimal implementation effort.

Implemented in C++, \textit{Q2NS} offers higher computational efficiency compared to Python-based alternatives, discussed in the related works. This efficiency stems not only from the use of a compiled language and optimizable state-representation backends, but also from ns-3's event-driven simulation core and its memory-optimized design. As demonstrated in Sec.~\ref{sec:3}, \textit{Q2NS} exhibits superior performance compared to representative state-of-the-art approaches in realistic networking scenarios, including entanglement-intensive workloads such as cluster-state preparation and entanglement-swapping chains. In addition, through the introduction of a dedicated visualization tool, \textit{Q2NSViz}, \textit{Q2NS} facilitates the development and intuitive understanding of quantum networking dynamics. Notably, \textit{Q2NSViz} provides a dual view of the physical network graph and entanglement-induced connectivity graph. By integrating entangled-state manipulation rules, \textit{Q2NSViz} enables researchers and educators to explore how entanglement resources are transformed and consumed throughout protocol execution. 

In summary, through its architectural principles and integration with ns-3, \textit{Q2NS} aims to provide a flexible and scalable platform, accessible to a broad community of researchers across both quantum and classical communities. 

The contributions of this work can be summarized as follows:
\begin{itemize}
    \item[i.] We introduce the \textit{Q2NS} architectural principles, design rationale, and core modules, with a plug-in architecture supporting heterogeneous quantum state representations;
    \item[ii.] We present \textit{Q2NSViz}, a visualization tool that captures both physical and entanglement-induced connectivity graphs with built-in entanglement-manipulation rules, broadening accessibility to quantum networking research;
    \item[iii.] We perform comprehensive benchmarks against qns-3, another quantum network simulator built on top of ns-3, showing that \textit{Q2NS} achieves superior computational efficiency even in entanglement-intensive scenarios, such as cluster state preparation and multi-hop entanglement distribution via swapping chains;
    \item[iv.] We validate our framework through two relevant case studies, illustrating \textit{Q2NS}'s ability to simulate realistic hybrid quantum-classical network scenarios.
\end{itemize}

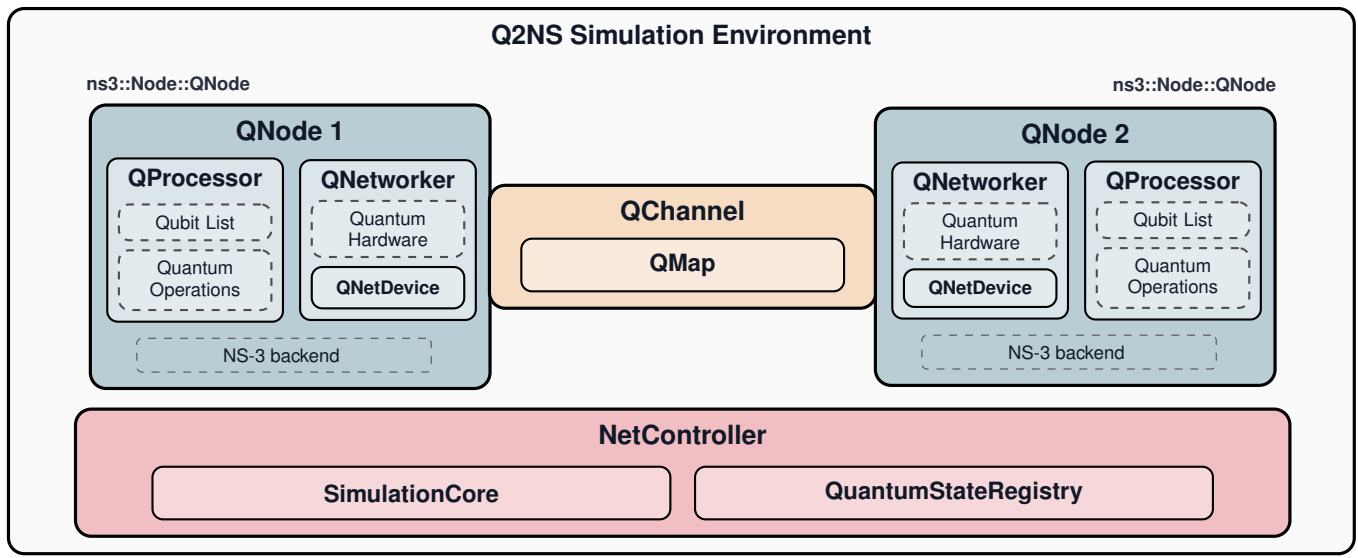
\begin{figure*}[t]
    \centering
    \input{Tikz/Fig1.tex}
    \caption{ 
        Graphical representation of the \textit{Q2NS} simulation environment, illustrating the main entities (NetController, QNodes, QChannels) and their components. The diagram highlights the simulator's modular architecture, the separation of concerns and the interactions between modules and their hierarchies within the simulation framework.
    }
    \label{fig:1}
    \hrulefill
\end{figure*}

\subsection{Related Works}
Although several quantum network simulators have emerged in recent years, existing tools do not provide both comprehensive physical accuracy and broad accessibility to the networking community. Most importantly, they often lack a native and faithful co-simulation capability that tightly couples quantum operations with the classical communication stack, required by realistic protocol execution. 

For instance, \textit{NetSquid} \cite{coopmans2021netsquid} offers detailed modeling of quantum operations and noise within an event-driven simulation framework. However, it is closed-source and abstracts classical communication, rather than explicitly modeling and simulating the underlying network stack or transport protocols. This makes it ill-suited for detailed studies of the interplay between the classical and Quantum Internet. 

\textit{SeQUeNCe} \cite{wu2021sequence} provides a modular quantum protocol stack and discrete-event control. However, it is coupled to specific architectural assumptions and lacks native integration with classical network layers.

Conversely, \textit{QKDNetSim} \cite{dervisevic2024qkdnetsim} and \textit{QKDNetSim+} \cite{soler2024qkdnetsimplus} extend \textit{ns-3} \cite{ns3} with QKD primitives. However, the base simulator and its improvements are narrow in focus and do not actually model quantum mechanics. 
The recently developed \textit{ns-3 Quantum Network Simulation Module}~\cite{ns3quantum2025} extends ns-3 more deeply, but is currently specialized for QKD rather than built for general-purpose quantum network simulation. In addition, no available work presents its architectural design. 

qns-3~\cite{Lin2025CFA} takes this further with a general-purpose, tensor-network-based extension of ns-3. Tensor networks can be powerful in low-entanglement regimes, but their runtime grows rapidly in entanglement-intensive scenarios, which are common and critical in quantum networking. The authors mitigate this with Control Flow Adaptation (CFA), replacing measurements with case-specific unitaries that maintain compact tensor networks. However, this approach is difficult to generalize and can depend on nonphysical abstractions, such as instantaneous global qubit access without classical signaling. A further exploration of when such abstractions are clear to derive and offer advantages would be valuable. Nevertheless, it remains unclear whether this is generalizable enough for a full classical-quantum network simulator. Notably, \textit{Q2NS}'s modularity and extensibility make implementing a tensor-network state representation relatively straightforward, enabling further exploration of tensor networks and CFA in quantum networking simulations. Table~\ref{tab:1} provides a summary of the above simulator comparison\footnote{A stable open-source release of \textit{Q2NS} developed within the scope of this work is planned following the completion of the peer-review process. The scripts required to reproduce the results can be made available to the reviewers upon request.}.

\section{Architectural Principles}
\label{sec:2}

\textit{Q2NS} is largely inspired by recent advances in quantum network architectural modeling \cite{CalCac-25}.
Indeed, the double ``N'' in \textit{Q2NS} is not merely nominal, it reflects a fundamental design principle introduced in \cite{CalCac-25}. Specifically, the effective management and utilization of entanglement resources demands not only a well-defined architecture, but also a distinct network-control layer that is clearly decoupled from in-network operations \cite{CalCac-25}. Accordingly, \textit{Q2NS} explicitly implements this decoupling, with the first ``N'' in the acronym referring to the network control and the second ``N'' to the in-network operations, within a unified architectural framework. Building upon this foundational principle, the \textit{Q2NS} architecture adheres to three key design tenets:

\begin{itemize}
    \item[-] \textit{Separation of concerns:} each module is responsible for a specific class of functionalities, and does not depend on internal workings of other modules.
    \item[-] \textit{Abstraction:} each module exposes a well-defined interface, enabling controlled interaction among components.
    \item[-] \textit{Flexibility:} the framework can be easily extended or modified, enabling new functionalities to be added or existing ones to be replaced as needed.
\end{itemize}

The following details how these architectural principles are instantiated within \textit{Q2NS} through its modular design coupled with a network-control layer and integration with ns-3.

\subsection{Entities and Modules}

The core of our architecture is modularity. The simulator is designed as a collection of self-contained modules that neatly interact via well-defined interfaces. Each module encapsulates a distinct logical functionality and can be easily extended or replaced without affecting the rest of the system. This separation of concerns establishes a clear boundary between logical and implementation-specific functionalities, enabling straightforward replacement of system modules or control mechanisms. 
The main entities are represented in Fig. \ref{fig:1} and consist of: \texttt{NetController}, \texttt{QNodes}, and \texttt{QChannels}. Each entity is composed of different modules with their own specific class of functionalities: 

\begin{itemize}
    \item \textbf{\texttt{NetController}}: is the super-entity overseeing the simulation and capturing the need for centralized network-control logic\footnote{The control logic is \textit{logically centralized} but, in real network deployments, may be \emph{physically distributed} via federated implementations, as discussed in \cite{CalCac-25}.}. Its functionality is divided into two modules: the \texttt{SimulationCore}, which manages the simulation environment and coordinates initialization and module interactions, and the \texttt{QStateRegistry}, which maintains and provides access to the single source of truth for all quantum states in the network. States are represented by \texttt{QState}, which provides a unified interface supporting multiple quantum-state representations. It currently integrates three backends: ket and density matrices using \textit{Quantum++}\cite{Gheorghiu2018Quantumpp} and stabilizers using \cite{de2022fast}'s quadratic form expressions approach. The design allows users to easily extend support to additional state representations. In essence, the \texttt{NetController} is the orchestrator of the network.
    
    \item \textbf{\texttt{QNode}}: is the entity modeling a network node, capable of performing quantum and classical operations. It comprises two modules: \texttt{QProcessor} and \texttt{QNetworker}, reflecting the separation of concerns described in the following section. \texttt{QNode} inherits from the classical ns-3 \texttt{Node} class, enabling native participation in the full classical protocol stack, and the use of ns-3's realistic channel models, without any additional integration layer. In essence, \texttt{QNode} acts as a task executor within the simulation framework.
    
    \item \textbf{\texttt{QChannel}}: models the quantum physical medium between nodes. It simulates realistic quantum-state transmission via the \texttt{QMap} module. This applies a Completely Positive Trace Preserving (CPTP) map to model the channel’s effects on the transmitted state. Importantly, this process can degrade key physical properties such as fidelity and entanglement.
\end{itemize}

\subsection{QProcessor and QNetworker}
As aforementioned, \texttt{QNode} is constituted by two main modules:
\begin{itemize}
    \item \textbf{\texttt{QProcessor}}: oversees the local quantum processing within a node. It performs typical local operations, including applying gates, measuring, and creating Bell pairs. In essence, the \texttt{QProcessor} is the brain of the \texttt{QNode}.
    
    \item \textbf{\texttt{QNetworker}}: oversees the quantum networking operations of the node. It is responsible for transmitting and receiving qubits using a \texttt{QNetDevice} interfaced with a \texttt{QChannel}. This procedure mirrors the classical packet transmission mechanism in ns-3, as both \texttt{QNetDevice} and \texttt{QChannel} classes inherit directly from their classical ns-3 counterparts. In essence, \texttt{QNetworker} acts as a communication functional block within the \texttt{QNode}, by mediating all qubit transmissions to and from the node. Internally, the \texttt{QNetworker} maintains a list of \texttt{QNetDevices} each connected to a \texttt{QChannel}. When a request is issued, it selects the appropriate \texttt{QNetDevice} for that transmission. The chosen \texttt{QNetDevice} transmits the qubit on its associated \texttt{QChannel}, which may apply a CPTP map, delay, or loss, as previously described. At the receiver node, the process is reversed: the qubit is first delivered to the receiving \texttt{QNetDevice}, then moved to the \texttt{QNetworker}, and lastly stored into the receiving \texttt{QProcessor}. 
\end{itemize}

Beyond its modularity and centralized control logic, \textit{Q2NS} leverages ns-3’s event-driven networking semantics to faithfully model classical signaling latency, including congestion-induced delays. This capability is essential for studying the scalability of quantum networks, where the tight coherence-time constraints of quantum memories must be reconciled with the variable delays induced by the classical network\cite{CacCalVan-20,IllCalMan-22}.

\subsection{Visualization Tool: Physical vs Entanglement Graph}
Another core aspect of \textit{Q2NS} is its ability to broaden access to quantum networking research. While leveraging ns-3 facilitates adoption within the classical networking community, we have developed a visualization tool, \textit{Q2NSViz}, to further broaden accessibility. This tool works seamlessly both within \textit{Q2NS} simulations and as a standalone application, enabling users with little to no programming background to create and explore quantum network animations. 

\begin{figure}[t]
    \centering
    \input{Tikz/Fig2.tex}
    \caption{Visualization Flow: \textit{Q2NSViz} $\rightarrow$ Trace File $\rightarrow$ Viewer. The generated logs correspond to an ``at-the-source'' entanglement generation scheme, where Alice creates an entangled pair and transmits one half to Bob.}
    \label{fig:2}
    \hrulefill
\end{figure}
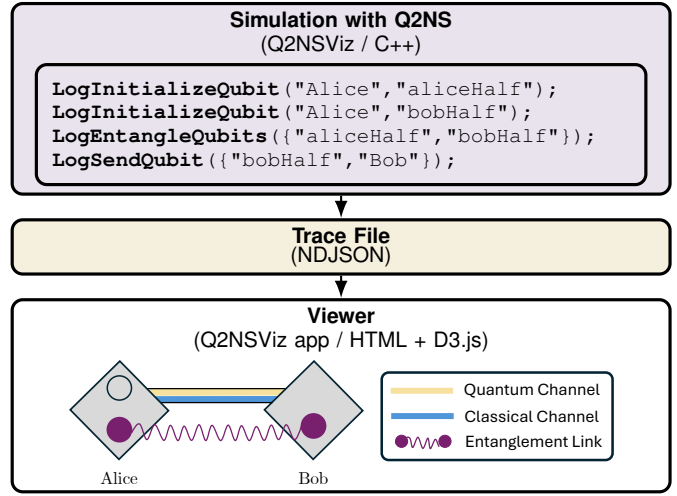

The visualization captures the physical network graph and tracks how quantum and classical bits propagate across links. It also includes the entanglement-induced connectivity graph \cite{IllCalMan-22, MazCalCac-25, CacPelIll-25} and incorporates built-in entanglement-manipulation rules, to explore how quantum correlations evolve, a crucial aspect of designing the Quantum Internet. This dual perspective enables researchers and educators to communicate phenomena ranging from the basics of quantum networking to the nuances of their current work.

We have developed a new and easily extensible trace format based on NDJSON, tailored to quantum-specific processes, such as qubit initialization, entanglement generation, measurement, and transmission. We also provide with the trace format a custom \textit{Q2NSViz} HTML/JavaScript animation application for interactive visualization. Traces can be generated programmatically via the \textit{Q2NSViz} API or created artificially via a dedicated command-line tool, enabling network animations without running a complete simulation and without manually crafting log files from scratch. The overall workflow is illustrated in Fig.~\ref{fig:2}. A version based on the widely used \textit{NetAnim}, designed to animate trace files for ns-3~\cite{ns3}, may be considered for future work. However, its legacy design, lack of maintenance, and limited extensibility make it less flexible for the expressive visualizations required in quantum networking.

Within the visualization tool, some traces and example scenarios are pre-loaded, providing recurring setups and representative use cases for entanglement-based networking. Remarkably, the animation app also embeds common entanglement-manipulation rules, correctly representing how BSMs, teleportation, and correction operations affect entanglement relations across network nodes. 
In particular, the latest version of \textit{Q2NSViz} includes built-in support for graph states, a family of multipartite entangled states widely studied and very interesting in quantum computation and communication for its rich entanglement structures and properties \cite{Hein2004GraphStates,Hein2006GraphStatesReview, MazCalCac-25,CheIllCac-25}.

Indeed, since graph states admit a unique correspondence to an associated graph, \textit{Q2NSViz} provides an intuitive visualization of such states as well as their behavior under local measurements, i.e., local complementation rules induced by single-qubit Pauli measurements \cite{Hein2006GraphStatesReview}. An example of such manipulation rules is provided in Sec.~\ref{sec:4}, where a simulation study of a quantum local area network (QLAN) is paired with its corresponding representation in the visualizer.

\section{Performance and Comparative Analysis}
\label{sec:3}

We present a comprehensive evaluation of \textit{Q2NS} through comparative benchmarks against qns-3, the closest openly available ns-3-based simulator with comparable classical-network integration and event-driven execution semantics. We do not include a direct performance comparison against \textit{NetSquid} \cite{coopmans2021netsquid}, as it is closed-source and provides limited native support for full classical protocol-stack dynamics, which makes a head-to-head benchmark with ns-3-based co-simulation frameworks methodologically non-uniform. For completeness, we note that the work introducing qns-3 reports a performance advantage over \textit{NetSquid} for the same entanglement-swapping workloads \cite{Lin2025CFA} considered in this section, where \textit{Q2NS} consistently and significantly outperforms qns-3.

Through entanglement-intensive scenarios, we assess \textit{Q2NS}'s extensibility and performance advantages in both the generation and manipulation of entangled states. Specifically, we provide:
\begin{enumerate}[label=\roman*.]
    \item an evaluation of \textit{Q2NS}'s extensibility and favorable performance relative to qns-3 for 1D and 2D cluster-state preparation;
    \item a comparative analysis against qns-3 for entanglement-swapping chains.
\end{enumerate}
The results in this section report wall-clock runtime and peak memory for each tested setup (i.e., for each parameter setting and backend), with summary statistics obtained from repeated executions to quantify measurement variability and compute confidence intervals on the mean. In particular, we repeated independent trials until the $95\%$ confidence interval for the mean value achieved a margin of error below $2\%$ over consecutive trials.

For completeness, in the conducted analysis, we distinguish between two phases in each ns-3 simulation. The first is a configuration phase, where the simulator builds the physical network graph, instantiates node and channel objects, installs applications, and schedules initial events. We refer to this duration as the \textit{configuration time}. During this phase, the simulation environment is allocated and initialized, but the discrete-event engine does not advance: no simulation time elapses and no physical mechanism is executed or modeled. The second phase starts when \texttt{Simulator::Run()} is invoked and ends when \texttt{Simulator::Stop()} is invoked or when the event queue becomes empty. This interval corresponds to the actual execution of the discrete-event loop, i.e., the processing of scheduled events and any associated modeling of the underlying physical mechanism. We refer to this duration as \textit{simulation time}. 

We measure both phases through the wall-clock time, by using the C++ standard library \texttt{std::chrono}. The choice of wall-clock runtime was based on two reasons: (i) it matches the metric used in the original qns-3 study~\cite{Lin2025CFA}, enabling a methodologically consistent comparison, and (ii) it reflects the realistic end-to-end execution time that users can expect in practice rather than an idealized compute-only metric. We measure memory usage with the simulator process’s maximum resident set size (RSS), i.e., the maximum amount of physical (resident) memory occupied by the process at any time during execution. This metric captures the worst-case memory footprint required for the simulation to complete successfully. 

It is important to observe that qns-3's main dependency, ExaTN/TAL-SH~\cite{10.3389/fams.2022.838601} -- a C++ library specialized in the efficient representation and contraction of tensor networks -- is primarily intended for Linux environments, while \textit{Q2NS} can run on both MacOS and Linux. To simplify installation and ensure a methodologically consistent comparison -- in particular, without disadvantaging qns-3 -- all experiments are conducted inside a Dockerized Ubuntu 22.04 LTS environment. The container was configured with 6 virtual CPU cores, 16\,GB of memory, and executed on a MacBook equipped with an Apple M3 processor and 24\,GB of unified memory. Although multiple virtual cores were allocated, we empirically observed that concurrent executions increased individual execution times, likely due to virtualization overhead and shared-resource contention within the container. Therefore, to obtain stable and comparable wall-clock measurements, all simulations were executed sequentially. Furthermore, we compiled and ran qns-3 and \textit{Q2NS} within the same containerized environment to ensure that environment-dependent effects (compiler, standard libraries, optimization flags, and memory layout) did not confound the measured performance differences\footnote{Indeed, the execution environment can influence wall-clock timing, changing how quickly the simulator processes events, \textit{without} altering the actual events the ns-3 engine schedules or the underlying physics simulated. Differences in compilers, standard libraries, optimization levels, and memory layout can also lead to variations in behavior, including segmentation faults or other crashes that may not occur in a different environment. These are the reasons behind our choice to compile and run qns-3 and \textit{Q2NS} within the same container}.

In light of these environment sensitivities, we also monitored robustness and failure modes across increasing problem sizes. We observed that some of qns-3 runs terminated with segmentation faults or backend-level limitations in the ExaTN/TAL-SH tensor-network stack upon reaching large enough problem sizes. In contrast, the only occurrences of failure observed for \textit{Q2NS} were segmentation faults at the known scalability limits of the \textit{Quantum++} backends \cite{Gheorghiu2018Quantumpp}. Notably, several qns-3 failures occurred despite remaining well within the time and memory budgets under which  \textit{Q2NS} and independent tests of resource-allocation completed successfully, suggesting that these terminations were not due to host resource exhaustion. While further studies would be necessary to confirm, this behavior could indicate toolchain- and integration-level sensitivities in the qns-3 build and configuration pipeline, which requires tighter coordination across external components than \textit{Q2NS}. If so, this suggests that qns-3 may be more prone to environment-dependent behavior than \textit{Q2NS}.

\begin{table}[b]
\centering
\caption{Asymptotic theoretical runtime scaling of the different backends for cluster-state simulations with $n$ qubits.}
\begin{tabular}{lll}

\toprule
\textbf{Simulator} & \textbf{State Representation} & \textbf{Theoretical Scaling} \\
\midrule
\textit{Q2NS} -- Ket &
State-vector ($\mathbb{C}^{2^n}$) &
$O(2^{n})$ \\

\textit{Q2NS} -- DM &
Density matrix ($\mathbb{C}^{2^n \times 2^n}$) &
$O(4^{n})$ \\

\textit{Q2NS} -- Stab &
Stabilizer &
$O(n^{2 - 4})$ \\

qns-3 &
Tensor network &
$\mathrm{poly}(n)\,2^{O(n)}$ \\
\bottomrule
\end{tabular}
\label{tab:2}
\end{table}

\subsection{qns-3 Comparison: Cluster States}
As mentioned above, we first consider a comparative performance benchmark based on cluster states. Cluster states are a canonical multipartite resource for entangled-based quantum networking~\cite{PirkerDuer2019,HahnPappaEisert2019,FreundPirkerVandreDuer2025} and measurement-based quantum computation~\cite{RaussendorfBriegel2001,RaussendorfBrowneBriegel2003}. Consequently, the ability to simulate them efficiently is a key requirement for any quantum network simulator. Moreover, cluster states provide a natural testbed for tensor-network-based approaches, such as those adopted in qns-3, since their entanglement properties are largely determined by the underlying lattice graph. Appendix~\ref{app:1} provides the necessary background on cluster states.

In this benchmark, we use cluster-state preparation and evaluation as a protocol-agnostic workload to stress elementary quantum operations on highly entangled resources, while avoiding dependence on a specific networking application. Furthermore, this use case showcases the modularity and flexibility of \textit{Q2NS}. Specifically, each quantum-state backend is implemented as an independent plug-in module and can be selected or replaced without changes to the rest of the simulator. This design has allowed us to incorporate and evaluate multiple, distinct quantum-state representations in this work and naturally supports further optimizations as well as the addition of entirely new backends. 

Appendix~\ref{app:1} also discusses the theoretical scaling of the considered quantum-state representation backends, including tensor-network-based approaches. This review has been included to further substantiate the reliability of the simulation results, which, as shown below, closely match the expected scaling behavior. The main takeaways are summarized in Table~\ref{tab:2}.

\textbf{Simulation setup.} To isolate backend-dependent effects (quantum-state representation and update rules) from network-level dynamics, we consider a minimal ns-3 configuration consisting of a single node and no classical or quantum channels. Given the simplicity of the ns-3 configuration and the dominance of quantum-state processing in this study, the configuration phase was empirically negligible. Accordingly, we report only the simulation-time component.

We adopt a backend-specific notion of \textit{state evaluation}, whose purpose is to trigger the dominant, worst-case computational cost of each representation on highly entangled states, while, as discussed above, remaining protocol-agnostic. Accordingly, \textit{evaluation} is not meant to produce a specific protocol-oriented output metric. Rather, it is a terminal operation applied after cluster-state preparation to stress the backend with a representative worst-case workload that might arise during protocol execution. For tensor-network simulation (qns-3), this terminal step corresponds to contracting the entire tensor network, which is well-known to dominate runtime in highly entangled regimes. For \textit{Q2NS} backends in (state-vector, density-matrix, or stabilizer), the analogous terminal step is implemented by measuring all qubits, which forces full state access and state-update operations in the selected representation. Although these operations are representation-specific, they serve the same methodological role of finalizing a worst-case workload, thereby enabling a fair end-to-end wall-clock comparison across simulators.

\begin{figure*}[t]
    \centering
    \begin{subfigure}{0.48\textwidth}
        \centering
        \input{Tikz/Fig3a}
        \caption{}
        \label{fig:3a}
    \end{subfigure}
    \hfill
    \begin{subfigure}{0.48\textwidth}
        \centering
        \input{Tikz/Fig3b}
        \caption{}
        \label{fig:3b}
    \end{subfigure}
    \caption{Simulation time (a) and memory (b) scaling of cluster state preparation and evaluation for \texttt{qns-3} versus \textit{Q2NS} (ket, density matrix, and stabilizer backends). Insets show the full range of the stabilizer backend and its best fit models, which extends beyond the axis limits that were chosen to clearly depict the other backends. Axes are logarithmic. Error bars represent $\pm 1$ standard deviation.}
    \label{fig:3}
    \hrulefill
\end{figure*}
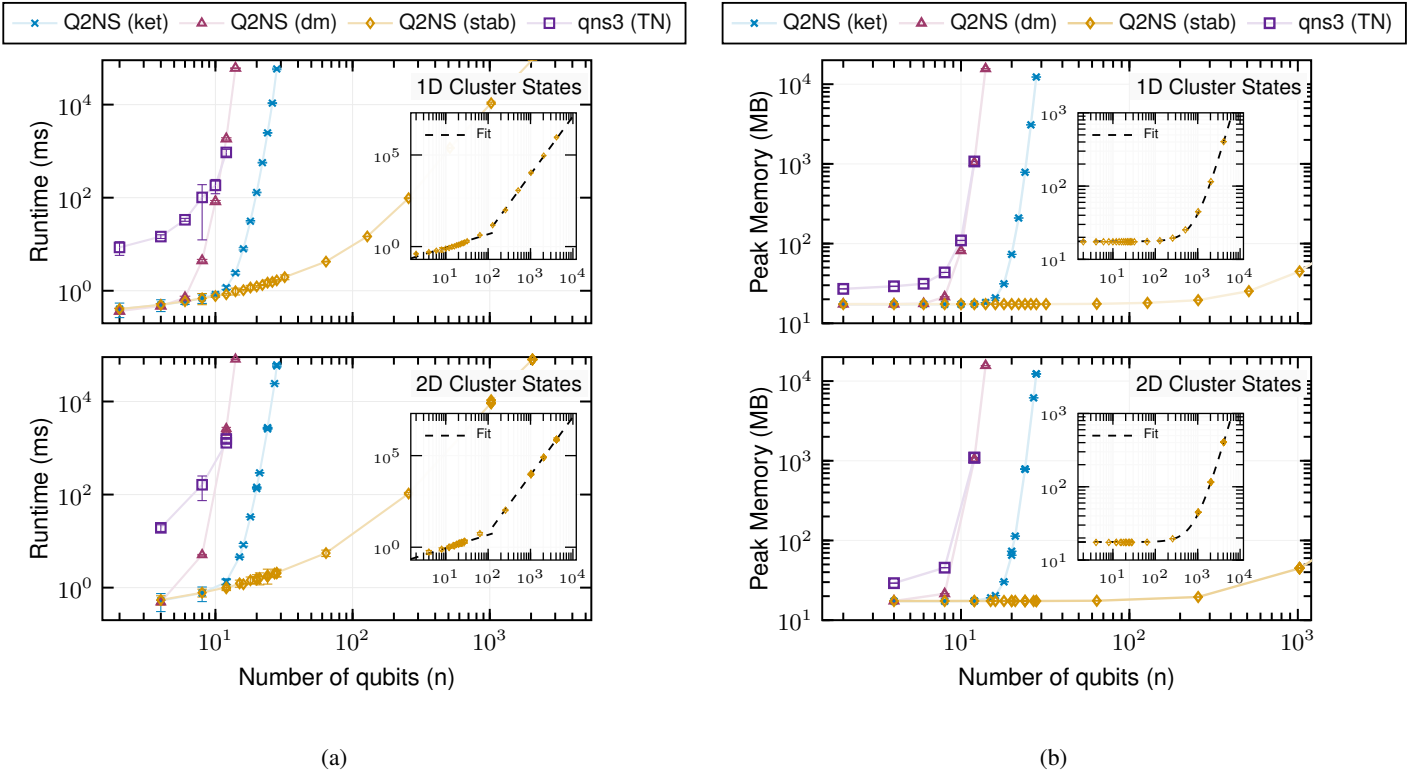

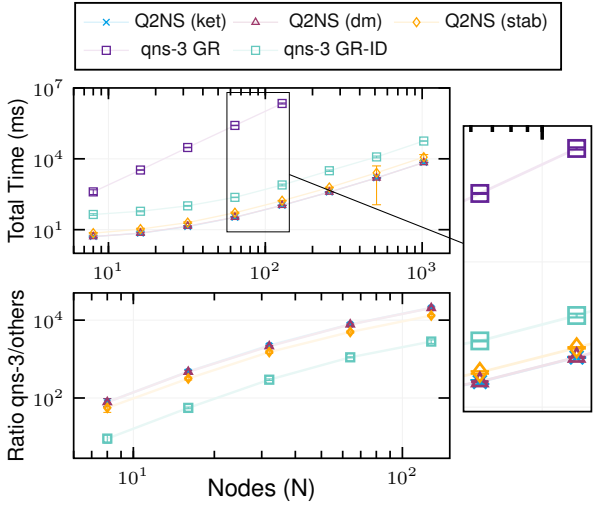
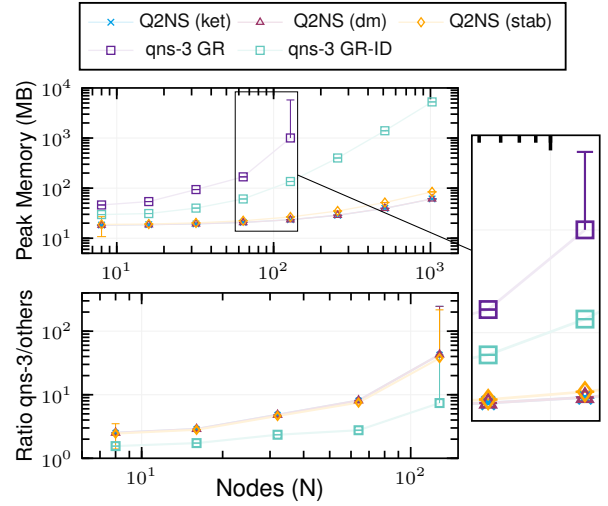
\begin{figure*}[t]
    \centering
    \begin{subfigure}{0.48\textwidth}
        \centering
        \input{Tikz/Fig4a}
        \caption{Total times for all simulators on top. The ratio of the qns-3 CFA total time to the rest on the bottom: the higher the value, the better the given simulation compares to qns-3 CFA.}
        \label{fig:4a}
    \end{subfigure}
     \hfill
    \begin{subfigure}{0.48\textwidth}
        \centering
        \input{Tikz/Fig4b}
        \caption{Peak memory usage for all simulators on top. The ratio of the qns-3 CFA total time to the rest on the bottom: the higher the value, the better the given simulation compares to qns-3 CFA.}
        \label{fig:4b}
    \end{subfigure}
    \caption{Comparison of qns-3 and \textit{Q2NS} for entanglement swapping chains as the number of network nodes increases, demonstrating \textit{Q2NS}'s higher performance. Axes are logarithmic. Error bars denote $\pm 1$ standard deviation. Note that qns-3 CFA-GR has a maximum size of $N = 128$ owing to the 1-hour time cutoff for single run data collection. For this reason the ratio plots on the bottom only extend to $N = 128$.}
    \label{fig:4}
    \hrulefill
\end{figure*}

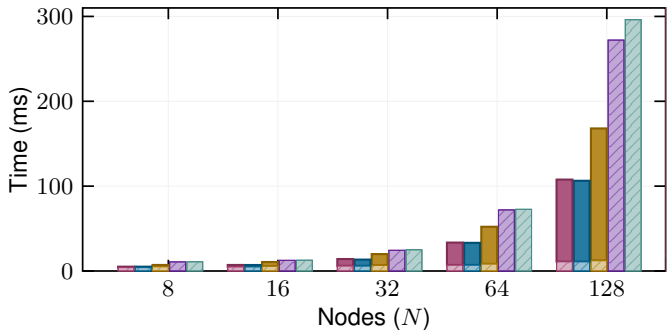
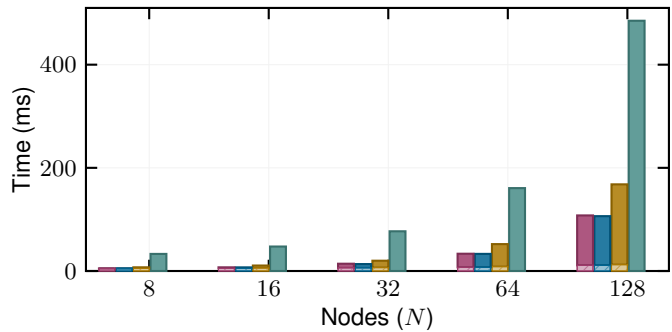
\begin{figure*}
    \centering
    
    \begin{subfigure}{0.48\textwidth}
        \centering
        \input{Tikz/Fig5a}
        \caption{qns-3 CFA-GR and qns-3 CFA-GR-ID configure times compared to stacked \textit{Q2NS} configuration and simulation times for each backend, demonstrating that \textit{Q2NS} completes full execution before any form of qns-3 has configured and can begin executing the simulation.}
        \label{fig:5a}
    \end{subfigure}
    \hfill
    \begin{subfigure}{0.48\textwidth}
        \centering
        \input{Tikz/Fig5b}
        \caption{qns-3 CFA-GR-ID simulation time compared to stacked \textit{Q2NS} configuration and simulation times for each backend, demonstrating that \textit{Q2NS} completes full execution before qns-3 CFA-GR-ID could, even ignoring its configuration time.}
        \label{fig:5b}
    \end{subfigure}
    
    \caption{Comparison of both versions of qns-3 CFA and \textit{Q2NS} for an entanglement swapping chain with increasing number of network nodes, broken down by mean configuration and simulation time. The x axes are logarithmic. Note that in (b), qns-3 CFA-GR data is not shown as it is orders of magnitude larger, making it impossible to plot on a linear scale with the rest of the data. qns-3 CFA-GR also has a maximum size of $N = 128$ owing to the 1-hour time cutoff for single run data collection.}
    \label{fig:5}
    \hrulefill
\end{figure*}

\textbf{Simulation results.} Fig.~\ref{fig:3a} shows simulation scaling for 1D and 2D cluster states. The qns-3 curves rise sharply and terminate at $n = 12$, reflecting the tensor-construction limits of the ExatN/TAL-SH backend, at least within the container's environment. qns-3 offers several contraction heuristics, which decide the order of tensor contraction based on different algorithms. The ``greedy'' option was consistently the most efficient when the different heuristics were used for the same cluster state creation. All options demonstrated a ceiling of at most $n = 12$, or even smaller in some cases, during the testing. The trend is consistent with exponential growth, but the small-$n$ ceiling makes a precise asymptotic fit difficult. The \textit{Q2NS}-density-matrix and ket backends follow the expected scaling trends of $4^n$ and $2^n$ respectively, as discussed in Appendix~\ref{app:1}. They reach maximum system sizes of $n = 14$ and $n = 28$ qubits, respectively, before exceeding the available memory limits. 

The most efficient is the \textit{Q2NS}-stab backend, exhibiting polynomial growth for both 1D and 2D cluster states. The observed behavior is expected by accounting for the $O(n^{1 - 2})$ operations required by state preparation combined with the $O(n^{2 - 4})$ all-qubit measurement procedure. The runtime data demonstrates a clear change in behavior when crossing from small to large problem sizes, which is reasonable given the mix of factors that can play into the runtime, including those discussed in Appendix~\ref{app:1}, each of which likely scale differently. As such, we performed fits for $n < 128$ and $n \geq 128$ separately, resulting in a scaling of $n^{0.723}$ for $n < 128$ and $n^{3.17}$ for $n \geq 128$. This piecewise fit had an overall $R^2 = 0.999$, where $R^2$ denotes the coefficient of determination quantifying the agreement between the fit and the data~\cite{MontgomeryRunger2018}. To ensure statistical rigor, we also tested a single global fit over the entire data set. This alternative performed worse, yielding a lower $R^2$ ($R^2_{\text{global}} = 0.963$) and a substantially higher leave-one-out RMSE. Our proposed fitting strategy reduces the leave-one-out RMSE by nearly one order of magnitude, indicating better predictive accuracy and robustness.

Furthermore, for all \textit{Q2NS} backends, there is no discernible difference between 1D- and 2D-cluster simulations for a given $n$, even for the largest system sizes. Accordingly, the data are best fit by treating $n$ as the sole independent variable, rather than the lattice dimensions $R$, $C$, or $\max(R,C)$, introduced in Appendix~\ref{app:1} where $n = RC$. 

On the other hand, for qns-3, the mean runtime for 2D cluster states were more than 1 standard deviation higher than the 1D clusters for $n=4$ and $n=12$. The 2D mean time was larger than the 1D mean time for $n = 8 \ (1\times8\ \text{lattice \, vs.}\ 2\times4\, \text{lattice})$ as well, but both had a large enough standard deviation to overshadow this difference. We note that when we compare 1D and 2D cluster-states at the same qubit number $n$, we instantiate the 2D case using all feasible lattice factorizations $R\times C$ such that $RC=n$ (with $R,C\geq 2$). For example, at $n=4$ the 1D cluster corresponds to a $1\times4$ lattice, whereas the 2D case is realized as $2\times2$. At $n=12$, the 1D cluster is $1\times12$, while the 2D realizations include $2\times6$ and $3\times4$. As also reflected in Fig.~\ref{fig:3}, 1D clusters are defined for any $n\geq 2$, whereas the smallest 2D cluster considered is $n=4$ (a $2\times2$ lattice). Given qns-3's limitation of simulating 12 qubits or less, only a small set of 2D lattice instances can be compared against the corresponding 1D cases, and all of them remain in the small-size regime. Consequently, these data do not support a robust assessment of asymptotic differences in performance between 1D and 2D clusters, nor of the overall scaling behavior. 
Nevertheless, the emerging trend -- together with the theoretical analysis reviewed in Appendix~\ref{app:1} -- support  that 2D cluster instances should exhibit exponentially worse scaling than 1D clusters, even under optimized contraction strategies.

Memory usage in Fig.~\ref{fig:3b} exhibits similar trends. qns-3 consistently requires more memory than any \textit{Q2NS} backend and demonstrates approximately exponential growth with $n$ before reaching the 12-qubit limitation. \textit{Q2NS}-density matrix exhibits the expected $O(4^n)$ scaling, initially consuming less memory than qns-3 but converging by the 12-qubit threshold. While both density matrices and kets inherently demonstrate computational inefficiencies, \textit{Q2NS}-kets maintain substantially lower memory consumption than qns-3 at comparable problem sizes, despite exhibiting the expected $O(2^n)$ growth. In contrast, \textit{Q2NS}-stab demonstrates more favorable memory scaling, exhibiting significantly slower growth compared to the exponential scaling of other backends even up to $n = 4096$ qubits. The backend is best fit by a constant-offset power law of the form $a + bN^k$, with an empirical scaling of $O(n^{1.97})$, a constant offset of $17.6$ MB, and with an $R^2 = 1.0$ in the measured range. Comparison to a piece-wise model, as used for the runtime, demonstrated negligible statistical difference while providing nearly identical results. So the simpler, global model was chosen. 

For qns-3, the mean memory consumption for 2D clusters exceeded the 1D mean by more than one standard deviation in all test-cases. While practical improvements to qns-3's contraction orderings may extend the observed limits, exact simulation of 2D and higher-dimensional cluster states will remain exponential in the minimal cut size~\cite{VerstraeteCirac2004PEPS,Schuch2007ComplexityOfPEPS}.

The performance results for this case-study are consistent with the theoretical scaling reported in Table~\ref{tab:2} and can be summarized as follows:

\begin{enumerate}
  \item \textbf{qns-3 (tensor-network backend)} exhibits rapidly increasing runtime and memory usage, consistent with an exponential dependence on $n$ due to the effective contraction width. In the container's environment, it fails to contract cluster states beyond $n = 12$ qubits.
  \item \textbf{\textit{Q2NS}-stab (stabilizer backend)} exhibits  a polynomial-time behavior,  with runtime well described by a $O(n^{3.17})$ scaling for larger problem sizes. Its peak memory footprint is significantly lower than that of all other tested backends, and follows an $O(n^{1.97})$ trend. There was no observed limit on the tractability of this backend for at least $n \leq 4096$ qubits.
  \item \textbf{\textit{Q2NS}-ket (state-vector backend)} follows the expected $O(2^n)$ runtime and memory scaling, remaining tractable for $n \leq 28$ qubits in our environment.
  \item \textbf{\textit{Q2NS}-dm (density matrix backend)} follows the expected $O(4^n)$ runtime and memory scaling, remaining tractable up for $n \leq 14$ qubits in our environment.
\end{enumerate}

\begin{table}[t]
\centering
\caption{Estimated scaling exponents $\beta$ for total time, configuration time, simulation time, and peak memory usage obtained from power-law fits of the form $Y \propto N^{\beta}$ in log--log space. We report fits over (i) the full range of tested $N$ and (ii) also for the largest-$N$ subset to mitigate finite-size effects. For \textit{Q2NS}, each entry reports the range across the ket, density-matrix, and stabilizer backends. Additional methodological details and fit results are provided in Appendix~\ref{app:2}.}
\setlength{\tabcolsep}{3.0pt}
\renewcommand{\arraystretch}{1.15}
\begin{tabular}{llcccc}
\toprule
\textbf{Simulator} & \textbf{Region} &
$\boldsymbol{\beta_{\text{total}}}$ &
$\boldsymbol{\beta_{\text{conf}}}$ &
$\boldsymbol{\beta_{\text{sim}}}$ &
$\boldsymbol{\beta_{\text{mem}}}$ \\
\midrule
\multirow{2}{*}{\textbf{\textit{Q2NS}}} 
  & High $N$ & 2.00--2.04       & 0.88       & 2.05--2.08 & 0.46--0.56 \\
  & All $N$  & 1.52--1.56      & 0.55--0.57 & 1.80--1.86 & 0.23--0.30 \\
\midrule
\multirow{2}{*}{\textbf{qns-3 CFA-GR}} 
  & Mid $N$  & 3.10       & 1.74       & 3.10       & 1.71       \\
  & All $N$  & 3.11       & 1.18       & 3.12       & 1.05       \\
\midrule
\multirow{2}{*}{\textbf{qns-3 CFA-GR-ID}} 
  & High $N$ & 2.05       & 2.07       & 2.03       & 1.76       \\
  & All $N$  & 1.51       & 1.66       & 1.45       & 1.08       \\
\bottomrule
\end{tabular}
\label{tab:3}
\end{table}

\subsection{qns-3 Comparison: Entanglement Swapping}
To complement the above comparison against qns-3, we consider an additional case study based on entanglement swapping over a linear chain of $N$ quantum nodes: Alice and Bob connected by $N-2$ intermediate repeater nodes. This canonical repeater-chain workload provides a standard benchmark in quantum networking and exercises both quantum-state evolution and classical communication. In both simulators, the scenario is implemented using the respective built-in entanglement-swapping functionality, that can be installed on quantum nodes. Before comparing time and memory consumption, we examine the level of abstraction used by each simulator, since this strongly affects the class of networking problems that can be reliably addressed by the given simulators.

\textbf{Simulation setup.} Given the increased complexity of the network configurations in this case study, we measured both configuration and simulation times. We use the same set of backends as the previous study: kets, density matrices, and stabilizers in \textit{Q2NS} and the required tensor-network backend in qns-3.
In both simulators, nodes are connected through a single shared IPv6 CSMA LAN with static routing, and classical messages (when modeled) are exchanged via 1024-byte UDPv6 packets. With the default CSMA data rate of 10\,Mbps, we observe packet drops that led to a failed attempt at end-to-end entanglement distribution for $N \geq 435$. While the focus of this case study is on the simulators' efficiency rather than simulated protocol success probability, we nonetheless increase the data rate to 10\,Gbps to avoid losses and ensure reproducible timing measurements. 

The two simulators differ substantially in how they model classical signaling and EPR distribution within the swapping chain workload. In \textit{Q2NS}, we implement an explicit and physically grounded workflow: EPR pairs are distributed over modeled quantum links, and each repeater transmits the outcome of its Bell-State measurement (BSM) to Bob over the classical network, so that Bob can apply the required Pauli corrections. In contrast, qns-3 provides three entanglement-swapping implementations -- a baseline version and two variants that incorporate CFA optimization~\cite{Lin2025CFA} -- none of which exchanges BSM outcomes via the classical network. Instead, all qns-3 CFA implementations avoid classical branching by coherently encoding BSM outcomes into a \emph{globally shared} auxiliary qubit register, thereby replacing explicit dissemination of measurement results with global coherent bookkeeping accessible from any node without modeled transmissions. In addition, only one CFA variant models classical communication during the initial EPR distribution step, while the other CFA variant further reduces communication overhead, by creating Bell pairs and assigning their halves to neighboring nodes without explicitly modeling their distribution. Consequently, the performance gains reported for CFA in \cite{Lin2025CFA} primarily reflect a reduction in communication-driven event scheduling within ns-3 (BSM-result dissemination, and in the most aggressive variant also EPR distribution), rather than a general acceleration under fully explicit hybrid classical--quantum co-simulation. While this approach is perfectly valid for studying the abstract theory of entanglement swapping, it significantly restricts the types of simulations that can be performed, particularly those where the impact of various communication protocols and configurations is of primary interest. Moreover, CFA as presented in~\cite{Lin2025CFA} does not appear to be a readily generalizable approach. Accordingly, CFA should be interpreted as an optimization tailored to this specific entanglement-swapping abstraction, and its performance may not be representative of qns-3 under broader network simulations where protocol dynamics and communication latencies are first-order effects. Nevertheless, since entanglement swapping is a core primitive in quantum networking,  the availability of these optimized variants in qns-3 is valuable. 

Finally, as reported in \cite{Lin2025CFA}, the qns-3 baseline implementation exhibits exponential scaling with the chain length, while the CFA variants demonstrate polynomial behavior for the considered workload. For example, for a 16-node chain, the baseline requires approximately $1027\, \mathrm{s}$ ($\approx 17\, \mathrm{min}$), whereas the CFA versions reduce this to $3$--$5\,\mathrm{s}$ \cite{Lin2025CFA}. Accordingly, to provide a conservative comparison in favor of qns-3, we benchmark \textit{Q2NS} only against the two optimized CFA variants, hereafter denoted as qns-3 CFA-GR and qns-3 CFA-GR-ID. These labels indicate the use of a Global Register (GR) and Instantaneous Distribution (ID) of EPR pairs. 

\textbf{Simulation results.} Results are reported in Fig.~\ref{fig:4} in terms of total time and peak memory usage, while Fig.~\ref{fig:5} splits the total time into configuration and simulation time. Detailed breakdowns of the fits for these metrics are provided in Table~\ref{tab:3}.

As shown in Fig.~\ref{fig:4}, qns-3 CFA-GR exhibits significantly worse performance than other approaches in both time and peak memory usage. Notably, its configuration time alone exceeds the total execution time of any \textit{Q2NS} backend, as demonstrated in Fig.~\ref{fig:5a}. Due to the 1-hour cutoff imposed on the total runtime, we collected qns-3 CFA-GR data only up to $N \leq 128$. Larger values appear feasible given sufficient runtime budget, until memory limits are reached. 

In contrast, qns-3 CFA-GR-ID achieves performance closer to \textit{Q2NS}, although it remains roughly 6-8 times slower, with configuration times that still exceed the total execution time of comparable \textit{Q2NS} runs. This result is noteworthy because qns-3 CFA-GR-ID benefits from strong idealizations -- it does not model classical messaging for disseminating BSM outcomes and it abstracts EPR distribution as instantaneous --thereby substantially reducing communication-driven event scheduling. Despite these advantages, \textit{Q2NS} still attains higher performance gains under a more physically grounded workflow. Moreover, qns-3 CFA-GR-ID exhibits significantly higher peak memory usage than \textit{Q2NS}, resulting in execution failures for $N \geq 1900$. 

The results across all \textit{Q2NS} backends are nearly identical, as the maximum number of qubits in any state never exceeds 4, rendering all backends effectively equivalent for this scenario. No execution failures were observed for \textit{Q2NS} within the tested range, including verification runs up to $N = 8192$. The stabilizer backend had a slightly worse overall factor for both time and memory, though never more than 2 times larger. This indicates an initial overhead that is noticeable for small state sizes where kets or density matrices are still reasonably efficient, but is easily overcome by large state simulations as demonstrated by the cluster state simulations. This is exactly why having multiple backends is a valuable aspect of \textit{Q2NS} as no one representation is best suited for all possible scenarios.

To further substantiate these observations, we performed a scaling analysis summarized in Table~\ref{tab:3} and detailed in Appendix~\ref{app:2}. The parameter $\beta$ denotes the fitted scaling exponent. Thus, smaller values of $\beta$ denote more favorable scaling behavior. The analysis confirms that qns-3 CFA-GR performs worse, not just in magnitude, but in scaling. In particular, its total runtime scales at least an additional $O(N)$ factor worse than any other approaches, while behaving similarly in memory scaling to qns-3 CFA-GR-ID, though with a higher overall magnitude as can be seen in Fig.~\ref{fig:4b}. Despite the fact that the total runtime scaling is largely dominated by simulation rather than configuration time, as indicated by Table~ref{tab:3}, qns-3 CFA-GR's configuration times alone far surpasses the combined \textit{Q2NS} total times in both magnitude and scaling as can be seen in Fig.~\ref{fig:5a}. It is important to note that the fits for qns-3 CFA-GR rely on less data due to the fact the runtime for $N \geq 128$ exceeded the aforementioned 1-hour cutoff used for data collection. Consequently, these results should be interpreted with greater caution than the results for qns-3 CFA-GR-ID or \textit{Q2NS}, especially in indicating potential asymptotic scaling.

The comparison between \textit{Q2NS} and qns-3 CFA-GR-ID reveals more nuanced performance characteristics. \textit{Q2NS} exhibits substantially more favorable configuration-time and peak-memory scaling, while total runtimes are comparable over the range of $N$ for which qns-3 CFA-GR-ID completes. Interpreting the total run-time trends requires care, however, because qns-3 CFA-GR-ID benefits from the idealizations discussed above, thereby reducing communication-driven event scheduling within ns-3. Within this abstraction level, qns-3 CFA-GR-ID shows a marginal simulation-time scaling advantage that is approximately $O(N^{0.3 - 0.4})$ better than \textit{Q2NS} when using the full $N$ dataset. However, this advantage becomes practically irrelevant as qns-3 CFA-GR-ID encounters execution failures for $N \geq 1900$, likely due to its higher memory footprint. And within the range that qns-3 CFA-GR-ID can run, the magnitude of its runtime is consistently higher as can be seen in Fig.~\ref{fig:4}. In fact, the magnitude of its configuration time and simulation time each individually surpass the total time of \textit{Q2NS} as can be seen in Fig.~\ref{fig:5}. Additionally, qns-3 CFA-GR-ID's small scaling benefits may reflect finite-size effects rather than true asymptotic behavior, since, when the analysis is restricted to $N \geq 128$, both simulators exhibit similar simulation-time scaling, approximately $O(N^{2 - 2.1})$.

Before concluding this section, we emphasize that the repeated executions used to obtain confidence intervals on wall-clock runtimes should not be conflated with Monte Carlo trials of the modeled physical process. Monte Carlo evaluation involves $T$ statistically independent trials and, in the worst case, increases wall-clock runtime linearly with $T$. The motivation is that statistical estimation error typically decreases as $O(1/\sqrt{T})$ as $T$ grows. However, due to the statistical independence of trials, parallelization is trivial. If the runtime of a single simulator run is $O(f(N))$ and one performs $T$ trials distributed across $W$ workers, then the overall run time will now scale as $O(\frac{T}{W}f(N))$, up to constant parallelization overhead. Therefore, the practical cost of Monte Carlo evaluation is primarily determined by available compute resources rather than by an intrinsic limitation of the simulator.

Finally, we highlight that in the noiseless entanglement-swapping chain considered here, a single Monte Carlo trial is sufficient, since the final logical outcome is invariant up to Pauli corrections: Alice and Bob share a Bell pair regardless of the intermediate BSM outcomes. Therefore, multiple Monte Carlo trials are not strictly necessary as they produce identical logical results. In contrast, and as observed above, noisy configurations would typically require multiple trials to obtain statistically meaningful estimates in \textit{Q2NS}. While this specific subroutine of qns-3 can, in principle, mitigate explicit Monte Carlo repetition by coherently encoding branching over BSM outcomes into a global register and error channels into the tensor-network representation. Without these abstractions, encoding all possible measurement outcomes into the tensor network would cause the time and memory to explode exponentially, which is exactly what was observed by the non-CFA implementation from qns-3. For instance, in ~\cite{Lin2025CFA}, non-CFA qns-3 implementations scaled exponentially in both studies where it was used, namely entanglement swapping and nested entanglement distillation.
Nevertheless, the potential Monte Carlo advantage of CFA-based qns-3 variants is tightly constrained in practice: 
\begin{itemize}
    \item qns-3 CFA-GR scales approximately $O(N)$ less efficiently and exhibits a substantially larger constant prefactor,  about $80 \times$ - $110 \times$ in our measurements. As a result, even if \textit{Q2NS} were to rely on explicit Monte Carlo trials, it could execute on the order of $80$-$110 \,O(N)$ trials while matching the wall-clock time of a \textit{single} qns-3 CFA-GR run at the same $N$. Parallelizing these trials would further strengthen \textit{Q2NS}'s position, as discussed above.
    \item qns-3 CFA-GR-ID exhibits substantially higher peak memory usage and ultimately terminates with execution failures beyond a modest chain length, $N \geq 1900$.
    \item These CFA optimizations are implemented in qns-3 as specialized, built-in swapping applications and are not readily extensible to broader and more complex network functionalities or to arbitrary hybrid quantum-classical workflows. Moreover, the associated abstractions limit the ability to study the impact of realistic classical signaling, especially qns-3 CFA-GR-ID. Consequently, any reduction in the need for explicit Monte Carlo trials offered by CFA should not be assumed to generalize beyond this simulator-specific swapping abstraction. In contrast, \textit{Q2NS} retains practical efficiency across a wider range of tractable network sizes while supporting physically grounded co-simulation and modular backend extension.
\end{itemize}

\begin{figure}[t]
    \centering
    \input{Tikz/Fig6}
    \caption{ 
        Fidelity of the teleported state $\ket{+}$ versus distance between Alice and Bob, for different values of $T_{dep}$ and with idle or congested classical channels. Error bars represent $\pm 1$ standard deviation.
    }
    \label{fig:6}
    \hrulefill
\end{figure}
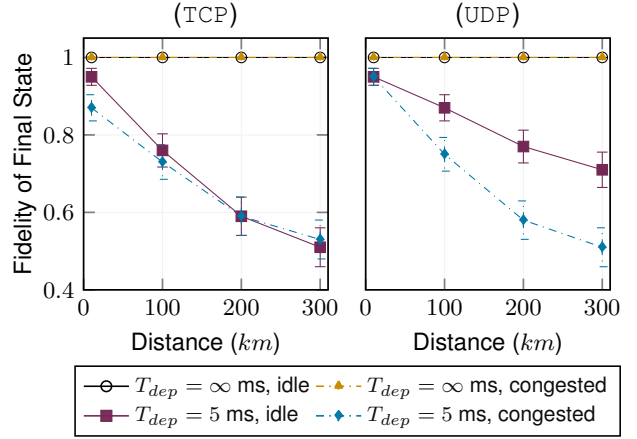

\begin{figure*}[t]
    \centering
      \begin{subfigure}{0.50\textwidth}
        \centering
        \includegraphics[width=0.95\linewidth]{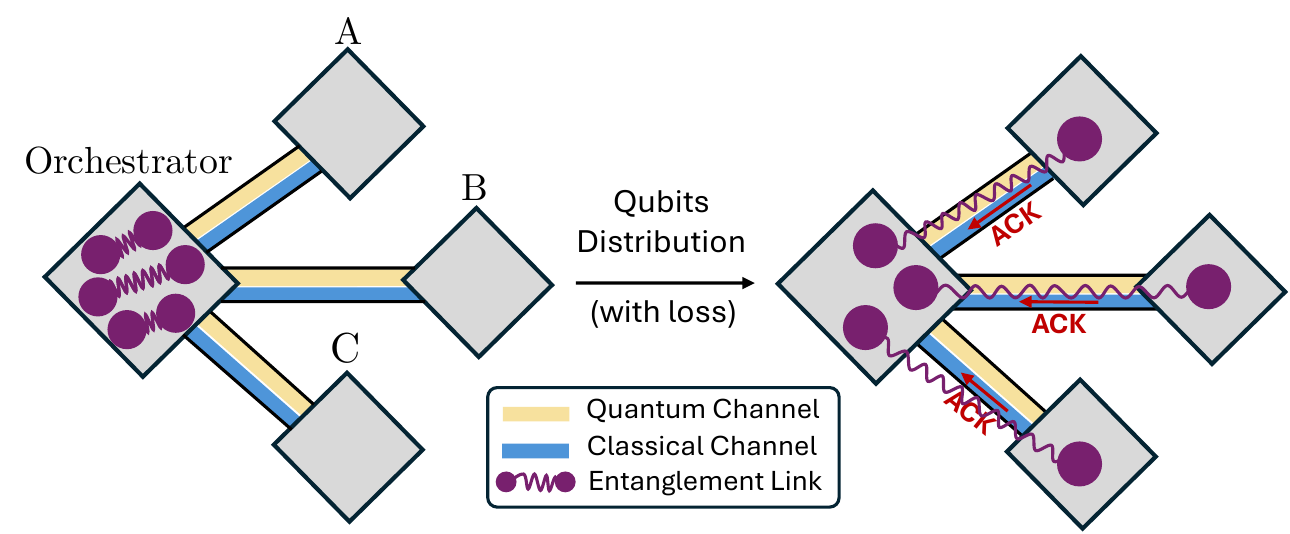}
        \caption{
            Distribution of Bell pairs from the orchestrator to clients over \textit{erasure} quantum channels, with classical acknowledgments handled by ns-3's  TCP/IP stack. 
        }
        \label{fig:7a}
    \end{subfigure}
        \hfill
    \begin{subfigure}{0.46\textwidth}
        \centering
        \includegraphics[width=\linewidth]{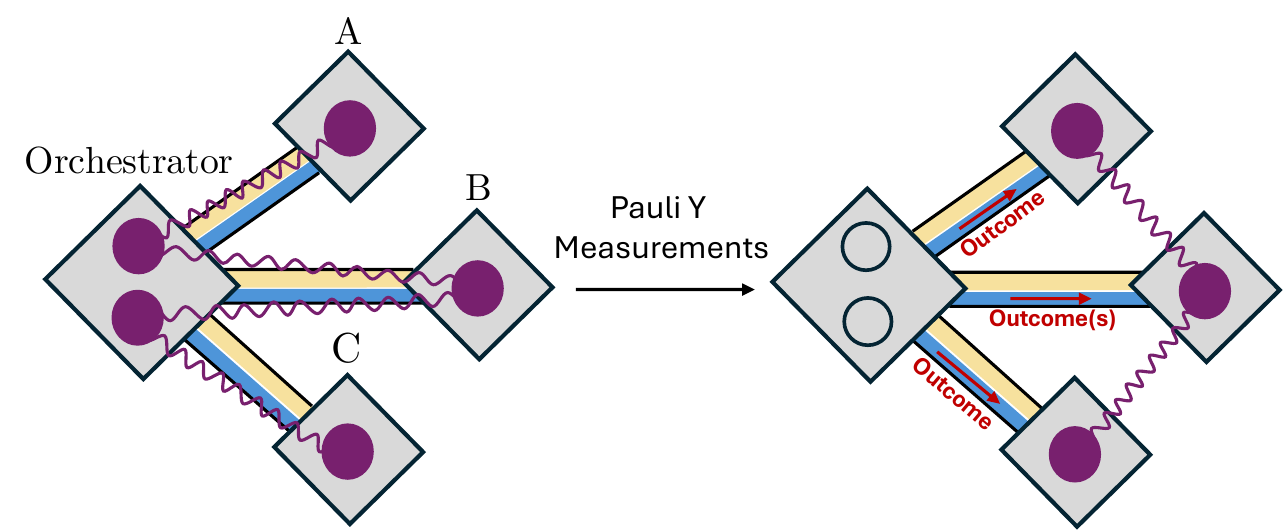}
        \caption{
            Pauli Y measurements on the orchestrator's qubits, with classical correction messages sent to clients by leveraging ns-3's TCP/IP stack.
        }
        \label{fig:7b}
    \end{subfigure}
    \caption{
        Viewer-like representation of a Quantum Local Area Network (QLAN) in \textit{Q2NS}, illustrating how the orchestrator engineers entanglement connectivity graphs via local operations and classical communication. The orchestrator distributes Bell pairs (a) to clients over lossy quantum channels and uses quantum teleportation to deliver qubits. The orchestrator performs local Pauli Y measurements (b) and sends classical correction messages to clients. 
    }
    \label{fig:7}
    \hrulefill
\end{figure*}

\section{Application-Level Case Studies}
\label{sec:4}

To demonstrate the practical capabilities and versatility of \textit{Q2NS} beyond synthetic benchmarks, we present two application-level case studies that showcase its ability to model complex hybrid quantum-classical scenarios:
\begin{enumerate}[label=\roman*.]
    \item a hybrid quantum-classical networking scenario naturally supported by \textit{Q2NS} architecture in subsection \ref{sec:4.1}; 
    \item an end-to-end multipartite entanglement distribution and manipulation workflow over constrained physical graphs, representative of Quantum Local Area Networks (QLANs) in subsection \ref{sec:4.2}.
\end{enumerate}

\subsection{Noisy-Congested Quantum Teleportation}
\label{sec:4.1}
First, we simulate the quantum teleportation protocol\footnote{Quantum teleportation enables the transfer of an arbitrary quantum state by leveraging a shared entangled pair, e.g., $\ket{\Phi^+} = \sfrac{1}{\sqrt{2}} (\ket{00} + \ket{11})$, and the exchange of two classical bits between Alice and Bob \cite{CirZolKim-97,CacCalVan-20}.} of a quantum state, e.g., the state $\ket{+}$, over varying distances between a sender (Alice) and a receiver (Bob), using either UDP or TCP for classical corrections.

\textbf{Simulation Setup.}
To highlight the joint quantum-classical capabilities of \textit{Q2NS}, two non-ideal conditions are modeled: (i) Pauli-depolarizing noise on Bob’s qubit and (ii) classical channel congestion. The depolarizing noise on Bob’s state $\rho$ acts as follows: $\mathcal{E}(\rho) = (1 - p)\rho + p\tfrac{\mathbb{I}}{2}$ \cite{NieChu-11}, with $p = 1 - e^{-t/T_{\text{dep}}}$. Here, $t$ represents the time taken by the classical correction packet to traverse the channel and complete the teleportation protocol, while $T_{\text{dep}}$ denotes the characteristic depolarization-time constant. Consequently, noise accumulates while Bob awaits Alice’s classical corrections, decreasing his qubit fidelity with the intended $|+\rangle$ state. This waiting time depends on both the link length and channel congestion. 

Simulations were performed for link lengths up to 300\,km, $T_{\text{dep}} \in \{5, \infty\}$\,ms, under both idle and congested channels.
To focus on demonstrating the capabilities of our simulator, we consider a worst-case scenario, in which no deferred measurement, error correction, or dynamical decoupling techniques~\cite{viola1998dd, viola1999dd, souza2011dd} are adopted.

The classical control channel is modeled as a point-to-point link with a capacity of 50 Mbps and a delay proportional to the link distance, assuming a propagation delay of 5\,$\mu$s/km (corresponding to light in optical fiber). Congestion is implemented using ns-3's \texttt{BulkSend} for TCP, without warm-up time, and the ns-3's \texttt{OnOffApplication} for UDP, with a fixed load of 100 Mbps.

These parameters are chosen to provide credible link settings, that clearly distinguish idle and congested behavior. In \textit{Q2NS}, these parameters can be easily adjusted, e.g., higher link rates beyond the gigabit range, modifying queue sizes or disciplines, or configuring richer both classical and quantum physical graphs, without altering the teleportation logic.

Data was collected for 1000 runs for each configuration.

\textbf{Simulation Results.}
The simulation results are shown in Fig.~\ref{fig:6}. Fidelities fit the theoretical value of the $|+\rangle$ state, $F_{+}(t) = \frac{1}{2}(1 + e^{-t/T_{dep}})$ with $R^2 \geq 0.997$. Qualitatively, three trends further confirm the simulation's accuracy:
\begin{itemize}
    \item[i.] In the noise-free scenario ($T_{dep} = \infty$), the fidelity remains constant at its maximum value of 1.0, regardless of the classical link length or congestion level.
    \item[ii.] In the noisy scenario ($T_{dep} = 5$ ms), the fidelity decreases with distance, with a significantly sharper drop under congested conditions, approaching the minimum value of 0.5 at long distances and high congestion levels.
    \item[iii.] TCP and UDP exhibit similar overall trends. However, TCP yields slightly lower fidelity in the idle-channel case. This is expected, due to the additional overhead of TCP and the absence of a warm-up phase, which introduces an inherent latency.
\end{itemize}
This simulation scenario highlights how classical network congestion directly impacts quantum fidelity due to increased decoherence during delayed classical communication. \textit{Q2NS} enables such an analysis with minimal effort, demonstrating its utility and effectiveness for studying the interplay between quantum and classical network infrastructure.

\subsection{Engineering Entanglement Graphs in QLANs}
\label{sec:4.2}
Our final case study further substantiates \textit{Q2NS}'s capability to model complex hybrid classical-quantum scenarios, including graph-state manipulations and channel imperfections, in a QLAN~\cite{MazCalCac-25, Mazza2024QCNCTwocolorable} -- a small-scale quantum network consisting of a centralized orchestrator and directly connected clients. 

The emulated QLAN follows the theoretical model in \cite{MazCalCac-25}, which features a centralized orchestrator that generates multipartite entangled states, distributes them to clients, by also leveraging fusion operations and quantum teleportation, and performs subsequent entanglement manipulations. By performing local Pauli measurements on its qubits and classically communicating the outcomes to the clients, the orchestrator induces an entanglement-activated connectivity graph over the client set. After this step, clients can realize any graph state that is reachable from the initially distributed resource through sequences of local complementations and vertex deletions, as mathematically characterized in ~\cite{Hein2004GraphStates, VanDenNest2004LC}. Consequently, the set of attainable final states is entirely determined by the initial resource state prepared and distributed by the orchestrator.

The aforementioned measurement-driven transformations of multipartite states, together with the associated classical signaling exchanges, are explicitly captured and visually rendered by the \textit{Q2NS} visualization tool, as illustrated in a representative example in Fig.~\ref{fig:7}. 

While the above QLAN model has been emulated in a prior work~\cite{Mazza2024SeQUeNCe}, those simulations lacked accurate modeling of quantum state distribution and exhibited pronounced scalability limitations as the size of the shared multipartite resource increased, failing beyond roughly $10$ client nodes. In contrast, \textit{Q2NS} addresses these shortcomings, by demonstrating both scalability and physical realism in modeling state distribution in QLAN architectures. In our experiments, we successfully simulated QLAN instances with more than $M=100$ clients without observing simulation failures. 

\begin{figure}[t]
    \centering
    \begin{subfigure}{0.48\textwidth}
        \centering 
        \input{Tikz/Fig8a}
        \caption{}
        \label{fig:8a}
    \end{subfigure}
    \begin{subfigure}{0.48\textwidth}
        \centering
        \input{Tikz/Fig8b}
        \caption{}
        \label{fig:8b}
    \end{subfigure}
    \caption{QLAN performance analysis showing (a) total runtime with increasing number of client nodes (stab backend), and (b) protocol completion time with $M=3$ clients across varying qubit-loss probabilities and channel lengths.}
    \label{fig:8}
    \hrulefill
\end{figure}
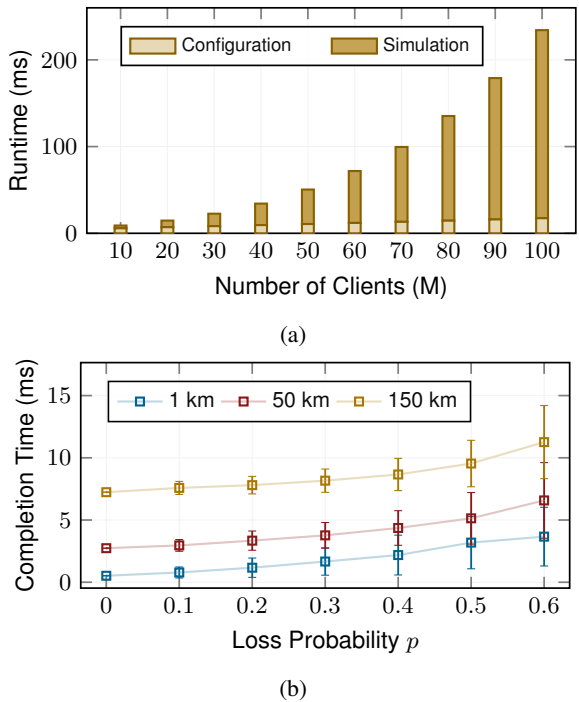

\textbf{Simulation setup.} As in the theoretical model of~\cite{MazCalCac-25}, the physical network is a star graph with a centralized orchestrator connected to $M$ client nodes. The same physical graph is used for both quantum communication (qubit transmission) and classical communication, leveraging the classical ns-3's TCP/IP stack.
The simulation proceeds in two main phases: (i) distribution of a 1D cluster state to the clients via quantum teleportation and (ii) Pauli-$Y$ measurements on the qubits retained by the orchestrator, followed by TCP delivery of the corresponding correction operations to the clients. To establish a shared entangled resource, the orchestrator initially distributes $M$ Bell pairs (one per client) over noisy quantum channels, modeled as an \textit{erasure channel} with a qubit loss-probability denoted as $p$. The orchestrator waits for the classical ack messages to confirm that each Bell pair has been successfully received without loss. This process is represented in Fig.~\ref{fig:7a} and can be nicely visualized through the \textit{Q2NSViz} tool, with particular emphasis on the distribution phase and the resulting entanglement graph obtained after the local Pauli measurements.

The orchestrator then generates an $N$ qubit 1D cluster state and distributes qubits to each client node by performing $M$ quantum teleportation protocols while retaining $N-M$ qubits locally. 
The orchestrator subsequently performs a sequence of Pauli $Y$ measurements on its retained $N-M$ qubits with resulting state: $\ket{G}'  = \ket{y,\pm}^{(a)} \otimes U_{y,\pm}^{(a)}\ket{\tau_a(G)-a}$, where $\tau(G)$ is the local complementation operation applied on vertex $a$ of the graph $G$ associated to the graph state $\ket{G}$ \cite{Hein2006GraphStatesReview}. 
The measurement outcomes, determining the correction operators $U_{y,\pm}^{(a)\dagger}$, are required to obtain the target graph state. As represented in Fig.~\ref{fig:7b}, the measurement results are then transmitted to the client nodes via TCP packets, enabling them to apply the corresponding correction unitaries to their respective qubits.

When the simulation concludes, the resulting distributed quantum state achieves maximum fidelity with respect to the expected target state -- a $M$-qubit 1D cluster state distributed across the clients. Gate operation timings follow superconducting qubit parameters (20\,ns for single-qubit gates, 40\,ns for two-qubit gates, 300\,ns for measurements) but are configurable together with the loss probability $p$, the channel length, and the thresholds for missing messages. The simulation includes comprehensive timing analytics to estimate the complete protocol completion time.

\textbf{Simulation results.} 
The results are presented in Fig.~\ref{fig:8}. Specifically, Fig.~\ref{fig:8a} demonstrates \textit{Q2NS}'s capability of simulating QLAN instances with an increasing number of clients, jointly capturing the manipulation of a $(2M-1)$-qubit multipartite resource state together with Bell-pair distribution and the associated classical coordination logic. The stabilizer backend exhibits favorable scaling, keeping the total runtime -- configuration plus simulation times -- largely below one second up to $M=100$ clients, with an observed near-quadratic growth consistent with theoretical expectations.

Fig.~\ref{fig:8b} presents a timing analysis of the QLAN protocol, characterizing the effects of qubit loss and channel length on the completion time for an $M=3$ client configuration.
The experiments carried out vary the physical distance between the nodes $d \in \{1\,\text{km}, 50\,\text{km}, 150\,\text{km}\}$ and different values of the probability of qubit loss for the Bell state distribution phase $p \in \{0,0.1,0.2,0.3,0.4,0.5,0.6\}$. The QLAN completion-time includes TCP connection establishment, Bell pair distribution with potential retransmissions, quantum teleportation, Pauli Y measurements, and correction(s) acknowledgments.

Results show that protocol completion-time scales linearly with distance due to increased round-trip delays for both quantum and classical communications. 
At 50\,km separation with a loss probability of $p=0.2$, typical completion times range from 2--3\,ms, dominated by quantum propagation delays (250\,$\mu$s per link) and TCP acknowledgment round-trips. Higher values of loss probability make the completion time dominated by the retransmission time, enlarging the standard deviations over multiple protocol instances. 
The proposed example highlights that a TCP-based classical communication provides reliable delivery of measurement corrections while quantum teleportation enables a reliable distribution even in the presence of high imperfections over the quantum channels. Timing analysis across local area networks (1--150\,km) suggests the protocol's suitability for realistic infrastructures.

\section{Conclusion}
We have introduced the architecture and design principles of \textit{Q2NS}, an open-source, modular quantum network simulator built on ns-3. \textit{Q2NS} addresses the challenges of simulating realistic quantum networks and their interplay with classical networks by providing a unified, flexible, and extensible framework that integrates both quantum and classical primitives.
The architectural modularity, separation of concerns, and centralized control logic enable rapid adaptation to emerging quantum networking paradigms and experimental requirements. The conducted performance evaluation highlighted several advantages of \textit{Q2NS}, including superior computational efficiency and straightforward extensibility. This is supported by comparative benchmarks against the tensor-network-based simulator qns-3 and by the ability of \textit{Q2NS} to integrate multiple quantum-state representations through a unified plug-in interface. We also demonstrated that \textit{Q2NS} can simulate hybrid quantum-classical networking scenarios with minimal effort, leveraging ns-3's event-driven semantics and its full classical stack primitives. In addition, we introduced \textit{Q2NSViz}, a dedicated visualization tool and trace format that supports intuitive inspection of both physical and entanglement-induced connectivity graphs, thereby facilitating debugging, analysis, and education for entanglement-based protocols.

We envision \textit{Q2NS} as a foundation for future research and experimentation in quantum networking -- bridging the gap between theoretical modeling and practical implementation, and accelerating progress toward scalable, interoperable Quantum Internet architectures.

\appendices

\section{Background and Backend Scaling} 
\label{app:1}

\subsection{Cluster State Background}
We describe the simulated states using the standard graph-state formalism~\cite{Hein2006GraphStatesReview,RaussendorfBriegel2001}. An $n = R C$ qubit  cluster state can be created by arranging its qubits on an $R \times C$ rectangular lattice with open boundary conditions. Lattice sites are labeled by the tuple $(x,y)$, where
$$
\qquad 0 \le x < R,\;\; 0 \le y < C.
$$

All qubits are initialized in the plus state,
\begin{equation}
    \ket{+}^{\otimes RC} =
    \bigotimes_{x=0}^{R-1}
    \bigotimes_{y=0}^{C-1}
    \ket{+}_{(x,y)} .
\end{equation}

The $R \times C$ cluster state is given by the application of Controlled-$\mathrm{Z}$ operations as follows:
\begin{equation}
    \label{eq:2}
    \ket{C}_{R,C} = \left(
    \prod_{((u,v),(u',v')) \in E}
    \mathrm{CZ}_{(u,v),(u',v')}
    \right)
    \ket{+}^{\otimes RC},
\end{equation}

where the edge set $E$ is given by horizontal and vertical nearest-neighbor couplings:
\begin{align}
    E_{\mathrm{h}} &= \big\{\,((x,y),(x+1,y)) : x+1 < R \,\big\}, \\
    E_{\mathrm{v}} &= \big\{\,((x,y),(x,y+1)) : y+1 < C \,\big\}, \\
    E &= E_{\mathrm{h}} \cup E_{\mathrm{v}}.
\end{align}

Clearly, when $R=1$ or $C=1$, Eq.~\eqref{eq:2} reduces to a 1D cluster state, i.e., a single chain in which each interior qubit has two neighbors. For $R,C>1$ we obtain a genuine 2D cluster state where interior qubits have up to four neighbors.

\subsection{Theoretical Scaling of Each Backend.}
Here, we present a brief theoretical overview of the scaling of each quantum-state representation used in this study.

\subsubsection{State-vector (ket)}
The default representation in \textit{Q2NS} relies on state vectors, built from kets in \textit{Quantum++}~\cite{Gheorghiu2018Quantumpp}. Generally, a state-vector representation stores all $2^n$ complex amplitudes of a pure $n$-qubit state. Gate application on state vectors cost $O(2^n)$ or $O(n\,2^n)$ depending on implementation details, and the memory footprint is also $O(2^n)$~\cite{NielsenChuang2010QuantumComputation,DeRaedt2007QCSimulation}. This exponential scaling is unavoidable for generic states and is what we expect to see from the \textit{Q2NS}-ket backend.

\subsubsection{Density matrix}
For the most comprehensive state representation, \textit{Q2NS} also offers density matrices via \textit{Quantum++}~\cite{Gheorghiu2018Quantumpp}. Generally, density matrices use a matrix with $2^n \times 2^n = 4^n$ complex amplitudes for an $n$-qubit state. As such, the scaling is more demanding than state vectors, with gate applications costing $O(4^n)$ or $O(n\,4^n)$ and a memory footprint of $O(4^n)$~\cite{NielsenChuang2010QuantumComputation,DeRaedt2007QCSimulation}. These costs are unavoidable for generic simulations and, as such, we expect the \textit{Q2NS}-density matrix backend to perform worse than others. Despite their intrinsic unfavorable scaling, density matrices offer a highly expressive representation, rather than an efficient one.

\subsubsection{Stabilizer representation}
When a simulation involves only Clifford states, such as cluster states, their stabilizer properties can be exploited for far more efficient representation. An $n$-qubit stabilizer state can be represented by independent $n$ Pauli generators, and Clifford updates can be performed in $O(n)$, although certain operations such as canonicalization, basis changes, or sign reconstruction can incur up to $O(n^{2-3})$ costs depending on the implementation~\cite{Gottesman1997StabilizerFormalism, AaronsonGottesman2004ImprovedSimulation, Gottesman1998HeisenbergPicture}.

\textit{Q2NS} relies on the \texttt{stab} library, developed by the same group as Quantum++. This implements the method described in ~\cite{Beaudrap2022faststabiliser}, which represents stabilizer states as quadratic form expansions. In this form, states exist in affine subspaces of $\mathbb{F}_2$ with additional phase information and gates act by affine transformations. 

Each single-qubit measurement:
(i) determines the measurement outcome based on commutation relations between the measured Pauli operator and the current stabilizer generators in $O(1)$,
(ii) updates the generator set to reflect post-measurement collapse which scales as $O(nr)$ where $r \leq n$ is the rank of the affine representation,
(iii) removing the measured qubit from the state and creating a new state object just for this qubit. This currently involves SWAP operations to move the qubit to the last index of the state and then dropping it, leading to an $O(n)$ cost, and
(iv) re-canonicalizing the stabilizer representation to maintain a valid affine description in line with the quadratic form expansion. This can scale anywhere from $O(n^{0 - 3})$ depending on the exact nature of the given state after the steps above, such as their locality. In the case of 1D and 2D cluster states, each stabilizer only acts on fewer than 2 or 5 connections per qubit, respectively, leading to bounded stabilizer degree. As such, $O(n^3)$ is the worst case in general scenarios.

Overall, state preparation relies on $O(n)$ operations that scale as $O(n)$. Then we perform $n$ measurements that are $O(n^{0 -3})$. Therefore, the overall scaling is expected to be upper-bounded by $O(n^{4})$, but likely behaving closer to $O(n^{2 - 3})$.

\subsubsection{Tensor-network simulation}
qns-3's architecture is specialized to a particularly notable state representation: tensor networks (TN). These provide a structured way to represent quantum states by decomposing a state into many, ideally small, tensors connected by contracted indices~\cite{Orus2014TNReview, Schollwoeck2011DMRG}. Contracting the network corresponds to summing over these shared indices to evaluate amplitudes, expectation values, or the effect of gates. The cost of contraction is fundamentally governed by the amount of entanglement in the state: low-entanglement states admit factorizations with small internal dimensions and therefore potentially efficient simulation, while highly entangled states require large intermediate tensors, leading to exponential contraction cost~\cite{Vidal2003SlightlyEntangled, Eisert2010}. 

As such, the runtime of tensor-network simulation is determined by the entanglement structure of the state and the order in which the tensors are contracted. For a network of $T$ tensors with contraction width $w$ (e.g., tree-width or a related width parameter), one obtains complexity
$$
\mathrm{poly}(T)\,2^{O(w)},
$$
as shown by Vidal’s analysis of slightly entangled states~\cite{Vidal2003SlightlyEntangled} and by Markov and Shi’s characterization of tensor-network contraction in terms of graph tree-width~\cite{MarkovShi2008TNContraction}. For graph states, including cluster states, the relevant width parameters (tree-width, branch-width, Schmidt-rank width) are determined by the underlying graph structure~\cite{Hein2006GraphStatesReview, VanDenNest2007ClassicalSimVsUniversality}.

In practice, many TN implementations rely on local or greedy contraction heuristics that do not explicitly minimize this width for a given problem structure: they choose the contraction order based on local cost estimates. Such heuristics are known to perform poorly on grid-like  networks~\cite{O_Gorman2019TNShape, Kourtis2019PractitionersGuide}, often producing intermediate tensors whose effective width grows linearly in the total number of qubits, $n = RC$ in this case. The resulting scaling,
$$
\mathrm{poly}(T)\,2^{\Theta(n)},
$$
is an unfavorable worst-case behavior. The contraction becomes effectively exponential in system size, even though a much better width-dependent scaling is theoretically achievable. Thus, even contracting a 1D cluster state can scale exponentially.

With proper optimizations, 1D cluster states can be efficiently simulated using matrix product states (MPS). These represent the system as chains of small tensors with limited bond dimension and are efficient for states whose bipartite entanglement entropy is bounded~\cite{PerezGarcia2007MPS,Schollwoeck2011DMRG}. A 1D cluster state on $n$ qubits is an MPS with fixed bond dimension $\chi = 2$~\cite{Hein2006GraphStatesReview,VanDenNest2007ClassicalSimVsUniversality} and therefore efficiently simulatable.

In higher dimensions, projected entangled-pair states (PEPS) generalize MPS to $2$D and beyond~\cite{VerstraeteCirac2004PEPS,Orus2014TNReview}. For an $R \times C$ grid, the treewidth scales as $w = \Theta(\min(R,C))$~\cite{MarkovShi2008TNContraction,Hein2006GraphStatesReview}. An optimal contraction therefore costs
$$
\mathrm{poly}(RC)\,2^{O(\min(R,C))},
$$
which is exponentially better than the naive $\mathrm{poly}(RC)\,2^{O(RC)}$ scaling that appears if the contraction order ignores the lattice structure, but is notably \textit{still} exponential. This limitation is fundamental: contracting generic PEPS is \#P-hard in the worst case~\cite{Schuch2007ComplexityOfPEPS}, and even approximating certain PEPS contractions remains computationally intractable under standard complexity assumptions~\cite{Haferkamp2023}.

There are many other forms of tensor networks, such as MERA, tree tensor networks, branching MPS, and string-bond states~\cite{Orus2014TNReview}. However, none of these can overcome the fundamental, exponential scaling of representing increasing entanglement width with tensor networks~\cite{Vidal2003SlightlyEntangled, Eisert2010}. 
Overall, tensor networks are an extremely powerful state representation in some cases, such as many low-entanglement regimes. However, their performance in higher-entanglement regimes is far more sensitive to contraction heuristics and can quickly degrade with increasingly complex entanglement structure. While tensor networks are an important tool in quantum simulations, it is not clear that they are a reliable tool for generalized entanglement-based quantum-networking simulations without significant specialization for each problem.

\section{Fitted Scaling Analysis}
\label{app:2}

\begin{table*}
\centering
\caption{Power-law scaling exponents $\beta$ and coefficient of determination $R^2$ for all backends, metrics, and regions. Fits are of the form $\log Y = \alpha + \beta \log N$, where $Y$ is mean total time, configuration time, simulation time, or peak memory usage.}
\label{tab:4}
\renewcommand{\arraystretch}{0.9}
\begin{tabular}{l lcccc}
\toprule
\textbf{Simulator} & \textbf{Region} &
$\boldsymbol{\beta_{\text{total}}}$ $(R^2)$ &
$\boldsymbol{\beta_{\text{configuration}}}$ $(R^2)$ &
$\boldsymbol{\beta_{\text{simulation}}}$ $(R^2)$ &
$\boldsymbol{\beta_{\text{memory}}}$ $(R^2)$ \\
\midrule

\multirow{4}{*}{\rotatebox[origin=c]{45}{\textit{Q2NS}} \rotatebox[origin=c]{45}{Ket}}
  & Low $N$  & 0.70 (0.965) & 0.28 (0.999) & 1.44 (0.973) & 0.04 (0.981) \\
  & Mid $N$  & 1.49 (0.995) & 0.43 (0.934) & 1.84 (1.000) & 0.14 (0.972) \\
  & High $N$ & 2.01 (0.999) & 0.88 (0.994) & 2.06 (0.999) & 0.46 (0.974) \\
  & All $N$  & 1.52 (0.967) & 0.56 (0.933) & 1.86 (0.995) & 0.23 (0.835) \\
\midrule

\multirow{4}{*}{\rotatebox[origin=c]{45}{\textit{Q2NS}} \rotatebox[origin=c]{45}{DM}}
  & Low $N$  & 0.73 (0.962) & 0.29 (1.000) & 1.50 (0.972) & 0.04 (0.963) \\
  & Mid $N$  & 1.47 (0.993) & 0.43 (0.928) & 1.80 (0.999) & 0.14 (0.964) \\
  & High $N$ & 2.00 (0.999) & 0.88 (0.993) & 2.05 (0.999) & 0.46 (0.975) \\
  & All $N$  & 1.52 (0.968) & 0.57 (0.932) & 1.85 (0.995) & 0.23 (0.836) \\
\midrule

\multirow{4}{*}{\rotatebox[origin=c]{45}{\textit{Q2NS}} \rotatebox[origin=c]{45}{Stab}}
  & Low $N$  & 0.74 (0.983) & 0.18 (0.998) & 1.37 (0.999) & 0.05 (0.937) \\
  & Mid $N$  & 1.54 (0.997) & 0.45 (0.986) & 1.77 (1.000) & 0.20 (0.976) \\
  & High $N$ & 2.04 (0.999) & 0.88 (0.995) & 2.08 (0.999) & 0.56 (0.984) \\
  & All $N$  & 1.56 (0.969) & 0.55 (0.921) & 1.80 (0.993) & 0.30 (0.861) \\
\midrule

\multirow{4}{*}{\rotatebox[origin=c]{45}{qns-3 CFA} \rotatebox[origin=c]{45}{GR}}
  & Low $N$  & 3.12 (1.000) & 0.59 (0.885) & 3.14 (1.000) & 0.51 (0.906) \\
  & Mid $N$  & 3.10 (1.000) & 1.74 (0.997) & 3.10 (1.000) & 1.71 (0.921) \\
  & High $N$ & ---          & ---          & ---          & ---          \\
  & All $N$  & 3.11 (1.000) & 1.18 (0.924) & 3.12 (1.000) & 1.05 (0.859) \\
\midrule

\multirow{4}{*}{\rotatebox[origin=c]{45}{qns-3 CFA} \rotatebox[origin=c]{45}{GR-ID}}
  & Low $N$  & 0.61 (0.978) & 0.60 (0.882) & 0.61 (0.993) & 0.21 (0.866) \\
  & Mid $N$  & 1.47 (0.988) & 1.78 (0.994) & 1.33 (0.987) & 0.88 (0.969) \\
  & High $N$ & 2.05 (0.999) & 2.07 (1.000) & 2.03 (0.998) & 1.76 (0.998) \\
  & All $N$  & 1.51 (0.958) & 1.66 (0.966) & 1.45 (0.954) & 1.08 (0.905) \\
\bottomrule
\end{tabular}
\end{table*}

\begin{figure*}
  \centering

  \input{Tikz/Fig9_legend}

  \begin{subfigure}{0.48\textwidth}
    \centering
    \input{Tikz/Fig9a}
    \caption{Total time, fit over all $N$.}
  \end{subfigure}
  \hfill
    \begin{subfigure}{0.48\textwidth}
    \centering
    \input{Tikz/Fig9b}
    \caption{Total time, fit over low, mid, and high $N$.}
  \end{subfigure}
  
  \begin{subfigure}{0.48\textwidth}
    \centering
    \input{Tikz/Fig9c}
    \caption{Peak memory, fit over all $N$.}
  \end{subfigure}
  \hfill
  \begin{subfigure}{0.48\textwidth}
    \centering
    \input{Tikz/Fig9d}
    \caption{Peak memory, fit over low, mid, and high $N$.}
  \end{subfigure}

  \caption{Swap protocol scaling with power-law fits. Left column: fits over all measured $N$. Right column: piecewise fits for low, mid, and high $N$. Top row: total time. Bottom row: peak memory.}
  \label{fig:9}
  \hrulefill
\end{figure*}
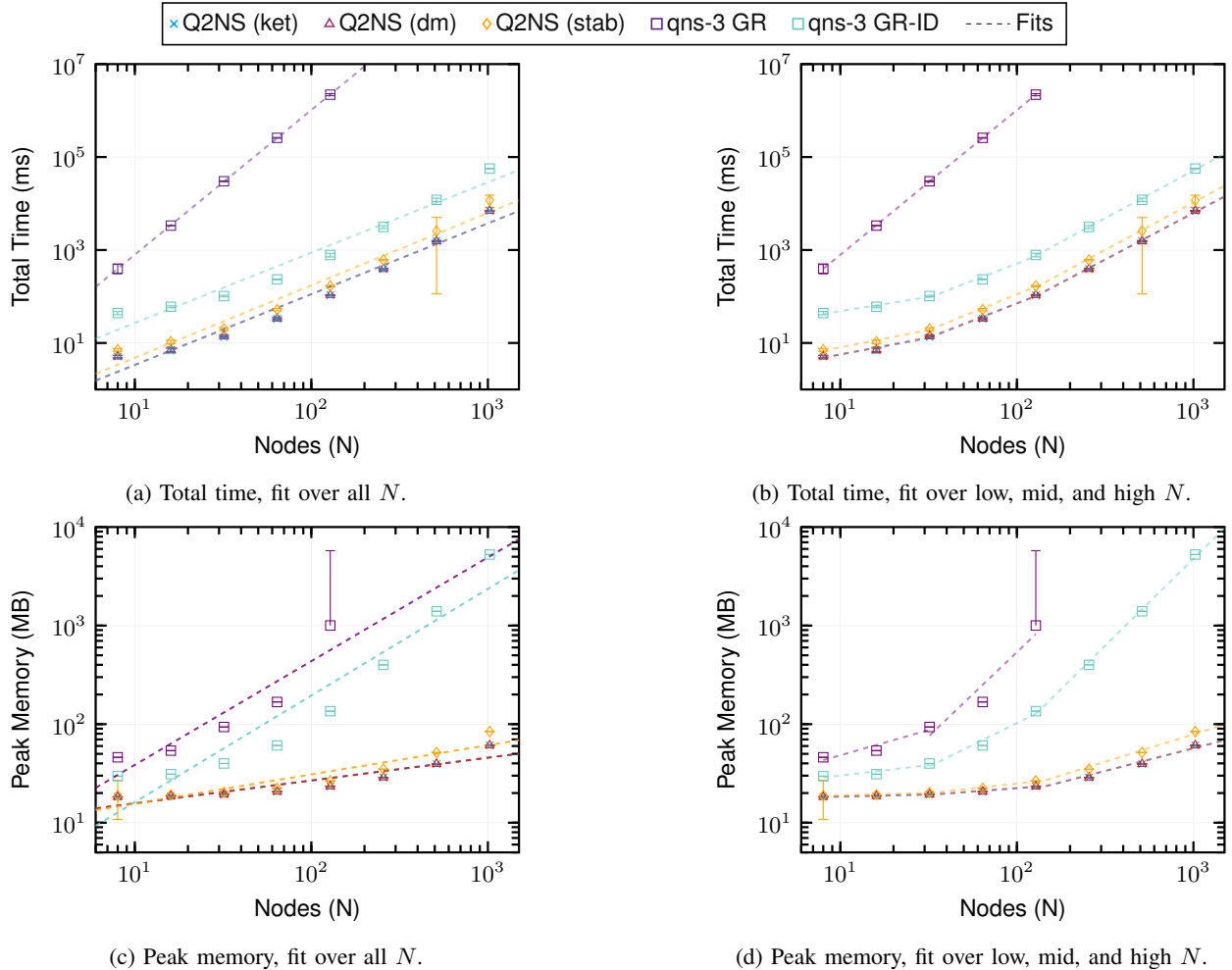

Here, we provide a detailed analysis of the fitted scaling trends observed in Sec.~\ref{sec:3}.
\paragraph{Scaling analysis}
To characterize the asymptotic behavior of simulation cost as a function of the number of nodes~$N$, we fit a power-law scaling model of the form
\begin{equation}
    Y(N) = C N^{\beta},
\end{equation}
where $Y$ denotes one of (i) total simulation time, (ii) configuration time, (iii) simulation time, or (iv) peak memory consumption. Taking logarithms yields a linear model
\begin{equation}
    \log Y = \alpha + \beta \log N,
\end{equation}
and the scaling exponent~$\beta$ is extracted by linear regression in log--log space. For each backend we perform four independent fits using:
(i) a low-$N$ region ($N=\{8,16,32\}$),
(ii) a mid-$N$ region ($N=\{32,64,128\}$),
(iii) a high-$N$ region ($N\geq 128$), except for qns-3 CFA-GR since this is not available, and
(iv) a global fit using all available~$N$ for the given dataset.
The coefficient of determination~$R^2$ is reported in all cases~\cite{MontgomeryRunger2018}.

Fits for multiple subregions are reported alongside global fits to account for regime-dependent behavior, given that fixed overheads and finite-size effects can play a stronger role at small system sizes and obscure asymptotic scaling. For this reason, the high-$N$ region provides the clearest evidence of the likely asymptotic behavior and fits of this regime are highlighted when available. 

To assess the quality of power-law fits beyond the reported $R^2$, exponential-in-$N$ fits were also evaluated. Both models were compared using log-likelihood and information criteria that penalize increased model complexity, namely Akaike and Bayesian information criteria ($\Delta$AIC and $\Delta$BIC)~\cite{BurnhamAnderson2002}.

For the time metrics, the power-law performed better in every region, simulator, or backend based on likelihood estimations and strong preference under information criteria. For peak memory usage, both models indicated more favorable scaling from \textit{Q2NS} versus either version of qns-3. In terms of comparing the two fit models, the comparison was more nuanced than time but nevertheless indicated a power-law fit was preferable for describing asymptotic behavior. At smaller system sizes, exponential fits generally performed better. However, in the high-$N$ regime, the relative quality of the exponential fits degraded, while the power-law fits strengthened across all models where this region was available--all except qns-3 CFA-GR.

This trend suggests that the seeming exponential characteristic of the low-$N$ memory consumption is not indicative of the asymptotic scaling. This higher sensitivity to system size for peak memory, where even the preferred functional form of the fit can change, is in line with the distinction between what time and peak memory measure. Namely, time metrics describe operational costs accumulated over an entire simulation run, whereas peak memory describes the cost at a single moment when  the memory simultaneously required by multiple objects with heterogeneous lifetimes reaches a maximum. This can render peak memory usage more sensitive to small size effects than time metrics.

Given these results, we focus on the power-law fits throughout the rest of this section and in the main text. Table~\ref{tab:4} provides the $\beta$ values representing the asymptotic scaling for each simulator or backend in each region. Fig.~\ref{fig:9} shows the fits in the all-$N$ and high-$N$ regions for the total time and peak memory.

\paragraph{Interpretation}
The exponent~$\beta$ describes the effective scaling of the simulator over the tested regime. For example, $\beta\approx 2$ corresponds to quadratic growth, while $\beta\approx 3$ indicates cubic scaling. A drift in~$\beta$ between regions indicates pre-asymptotic behavior or crossover between different computational regimes. This effect is visible in the \textit{Q2NS} backends, where memory usage follows a shallow power law ($\beta\approx 0.2$ globally) but increases to $\beta\approx 0.45$ in the high-$N$ regime, producing the mild curvature observed in log--log plots.

\paragraph{Use of mean values}
We also repeated all fits using two uncertainty-aware procedures:
(i) weighted regression using weights $w_i=1/\sigma_i^2$, and
(ii) upper/lower envelope fits using $(\mathrm{mean}\pm \mathrm{std})$.
Both approaches produced scaling exponents consistent with the mean-based fits. In particular, the envelope fits yielded very narrow ranges for~$\beta$. Weighted fits were numerically unstable for the \textit{Q2NS} time metrics due to very small standard deviations at low~$N$, which forced the regression to overweight pre-asymptotic points. Since the qualitative scaling behavior was unchanged and the mean-based regressions were numerically stable across all metrics, we report only the mean-based exponents in the main tables.

\bibliography{biblio}

\end{document}

%% file: preamble.tex
\usepackage[utf8]{inputenc}
\usepackage[T1]{fontenc}
\usepackage{cite}
\usepackage{amsmath,amssymb,amsfonts,amsthm}
\usepackage{graphicx}
\usepackage{blindtext}
\usepackage{textcomp}
\usepackage{braket}
\usepackage{hyperref}
\usepackage{subcaption}
\usepackage{float}
\usepackage{tikz}
\usepackage{quantikz}
\usepackage{pgfplots}
\usepackage{xfrac}
\usepgfplotslibrary{groupplots}

\usepackage{enumitem}
\usepackage{xstring}
\usepackage{pgfplotstable}
\usepackage{booktabs,multirow}
\usepackage[most]{tcolorbox}

\usetikzlibrary{arrows.meta,positioning,calc}
\usepackage{graphicx}
\definecolor{accentblue}{RGB}{45,95,140}   
\definecolor{dmgreen}{RGB}{0,158,115}     
\definecolor{stimorange}{RGB}{230,159,0}   
\definecolor{qnscolor}{RGB}{148,103,189}   
\definecolor{qnslocblue}{RGB}{100,150,180} 
\definecolor{simcolor}{HTML}{F7F9FF}
\definecolor{tracecolor}{HTML}{FAEDEA}

\usepackage{pgfplots}
\usepgfplotslibrary{groupplots}
\pgfplotsset{compat=1.18}
\usetikzlibrary{spy, calc, patterns}

\usepackage{xcolor}

\usepackage{hyperref}
\hypersetup{
    colorlinks = true,
    linkcolor = blue!50!black,
    citecolor = blue!50!black,
    filecolor = blue!50!black,
    urlcolor = blue!50!black,
}

\newcommand{\RN}[1]{
    \textup{\uppercase\expandafter{\romannumeral#1}}
}

\def\BibTeX{{\rm B\kern-.05em{\sc i\kern-.025em b}\kern-.08em
    T\kern-.1667em\lower.7ex\hbox{E}\kern-.125emX}}

%% file: Tikz/Fig1.tex
\definecolor{pastelblue}{RGB}{185,205,212}
\definecolor{pastelred}{RGB}{240,190,195}
\definecolor{pastelorange}{RGB}{245,220,195}
\definecolor{envH}{RGB}{250,250,250}

\definecolor{ink}{HTML}{111827}
\colorlet{controllerFill}{pastelred}
\colorlet{controllerLine}{black}
\colorlet{qnodeFill}{pastelblue}
\colorlet{qnodeLine}{black}
\colorlet{qchannelFill}{pastelorange}
\colorlet{qchannelLine}{black}

\tikzset{
  mainbox/.style={
    rectangle, rounded corners=6pt, very thick,
    minimum height=1.0cm,
    inner sep=6pt, text=ink, align=center,
    font=\sffamily\footnotesize
  },
  modulebox/.style={
    rectangle, rounded corners=4pt, thick,
    minimum height=0.7cm,
    inner sep=4pt, text=ink, align=center,
    font=\sffamily\scriptsize
  }
}

\begin{tikzpicture}[
  font=\sffamily\footnotesize,
  >=Latex
]

\node[mainbox, fill=envH, draw=black, very thick, text width=0.95\linewidth, inner sep=6pt] (environment) {%
  {\normalsize \textbf{Q2NS Simulation Environment}}\\[0.5pt]
  \begin{tikzpicture}
    \node[mainbox, fill=qnodeFill, draw=qnodeLine, text width=0.28\linewidth, minimum height=3cm] (qnode1) {%
      {\normalsize \textbf{QNode 1}}\\[6pt]
      \begin{tikzpicture}
        \node[modulebox, fill=qnodeFill!50, draw=qnodeLine, text width=0.42\linewidth] (qproc1) {%
          {\small \textbf{QProcessor}}\\[6pt]
          \begin{tikzpicture}
            \node[modulebox, fill=qnodeFill!40, draw=qnodeLine!70, dashed, text width=0.85\linewidth, minimum height=0.4cm, inner sep=4pt] {%
              {\scriptsize Qubit List}
            };
          \end{tikzpicture}
          \\[3pt]
          \begin{tikzpicture}
            \node[modulebox, fill=qnodeFill!40, draw=qnodeLine!70, dashed, text width=0.85\linewidth, minimum height=0.4cm, inner sep=4pt] {%
              {\scriptsize Quantum Operations}
            };
          \end{tikzpicture}
        };
        \node[modulebox, fill=qnodeFill!50, draw=qnodeLine, text width=0.42\linewidth, right=2mm of qproc1] (qnet1) {%
          {\small \textbf{QNetworker}}\\[4pt]
          \begin{tikzpicture}
            \node[modulebox, fill=qnodeFill!40, draw=qnodeLine!70, dashed, text width=0.85\linewidth, minimum height=0.4cm, inner sep=4pt] {%
              {\scriptsize Quantum Hardware}
            };
          \end{tikzpicture}
          \\[3pt]
          \begin{tikzpicture}
            \node[modulebox, fill=qnodeFill!40, draw=qnodeLine, text width=0.85\linewidth, minimum height=0.55cm, inner sep=4pt] {%
              {\scriptsize \textbf{QNetDevice}}
            };
          \end{tikzpicture}
        };
        \node[below=2mm of qproc1.south east, anchor=north, font=\scriptsize, text=ink, dashed, draw=qnodeLine!70, rounded corners=2pt, inner sep=2pt, minimum width=0.8\linewidth, minimum height=0.45cm] {%
          NS-3 backend
        };
      \end{tikzpicture}
    };
    \node[above right=-0.35cm of qnode1.north west, font=\scriptsize, text=ink!90] {\textbf{ns3::Node::QNode}};
    
    \node[mainbox, fill=qchannelFill, draw=qchannelLine, text width=0.27\linewidth, right=-0.5mm of qnode1] (qchannel) {%
      {\normalsize \textbf{QChannel}}\\[6pt]
      \begin{tikzpicture}
        \node[modulebox, fill=qchannelFill!60, draw=qchannelLine, text width=0.85\linewidth] {%
          {\small \textbf{QMap}}
        };
      \end{tikzpicture}
    };
    
    \node[mainbox, fill=qnodeFill, draw=qnodeLine, text width=0.28\linewidth, right=-0.50mm of qchannel, minimum height=3cm] (qnode2) {%
      {\normalsize \textbf{QNode 2}}\\[6pt]
      \begin{tikzpicture}
        \node[modulebox, fill=qnodeFill!50, draw=qnodeLine, text width=0.42\linewidth] (qnet2) {%
          {\small \textbf{QNetworker}}\\[4pt]
          \begin{tikzpicture}
            \node[modulebox, fill=qnodeFill!40, draw=qnodeLine!70, dashed, text width=0.85\linewidth, minimum height=0.4cm, inner sep=4pt] {%
              {\scriptsize Quantum Hardware}
            };
          \end{tikzpicture}
          \\[3pt]
          \begin{tikzpicture}
            \node[modulebox, fill=qnodeFill!40, draw=qnodeLine, text width=0.85\linewidth, minimum height=0.5cm, inner sep=4pt] {%
              {\scriptsize \textbf{QNetDevice}}
            };
          \end{tikzpicture}
        };
        \node[modulebox, fill=qnodeFill!50, draw=qnodeLine, text width=0.42\linewidth, right=2mm of qnet2] (qproc2) {%
          {\small \textbf{QProcessor}}\\[4pt]
          \begin{tikzpicture}
            \node[modulebox, fill=qnodeFill!40, draw=qnodeLine!70, dashed, text width=0.85\linewidth, minimum height=0.4cm, inner sep=4pt] {%
              {\scriptsize Qubit List}
            };
          \end{tikzpicture}
          \\[3pt]
          \begin{tikzpicture}
            \node[modulebox, fill=qnodeFill!40, draw=qnodeLine!70, dashed, text width=0.85\linewidth, minimum height=0.4cm, inner sep=4pt] {%
              {\scriptsize Quantum Operations}
            };
          \end{tikzpicture}
        };
        \node[below=2mm of qnet2.south east, anchor=north, font=\scriptsize, text=ink, dashed, draw=qnodeLine!70, rounded corners=2pt, inner sep=2pt, minimum width=0.8\linewidth, minimum height=0.45cm] {%
          NS-3 backend
        };
      \end{tikzpicture}
    };
    \node[above left=-0.3cm of qnode2.north east, font=\scriptsize, text=ink!90] {\textbf{ns3::Node::QNode}};
    
    \node[mainbox, fill=controllerFill, draw=controllerLine, text width=0.9\linewidth, below=13mm of qchannel] (netcontroller) {%
      {\normalsize \textbf{NetController}}\\[8pt]
      \begin{tikzpicture}
        \node[modulebox, fill=controllerFill!60, draw=controllerLine, text width=0.42\linewidth] (simcore) {%
          {\small \textbf{SimulationCore}}
        };
        \node[modulebox, fill=controllerFill!60, draw=controllerLine, text width=0.42\linewidth, right=3mm of simcore] (qstatereg) {%
          {\small \textbf{QuantumStateRegistry}}
        };
      \end{tikzpicture}
    };
  \end{tikzpicture}
};

\end{tikzpicture}

%% file: Tikz/Fig2.tex
\definecolor{simH}{RGB}{240,242,250} 
\definecolor{simH2}{RGB}{232,227,237} 
\definecolor{traceH}{RGB}{241,230,227} 
\definecolor{traceH2}{RGB}{245,239,221}
\definecolor{viewH}{RGB}{248,248,248} 
\definecolor{accentblue}{RGB}{45,95,140}      
\definecolor{dmgreen}{RGB}{0,158,115}        
\definecolor{stimorange}{RGB}{230,159,0}     
\definecolor{qnscolor}{RGB}{148,103,189}     
\definecolor{qnslocblue}{RGB}{100,150,180}    

\definecolor{ink}{HTML}{111827}    
\colorlet{simFill}{simH2}
\colorlet{simLine}{black}
\colorlet{traceFill}{traceH2}
\colorlet{traceLine}{black}
\colorlet{viewFill}{viewH!5}
\colorlet{viewLine}{black}
\colorlet{arrowInk}{black}

\tikzset{
  stage/.style={
    rectangle, rounded corners=6pt, very thick,
    minimum height=1.0cm,
    inner sep=4pt, text=ink, align=center
  }
}

\begin{tikzpicture}[
  font=\sffamily\footnotesize,
  >=Latex,
  stage/.style={
    rectangle, rounded corners, draw=black, very thick,
    text width=0.95\linewidth,
    minimum height=1.2cm,
    align=center, inner sep=3pt
  }
]

\node[stage, fill=simFill, draw=simLine] (sim) {%
  \textbf{\textcolor{simLine}{Simulation with Q2NS}}\\
  (Q2NSViz / C++)\\[3pt]
  \begin{tikzpicture}
    \node[stage, draw=simLine, fill=simFill, very thick, text width=0.9\linewidth, align=left, inner sep=6pt]{
      \ttfamily
        \textbf{LogInitializeQubit}("Alice","aliceHalf");\\
      \textbf{LogInitializeQubit}("Alice","bobHalf");\\
      \textbf{LogEntangleQubits}(\{"aliceHalf","bobHalf"\});\\
      \textbf{LogSendQubit}(\{"bobHalf","Bob"\});\\
    };
  \end{tikzpicture}
};

\node[
  stage, fill=traceFill, draw=traceLine,
  minimum height=0.55cm, inner sep=3pt, below=3mm of sim
] (tracefile) {%
  \textbf{\textcolor{traceLine}{Trace File}}\\[-2pt]
  (NDJSON)
};

\node[stage, fill=viewFill, draw=viewLine, below=3mm of tracefile] (viewer) {
  \textbf{\textcolor{viewLine}{Viewer}}\\
  (Q2NSViz app / HTML + D3.js)\\[3pt]
  \includegraphics[width=0.85\linewidth, trim=0 19pt 0 0pt, clip]{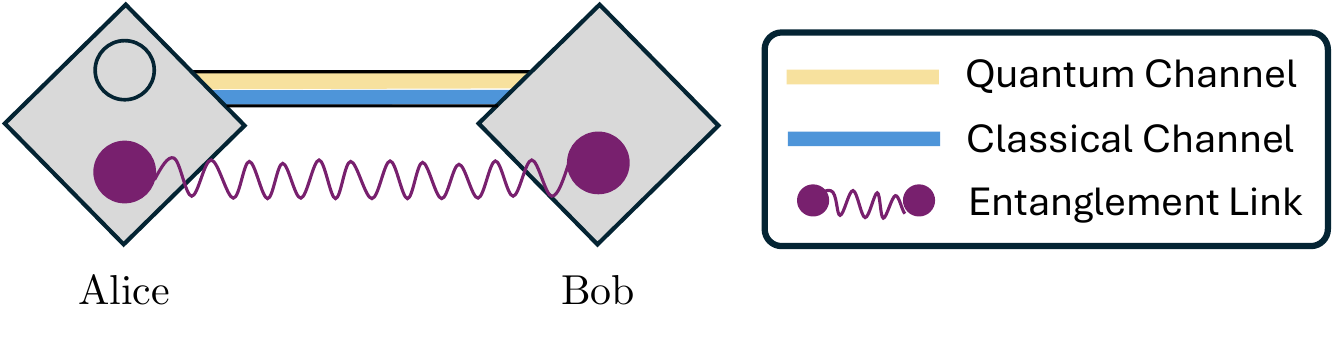}
};

\draw[->, line width=0.8pt, draw=arrowInk] (sim.south) -- (tracefile.north);
\draw[->, line width=0.8pt, draw=arrowInk] (tracefile.south) -- (viewer.north);

\end{tikzpicture}

%% file: Tikz/Fig3a.tex
\begin{tikzpicture}

\definecolor{accentblue}{RGB}{0,130,190}      
\definecolor{dmgreen}{RGB}{180,30,30}        
\definecolor{staborange}{RGB}{210,145,0}      
\definecolor{softgray}{RGB}{220,220,220}      
\definecolor{graygrid}{RGB}{230,230,230}
\definecolor{qnattynet-magenta}{RGB}{158, 57, 106}

\definecolor{qnscolor}{RGB}{100,35,150}     

\def\ClusterDmOneD{Data/cluster/cluster_dm_1D.csv}
\def\ClusterDmTwoD{Data/cluster/cluster_dm_2D.csv}
\def\ClusterKetOneD{Data/cluster/cluster_ket_1D.csv}
\def\ClusterKetTwoD{Data/cluster/cluster_ket_2D.csv}
\def\ClusterStabOneD{Data/cluster/cluster_stab_1D.csv}
\def\ClusterStabTwoD{Data/cluster/cluster_stab_2D.csv}
\def\ClusterQnsOneD{Data/cluster/cluster_qns3_1D.csv}
\def\ClusterQnsTwoD{Data/cluster/cluster_qns3_2D.csv}

\newcommand{\PlotKetOneD}[1][]{%
\addplot+[
    #1,
    mark=x, 
    mark size=1.8pt,
    line width=0.8pt,
    draw=accentblue!15,
    mark options={draw=accentblue},
    error bars/.cd,
        y dir=both,
        y explicit,
        error bar style={accentblue},
    /pgfplots/.cd
] table[
    x=N,
    y expr=\thisrow{mean_time_s}*1000,
    y error expr=\thisrow{std_time_s}*1000,
    col sep=comma,
]{\ClusterKetOneD};}

\newcommand{\PlotKetTwoD}[1][]{%
\addplot+[
    #1,
    mark=x, 
    mark size=1.8pt,
    line width=0.8pt,
    draw=accentblue!15,
    mark options={draw=accentblue},
    error bars/.cd,
        y dir=both,
        y explicit,
        error bar style={accentblue},
    /pgfplots/.cd
] table[
    x=N,
    y expr=\thisrow{mean_time_s}*1000,
    y error expr=\thisrow{std_time_s}*1000,
    col sep=comma,
]{\ClusterKetTwoD};}

\newcommand{\PlotDmOneD}[1][]{%
\addplot+[
    #1,
    mark=triangle, 
    mark size=2pt,
    line width=0.8pt,
    draw=qnattynet-magenta!15,
    mark options={draw=qnattynet-magenta},
    error bars/.cd,
        y dir=both,
        y explicit,
        error bar style={qnattynet-magenta},
    /pgfplots/.cd
] table[
    x=N,
    y expr=\thisrow{mean_time_s}*1000,
    y error expr=\thisrow{std_time_s}*1000,
    col sep=comma,
]{\ClusterDmOneD};}

\newcommand{\PlotDmTwoD}[1][]{%
\addplot+[
    #1,
    mark=triangle, 
    mark size=1.8pt,
    line width=0.8pt,
    draw=qnattynet-magenta!15,
    mark options={draw=qnattynet-magenta},
    error bars/.cd,
        y dir=both,
        y explicit,
        error bar style={qnattynet-magenta},
    /pgfplots/.cd
] table[
    x=N,
    y expr=\thisrow{mean_time_s}*1000,
    y error expr=\thisrow{std_time_s}*1000,
    col sep=comma,
]{\ClusterDmTwoD};}

\newcommand{\PlotStabOneD}[1][]{%
\addplot+[
    #1,
    mark=diamond, 
    mark size=1.8pt,
    line width=1pt,
    draw=staborange!15,
    mark options={draw=staborange},
    error bars/.cd,
        y dir=both,
        y explicit,
        error bar style={staborange},
    /pgfplots/.cd
] table[
    x=N,
    y expr=\thisrow{mean_time_s}*1000,
    y error expr=\thisrow{std_time_s}*1000,
    col sep=comma,
]{\ClusterStabOneD};}

\newcommand{\PlotStabTwoD}[1][]{%
\addplot+[
    #1,
    mark=diamond, 
    mark size=1.8pt,
    line width=1pt,
    draw=staborange!15,
    mark options={draw=staborange},
    error bars/.cd,
        y dir=both,
        y explicit,
        error bar style={staborange},
    /pgfplots/.cd
] table[
    x=N,
    y expr=\thisrow{mean_time_s}*1000,
    y error expr=\thisrow{std_time_s}*1000,
    col sep=comma,
]{\ClusterStabTwoD};}

\newcommand{\PlotQnsOneD}[1][]{%
\addplot+[
    #1,
    mark=square,
    mark size=1.8pt,
    line width=0.8pt,
    draw=qnscolor!15,
    mark options={draw=qnscolor},
    error bars/.cd,
        y dir=both,
        y explicit,
        error bar style={qnscolor},
    /pgfplots/.cd
] table[
    x=N,
    y expr=\thisrow{mean_time_s}*1000,
    y error expr=\thisrow{std_time_s}*1000,
    col sep=comma,
]{\ClusterQnsOneD};}

\newcommand{\PlotQnsTwoD}[1][]{%
\addplot+[
    #1,
    mark=square, 
    mark size=1.8pt,
    line width=0.8pt,
    draw=qnscolor!15,
    mark options={draw=qnscolor},
    error bars/.cd,
        y dir=both,
        y explicit,
        error bar style={qnscolor},
    /pgfplots/.cd
] table[
    x=N,
    y expr=\thisrow{mean_time_s}*1000,
    y error expr=\thisrow{std_time_s}*1000,
    col sep=comma,
]{\ClusterQnsTwoD};}

\begin{groupplot}[
    group style={
        group size=1 by 2,
        vertical sep=0.45cm,
        group name=plots
    },
    width=6.5cm,
    height=3.5cm,
    scale only axis,
    xmajorgrids,
    ymajorgrids,
    x grid style={very thin, gray!15},
    y grid style={very thin, gray!15},
    axis line style={line width=0.75pt, color=black},
    tick style={line width=0.75pt, color=black},
    xmode=log,
    ymode=log,
    xmin=1.5, xmax=5500,
    ymin=2e-1, ymax=90000,   
    xlabel={Number of qubits (n)},
    ylabel={Runtime (ms)},
    ylabel style={
        font=\sffamily\small,
        yshift=-0.5em
        },
    xlabel style={font=\sffamily\small},
    ticklabel style={font=\sffamily\small},
    title style={font=\sffamily\normalsize, yshift=-1.25ex},
    legend cell align={left},
    legend style={
        at={(0.495,2.35)},
        anchor=north,
        draw=black,
        line width=0.7pt,
        fill=white,
        fill opacity=0.95,
        font=\sffamily\footnotesize,
        legend columns=4,
        /tikz/every even column/.append style={column sep=0.05cm}
    },
]

\nextgroupplot[
    ylabel={Runtime (ms)},
    xticklabels={},
    xlabel={},
]

\PlotKetOneD
\PlotDmOneD

\addplot+[
    mark=diamond, 
    mark size=1.8pt,
    line width=0.8pt,
    draw=staborange!30,
    mark options={draw=staborange},
    error bars/.cd,
        y dir=both,
        y explicit,
        error bar style={staborange},
    /pgfplots/.cd
] table[
    x=N,
    y expr=\thisrow{mean_time_s}*1000,
    y error expr=\thisrow{std_time_s}*1000,
    col sep=comma,
]{\ClusterStabOneD};

\PlotQnsOneD

\node[
    anchor=west,
    fill=gray!5,
    inner sep=2pt,
    font=\sffamily\footnotesize
] at (rel axis cs:0.63,0.9) {1D Cluster States};

\nextgroupplot[
]

\PlotKetTwoD
\addlegendentry{Q2NS (ket)}
\PlotDmTwoD
\addlegendentry{Q2NS (dm)}

\addplot+[
    mark=diamond, 
    mark size=1.8pt,
    line width=0.8pt,
    draw=staborange!30,
    mark options={draw=staborange},
    error bars/.cd,
        y dir=both,
        y explicit,
        error bar style={staborange},
    /pgfplots/.cd
] table[
    x=N,
    y expr=\thisrow{mean_time_s}*1000,
    y error expr=\thisrow{std_time_s}*1000,
    col sep=comma,
]{\ClusterStabTwoD};
\addlegendentry{Q2NS (stab)}

\PlotQnsTwoD
\addlegendentry{qns3 (TN)}

\node[
    anchor=west,
    fill=gray!5,
    inner sep=2pt,
    font=\sffamily\footnotesize
] at (rel axis cs:0.63,0.9) {2D Cluster States};

\end{groupplot}

\begin{axis}[
    at={($(plots c1r2.north east)+(-0.2cm,3.25cm)$)}, 
    anchor=north east,
    width=2.2cm,
    height=1.95cm,
    scale only axis,
    xmode=log,
    ymode=log,
    xmin=1.5, xmax=12000,
    ymin=2e-1, ymax=2e7,
    axis background/.style={fill=white, fill opacity=0.95},
    axis line style={line width=0.5pt, color=black},
    tick style={line width=0.5pt, color=black},
    ticklabel style={font=\sffamily\tiny, color=black},
    xlabel={},
    ylabel={},
    grid=both,
    minor grid style={gray!5},
    major grid style={gray!5},
    legend style={
        at={(0.03,0.97)},
        anchor=north west,
        font=\sffamily\tiny,
        draw=none,
        fill=none
    },
    legend cell align=left
]

\addplot[
    dashed,
    line width=0.7pt,
    draw=black,
    domain=1.5:128,
    samples=200
]{0.169 * x^(0.723)};

\addplot[
    dashed,
    line width=0.7pt,
    draw=black,
    domain=128:12000,
    samples=200
]{2.695e-06 * x^(3.17)};
\addlegendentry{Fit}

\addplot+[
    only marks,
    mark=diamond, 
    mark size=1pt,
    draw=staborange,
    mark options={draw=staborange, fill=staborange},
    error bars/.cd,
        y dir=both,
        y explicit,
        error bar style={staborange, line width=0.4pt},
    /pgfplots/.cd
] table[
    x=N,
    y expr=\thisrow{mean_time_s}*1000,
    y error expr=\thisrow{std_time_s}*1000,
    col sep=comma,
]{\ClusterStabOneD};

\end{axis}

\begin{axis}[
    at={($(plots c1r2.north east)+(-0.2cm,-0.75cm)$)},
    anchor=north east,
    width=2.2cm,
    height=1.95cm,
    scale only axis,
    xmode=log,
    ymode=log,
    xmin=1.5, xmax=12000,
    ymin=2e-1, ymax=2e7,
    axis background/.style={fill=white, fill opacity=0.95},
    axis line style={line width=0.5pt, color=black},
    tick style={line width=0.5pt, color=black},
    ticklabel style={font=\sffamily\tiny, color=black},
    xlabel={},
    ylabel={},
    grid=both,
    minor grid style={gray!5},
    major grid style={gray!5},
    legend style={
        at={(0.03,0.97)},
        anchor=north west,
        font=\sffamily\tiny,
        draw=none,
        fill=none
    },
    legend cell align=left
]

\addplot[
    dashed,
    line width=0.7pt,
    draw=black,
    domain=1.5:128,
    samples=200
]{0.169 * x^(0.723)};

\addplot[
    dashed,
    line width=0.7pt,
    draw=black,
    domain=128:12000,
    samples=200
]{2.695e-06 * x^(3.17)};
\addlegendentry{Fit}

\addplot+[
    only marks,
    mark=diamond, 
    mark size=1.2pt,
    draw=staborange,
    mark options={draw=staborange, fill=staborange},
    error bars/.cd,
        y dir=both,
        y explicit,
        error bar style={staborange, line width=0.4pt},
    /pgfplots/.cd
] table[
    x=N,
    y expr=\thisrow{mean_time_s}*1000,
    y error expr=\thisrow{std_time_s}*1000,
    col sep=comma,
]{\ClusterStabTwoD};

\end{axis}

\end{tikzpicture}

%% file: Tikz/Fig3b.tex
\begin{tikzpicture}

\definecolor{accentblue}{RGB}{0,130,190}      
\definecolor{dmgreen}{RGB}{180,30,30}        
\definecolor{staborange}{RGB}{210,145,0}      
\definecolor{softgray}{RGB}{220,220,220}      
\definecolor{graygrid}{RGB}{230,230,230}   
\definecolor{qnattynet-magenta}{RGB}{158, 57, 106}

\definecolor{qnscolor}{RGB}{100,35,150}   

\def\ClusterDmOneD{Data/cluster/cluster_dm_1D.csv}
\def\ClusterDmTwoD{Data/cluster/cluster_dm_2D.csv}
\def\ClusterKetOneD{Data/cluster/cluster_ket_1D.csv}
\def\ClusterKetTwoD{Data/cluster/cluster_ket_2D.csv}
\def\ClusterStabOneD{Data/cluster/cluster_stab_1D.csv}
\def\ClusterStabTwoD{Data/cluster/cluster_stab_2D.csv}
\def\ClusterQnsOneD{Data/cluster/cluster_qns3_1D.csv}
\def\ClusterQnsTwoD{Data/cluster/cluster_qns3_2D.csv}

\newcommand{\PlotKetOneD}[1][]{%
\addplot+[
    #1,
    mark=x, 
    mark size=1.8pt,
    line width=0.9pt,
    draw=accentblue!15,
    mark options={draw=accentblue},
    error bars/.cd,
        y dir=both,
        y explicit,
        error bar style={accentblue},
    /pgfplots/.cd
] table[
    x=N,
    y expr=\thisrow{mean_mem_mb},
    y error expr=\thisrow{std_mem_mb},
    col sep=comma,
]{\ClusterKetOneD};}

\newcommand{\PlotKetTwoD}[1][]{%
\addplot+[
    #1,
    mark=x, 
    mark size=1.8pt,
    line width=0.9pt,
    draw=accentblue!15,
    mark options={draw=accentblue},
    error bars/.cd,
        y dir=both,
        y explicit,
        error bar style={accentblue},
    /pgfplots/.cd
] table[
    x=N,
    y expr=\thisrow{mean_mem_mb},
    y error expr=\thisrow{std_mem_mb},
    col sep=comma,
]{\ClusterKetTwoD};}

\newcommand{\PlotDmOneD}[1][]{%
\addplot+[
    #1,
    mark=triangle, 
    mark size=1.8pt,
    line width=0.9pt,
    draw=qnattynet-magenta!15,
    mark options={draw=qnattynet-magenta},
    error bars/.cd,
        y dir=both,
        y explicit,
        error bar style={qnattynet-magenta},
    /pgfplots/.cd
] table[
    x=N,
    y expr=\thisrow{mean_mem_mb},
    y error expr=\thisrow{std_mem_mb},
    col sep=comma,
]{\ClusterDmOneD};}

\newcommand{\PlotDmTwoD}[1][]{%
\addplot+[
    #1,
    mark=triangle, 
    mark size=1.8pt,
    line width=0.9pt,
    draw=qnattynet-magenta!15,
    mark options={draw=qnattynet-magenta},
    error bars/.cd,
        y dir=both,
        y explicit,
        error bar style={qnattynet-magenta},
    /pgfplots/.cd
] table[
    x=N,
    y expr=\thisrow{mean_mem_mb},
    y error expr=\thisrow{std_mem_mb},
    col sep=comma,
]{\ClusterDmTwoD};}

\newcommand{\PlotStabOneD}[1][]{%
\addplot+[
    #1,
    mark=diamond, 
    mark size=1.8pt,
    line width=0.9pt,
    draw=staborange!15,
    mark options={draw=staborange},
    error bars/.cd,
        y dir=both,
        y explicit,
        error bar style={staborange},
    /pgfplots/.cd
] table[
    x=N,
    y expr=\thisrow{mean_mem_mb},
    y error expr=\thisrow{std_mem_mb},
    col sep=comma,
]{\ClusterStabOneD};}

\newcommand{\PlotStabTwoD}[1][]{%
\addplot+[
    #1,
    mark=diamond, 
    mark size=1.8pt,
    line width=0.9pt,
    draw=staborange!40,
    mark options={draw=staborange},
    error bars/.cd,
        y dir=both,
        y explicit,
        error bar style={staborange},
    /pgfplots/.cd
] table[
    x=N,
    y expr=\thisrow{mean_mem_mb},
    y error expr=\thisrow{std_mem_mb},
    col sep=comma,
]{\ClusterStabTwoD};}

\newcommand{\PlotQnsOneD}[1][]{%
\addplot+[
    #1,
    mark=square, 
    mark size=1.8pt,
    line width=0.9pt,
    draw=qnscolor!15,
    mark options={draw=qnscolor},
    error bars/.cd,
        y dir=both,
        y explicit,
        error bar style={qnscolor},
    /pgfplots/.cd
] table[
    x=N,
    y expr=\thisrow{mean_mem_mb},
    y error expr=\thisrow{std_mem_mb},
    col sep=comma,
]{\ClusterQnsOneD};}

\newcommand{\PlotQnsTwoD}[1][]{%
\addplot+[
    #1,
    mark=square, 
    mark size=1.8pt,
    line width=0.9pt,
    draw=qnscolor!15,
    mark options={draw=qnscolor},
    error bars/.cd,
        y dir=both,
        y explicit,
        error bar style={qnscolor},
    /pgfplots/.cd
] table[
    x=N,
    y expr=\thisrow{mean_mem_mb},
    y error expr=\thisrow{std_mem_mb},
    col sep=comma,
]{\ClusterQnsTwoD};}

\begin{groupplot}[
    group style={
        group size=1 by 2,
        vertical sep=0.45cm,
        group name=plots
    },
    width=6.5cm,
    height=3.5cm,
    scale only axis,
    xmajorgrids,
    ymajorgrids,
    x grid style={very thin, gray!10},
    y grid style={very thin, gray!10},
    axis line style={line width=0.75pt, color=black},
    tick style={line width=0.75pt, color=black},
    xmode=log,
    ymode=log,
    xmin=1.5, xmax=1200,
    ymin=1e1, ymax=2e4,
    xlabel={Number of qubits (n)},
    ylabel={Peak Memory (MB)},
    ylabel style={
        font=\sffamily\small,
        yshift=-0.5em
        },
    xlabel style={font=\sffamily\small},
    ticklabel style={font=\sffamily\small},
    title style={font=\sffamily\normalsize, yshift=-1.25ex},
    legend cell align={left},
    legend style={
        at={(0.495,2.35)},
        anchor=north,
        draw=black,
        line width=0.7pt,
        fill=white,
        fill opacity=0.95,
        font=\sffamily\footnotesize,
        legend columns=4,
        /tikz/every even column/.append style={column sep=0.05cm}
    },
]

\nextgroupplot[
    ylabel={Peak Memory (MB)},
    xticklabels={},
    xlabel={},
]

\PlotKetOneD
\PlotDmOneD
\PlotStabOneD
\PlotQnsOneD

\node[
    anchor=west,
    fill=gray!5,
    inner sep=2pt,
    font=\sffamily\footnotesize
] at (rel axis cs:0.63,0.9)  {1D Cluster States};

\nextgroupplot[
]

\PlotKetTwoD
\addlegendentry{Q2NS (ket)}
\PlotDmTwoD
\addlegendentry{Q2NS (dm)}
\PlotStabTwoD
\addlegendentry{Q2NS (stab)}
\PlotQnsTwoD
\addlegendentry{qns3 (TN)}

\node[
    anchor=west,
    fill=gray!5,
    inner sep=2pt,
    font=\sffamily\footnotesize
] at (rel axis cs:0.63,0.9) {2D Cluster States};

\end{groupplot}

\begin{axis}[
    at={($(plots c1r2.north east)+(-0.9cm,3.25cm)$)}, 
    anchor=north east,
    width=2.2cm,
    height=1.95cm,
    scale only axis,
    xmode=log,
    ymode=log,
    xmin=1.5, xmax=12000,
    ymin=1e1, ymax=1000, 
    axis background/.style={fill=white, fill opacity=0.95},
    axis line style={line width=0.5pt, color=black},
    tick style={line width=0.5pt, color=black},
    ticklabel style={font=\sffamily\tiny, color=black},
    xlabel={},
    ylabel={},
    grid=both,
    minor grid style={gray!5},
    major grid style={gray!5},
        legend style={
        at={(0.03,0.97)},
        anchor=north west,
        font=\sffamily\tiny,
        draw=none,
        fill=none
    },
    legend cell align=left
]
\addplot[
    dashed,
    line width=0.7pt,
    draw=black,
    domain=1.5:12000,
    samples=200
]{17.619 + 3.054e-05 * x^(1.966)};
\addlegendentry{Fit}

\addplot+[
    only marks,
    mark=diamond, 
    mark size=1.2pt,
    draw=staborange,
    mark options={draw=staborange, fill=staborange},
    error bars/.cd,
        y dir=both,
        y explicit,
        error bar style={staborange, line width=0.4pt},
    /pgfplots/.cd
] table[
    x=N,
    y expr=\thisrow{mean_mem_mb},
    y error expr=\thisrow{std_mem_mb},
    col sep=comma,
]{\ClusterStabOneD};

\end{axis}

\begin{axis}[
    at={($(plots c1r2.north east)+(-0.9cm,-0.75cm)$)},
    anchor=north east,
    width=2.2cm,
    height=1.95cm,
    scale only axis,
    xmode=log,
    ymode=log,
    xmin=1.5, xmax=12000,
    ymin=1e1, ymax=1000,
    axis background/.style={fill=white, fill opacity=0.95},
    axis line style={line width=0.5pt, color=black},
    tick style={line width=0.5pt, color=black},
    ticklabel style={font=\sffamily\tiny, color=black},
    xlabel={},
    ylabel={},
    grid=both,
    minor grid style={gray!5},
    major grid style={gray!5},
    legend style={
        at={(0.03,0.97)},
        anchor=north west,
        font=\sffamily\tiny,
        draw=none,
        fill=none
    },
    legend cell align=left
]
\addplot[
    dashed,
    line width=0.7pt,
    draw=black,
    domain=1.5:12000,
    samples=200
]{17.619 + 3.054e-05 * x^(1.966)};
\addlegendentry{Fit}

\addplot+[
    only marks,
    mark=diamond, 
    mark size=1.2pt,
    draw=staborange,
    mark options={draw=staborange, fill=staborange},
    error bars/.cd,
        y dir=both,
        y explicit,
        error bar style={staborange, line width=0.4pt},
    /pgfplots/.cd
] table[
    x=N,
    y expr=\thisrow{mean_mem_mb},
    y error expr=\thisrow{std_mem_mb},
    col sep=comma,
]{\ClusterStabTwoD};

\end{axis}

\end{tikzpicture}

%% file: Tikz/Fig4a.tex
\begin{tikzpicture}[
    spy using outlines={rectangle, magnification=2.05, width=1.7cm, height=3.8cm, connect spies}
]

\definecolor{accentblue}{RGB}{0,160,230}      
\definecolor{softgray}{RGB}{220,220,220}
\definecolor{graygrid}{RGB}{230,230,230}
\definecolor{dmgreen}{RGB}{220,20,20}       
\definecolor{staborange}{RGB}{255,165,0}     
\definecolor{qnslocteal}{RGB}{100,200,190}    

\definecolor{qnattynet-magenta}{RGB}{158, 57, 106}

\definecolor{qnscolor}{RGB}{100,35,150} 

\def\SwapKetFile{Data/swap/swap_ket_lowconfig.csv}
\def\SwapDmFile{Data/swap/swap_dm_lowconfig.csv}
\def\SwapStabFile{Data/swap/swap_stab_lowconfig.csv}
\def\SwapQnsFile{Data/swap/swap_qns3.csv}
\def\SwapQnsLocFile{Data/swap/swap_qns3_local.csv}

\pgfplotstableread[col sep=comma]{\SwapKetFile}\swapket
\pgfplotstableread[col sep=comma]{\SwapDmFile}\swapdm
\pgfplotstableread[col sep=comma]{\SwapStabFile}\swapstab
\pgfplotstableread[col sep=comma]{\SwapQnsFile}\swapqns
\pgfplotstableread[col sep=comma]{\SwapQnsLocFile}\swapqnsloc


\newcommand{\MakeRatioCols}[3]{
  \pgfplotstablecreatecol[
    copy column from table={\swapqns}{mean_total_ms}
  ]{qns_mean}{#1}

  \pgfplotstablecreatecol[
    copy column from table={\swapqns}{std_total_ms}
  ]{qns_std}{#1}

  \pgfplotstablecreatecol[
    create col/expr={\thisrow{qns_mean} / \thisrow{mean_total_ms}}
  ]{#2}{#1}

  \pgfplotstablecreatecol[
    create col/expr={ %
      (\thisrow{qns_mean} / \thisrow{mean_total_ms}) *
      sqrt( ( (\thisrow{qns_std} / \thisrow{qns_mean})^2 ) +
            ( (\thisrow{std_total_ms} / \thisrow{mean_total_ms})^2 ) ) }
  ]{#3}{#1}
}

\MakeRatioCols{\swapket}{ratioket}{ratioket_std}
\MakeRatioCols{\swapdm}{ratiodm}{ratiodm_std}
\MakeRatioCols{\swapstab}{ratiostab}{ratiostab_std}
\MakeRatioCols{\swapqnsloc}{ratioqnsloc}{ratioqnsloc_std}

\begin{groupplot}[
    group style={
        group size=1 by 2,
        vertical sep=15pt,
        group name=swapplots
    },
    width=5cm, height=2.2cm,
    scale only axis,
    xmajorgrids,
    ymajorgrids,
    x grid style={very thin, gray!10},
    y grid style={very thin, gray!10},
    axis line style={line width=0.65pt, color=black},
    tick style={line width=0.65pt, color=black},
    tick label style={font=\sffamily\scriptsize, color=black},
    label style={font=\sffamily\small, color=black},
    legend style={
        at={(0.65,1.52)},
        anchor=north,
        draw=black,
        fill=white,
        line width=0.7pt,
        fill opacity=0.95,
        legend columns=3,
        /tikz/every even column/.append style={column sep=0.001cm},
        font=\sffamily\scriptsize
    },
]

\nextgroupplot[
    ylabel={Total Time (ms)},
    ylabel style={
        yshift=-0.5em,
        font=\sffamily\footnotesize,
        color=black
    },
    xmin=6,
    xmax=1500,
    ymin=1,
    ymax=1e7,
    xmode=log,
    ymode=log
]

\addplot+[
    color=accentblue!15,
    mark=x,
    mark options={draw=accentblue},
    mark size=1.75pt,
    line width=0.5pt,
    error bars/.cd,
        y dir=both,
        y explicit,
        error bar style={accentblue, line width=0.4pt}
] table[
    x=N,
    y=mean_total_ms,
    y error=std_total_ms,
]{\swapket};
\addlegendentry{Q2NS (ket)}

\addplot+[
    color=qnattynet-magenta!15,
    mark=triangle,
    mark options={draw=qnattynet-magenta},
    mark size=1.6pt,
    line width=0.5pt,
    error bars/.cd,
        y dir=both,
        y explicit,
        error bar style={qnattynet-magenta, line width=0.4pt}
] table[
    x=N,
    y=mean_total_ms,
    y error=std_total_ms,
]{\swapdm};
\addlegendentry{Q2NS (dm)}

\addplot+[
    color=staborange!15,
    mark options={draw=staborange},
    mark=diamond,
    mark size=1.7pt,
    line width=0.5pt,
    error bars/.cd,
        y dir=both,
        y explicit,
        error bar style={staborange, line width=0.4pt}
] table[
    x=N,
    y=mean_total_ms,
    y error=std_total_ms,
]{\swapstab};
\addlegendentry{Q2NS (stab)}

\addplot+[
    color=violet!10,
    mark options={draw=qnscolor},
    mark=square,
    mark size=1.5pt,
    line width=0.5pt,
    error bars/.cd,
        y dir=both,
        y explicit,
        error bar style={qnscolor, line width=0.4pt}
] table[
    x=N,
    y=mean_total_ms,
    y error=std_total_ms,
]{\swapqns};
\addlegendentry{qns-3 GR}

\addplot+[
    mark=square, 
    mark size=1.5pt,
    line width=0.5pt,
    color=qnslocteal!15,
    mark options={draw=qnslocteal},
    error bars/.cd,
        y dir=both,
        y explicit,
        error bar style={qnslocteal, line width=0.4pt}
] table[
    x=N,
    y=mean_total_ms,
    y error=std_total_ms,
]{\swapqnsloc};
\addlegendentry{qns-3 GR-ID}

\coordinate (spy point) at (axis cs:90,7200);
\spy [black, line width=0.5pt] on (spy point)
      in node [fill=white, draw=black, line width=0.65pt] at (rel axis cs:0.65,-0.2);

\nextgroupplot[
    xlabel={Nodes (N)},
    ylabel={Ratio qns-3/others},
    xlabel style={
        yshift=0.9em,
        font=\sffamily\small,
        color=black
    },
    ylabel style={
        yshift=-0.5em,
        font=\sffamily\footnotesize,
        color=black
    },
    xmin=6,
    xmax=150,
    ymax=5e4,
    xmode=log,
    ymode=log,
]

\addplot[
    mark=x, 
    mark size=1.5pt,
    line width=0.9pt,
    color=accentblue!10,
    mark options={draw=accentblue},
    error bars/.cd,
        y dir=both,
        y explicit,
        error bar style={accentblue, line width=0.4pt}
] table[
    x=N,
    y=ratioket,
    y error=ratioket_std,
]{\swapket};

\addplot[
    mark=triangle, 
    mark size=1.5pt,
    line width=0.9pt,
    color=qnattynet-magenta!10,
    mark options={draw=qnattynet-magenta},
    error bars/.cd,
        y dir=both,
        y explicit,
        error bar style={qnattynet-magenta, line width=0.4pt}
] table[
    x=N,
    y=ratiodm,
    y error=ratiodm_std,
]{\swapdm};

\addplot[
    mark=diamond, 
    mark size=1.5pt,
    line width=0.9pt,
    color=staborange!10,
    mark options = {draw=staborange},
    error bars/.cd,
        y dir=both,
        y explicit,
        error bar style={staborange, line width=0.4pt}
] table[
    x=N,
    y=ratiostab,
    y error=ratiostab_std,
]{\swapstab};

\addplot[
    mark=square, 
    mark size=1.5pt,
    line width=0.9pt,
    color=qnslocteal!15,
    mark options={draw=qnslocteal},
    error bars/.cd,
        y dir=both,
        y explicit,
        error bar style={qnslocteal, line width=0.4pt}
] table[
    x=N,
    y=ratioqnsloc,
    y error=ratioqnsloc_std,
]{\swapqnsloc};

\end{groupplot}

\end{tikzpicture}

%% file: Tikz/Fig4b.tex
\begin{tikzpicture}[
    spy using outlines={rectangle, magnification=2.05, width=1.7cm, height=3.8cm, connect spies}
]

\definecolor{accentblue}{RGB}{0,160,230}      
\definecolor{softgray}{RGB}{220,220,220}
\definecolor{graygrid}{RGB}{230,230,230}
\definecolor{dmgreen}{RGB}{220,20,20}        
\definecolor{staborange}{RGB}{255,165,0}      
\definecolor{qnslocteal}{RGB}{100,200,190}  

\definecolor{qnattynet-magenta}{RGB}{158, 57, 106}
\definecolor{qnscolor}{RGB}{100,35,150} 

\def\SwapKetFile{Data/swap/swap_ket_lowconfig.csv}
\def\SwapDmFile{Data/swap/swap_dm_lowconfig.csv}
\def\SwapStabFile{Data/swap/swap_stab_lowconfig.csv}
\def\SwapQnsFile{Data/swap/swap_qns3.csv}
\def\SwapQnsLocFile{Data/swap/swap_qns3_local.csv}

\pgfplotstableread[col sep=comma]{\SwapKetFile}\swapket
\pgfplotstableread[col sep=comma]{\SwapDmFile}\swapdm
\pgfplotstableread[col sep=comma]{\SwapStabFile}\swapstab
\pgfplotstableread[col sep=comma]{\SwapQnsFile}\swapqns
\pgfplotstableread[col sep=comma]{\SwapQnsLocFile}\swapqnsloc


\newcommand{\MakeRatioCols}[3]{
  \pgfplotstablecreatecol[
    copy column from table={\swapqns}{mean_peak_kb}
  ]{qns_mean}{#1}

  \pgfplotstablecreatecol[
    copy column from table={\swapqns}{std_peak_kb}
  ]{qns_std}{#1}

  \pgfplotstablecreatecol[
    create col/expr={\thisrow{qns_mean} / \thisrow{mean_peak_kb}}
  ]{#2}{#1}

  \pgfplotstablecreatecol[
    create col/expr={ %
      (\thisrow{qns_mean} / \thisrow{mean_peak_kb}) *
      sqrt( ( (\thisrow{qns_std} / \thisrow{qns_mean})^2 ) +
            ( (\thisrow{std_peak_kb} / \thisrow{mean_peak_kb})^2 ) ) }
  ]{#3}{#1}
}

\MakeRatioCols{\swapket}{ratioket}{ratioket_std}
\MakeRatioCols{\swapdm}{ratiodm}{ratiodm_std}
\MakeRatioCols{\swapstab}{ratiostab}{ratiostab_std}
\MakeRatioCols{\swapqnsloc}{ratioqnsloc}{ratioqnsloc_std}

\begin{groupplot}[
    group style={
        group size=1 by 2,
        vertical sep=15pt,
        group name=swapplots
    },
    width=5cm, height=2.2cm,
    scale only axis,
    xmajorgrids,
    ymajorgrids,
    x grid style={very thin, gray!10},
    y grid style={very thin, gray!10},
    axis line style={line width=0.65pt, color=black},
    tick style={line width=0.65pt, color=black},
    tick label style={font=\sffamily\scriptsize, color=black},
    label style={font=\sffamily\small, color=black},
    legend style={
        at={(0.64,1.52)},
        anchor=north,
        draw=black,
        fill=white,
        line width=0.7pt,
        fill opacity=0.95,
        legend columns=3,
        /tikz/every even column/.append style={column sep=0.001cm},
        font=\sffamily\scriptsize
    },
]

\nextgroupplot[
    ylabel={Peak Memory (MB)},
    ylabel style={
        yshift=-0.5em,
        font=\sffamily\footnotesize,
        color=black
    },
    xmin=6,
    xmax=1500,
    ymin=5,
    ymax=1e4,
    xmode=log,
    ymode=log
]

\addplot+[
    color=accentblue!15,
    mark options={draw=accentblue},
    mark=x,
    mark size=1.75pt,
    line width=0.5pt,
    error bars/.cd,
        y dir=both,
        y explicit,
        error bar style={accentblue, line width=0.4pt}
] table[
    x=N,
    y expr=\thisrow{mean_peak_kb}/1024,
    y error expr=\thisrow{std_peak_kb}/1024,
]{\swapket};
\addlegendentry{Q2NS (ket)}

\addplot+[
    color=qnattynet-magenta!15,
    mark options={draw=qnattynet-magenta},
    mark=triangle,
    mark size=1.6pt,
    line width=0.5pt,
    error bars/.cd,
        y dir=both,
        y explicit,
        error bar style={qnattynet-magenta, line width=0.4pt}
] table[
    x=N,
    y expr=\thisrow{mean_peak_kb}/1024,
    y error expr=\thisrow{std_peak_kb}/1024,
]{\swapdm};
\addlegendentry{Q2NS (dm)}

\addplot+[
    color=staborange!15,
    mark options={draw=staborange},
    mark=diamond,
    mark size=1.7pt,
    line width=0.5pt,
    error bars/.cd,
        y dir=both,
        y explicit,
        error bar style={staborange, line width=0.4pt}
] table[
    x=N,
    y expr=\thisrow{mean_peak_kb}/1024,
    y error expr=\thisrow{std_peak_kb}/1024,
]{\swapstab};
\addlegendentry{Q2NS (stab)}

\addplot+[
    color=qnscolor!10,
    mark options={draw=qnscolor},
    mark=square,
    mark size=1.5pt,
    line width=0.5pt,
    error bars/.cd,
        y dir=both,
        y explicit,
        error bar style={qnscolor, line width=0.4pt}
] table[
    x=N,
    y expr=\thisrow{mean_peak_kb}/1024,
    y error expr=\thisrow{std_peak_kb}/1024,
]{\swapqns};
\addlegendentry{qns-3 GR}

\addplot+[
    color=qnslocteal!15,
    mark options={draw=qnslocteal},
    mark=square,
    mark size=1.5pt,
    line width=0.5pt,
    error bars/.cd,
        y dir=both,
        y explicit,
        error bar style={qnslocteal, line width=0.4pt}
] table[
    x=N,
    y expr=\thisrow{mean_peak_kb}/1024,
    y error expr=\thisrow{std_peak_kb}/1024,
]{\swapqnsloc};
\addlegendentry{qns-3 GR-ID}

\coordinate (spy point) at (axis cs:90,340);
\spy [black, line width=0.5pt] on (spy point)
      in node [fill=white, draw=black, line width=0.65pt] at (rel axis cs:0.65,-0.15);

\nextgroupplot[
    xlabel={Nodes (N)},
    ylabel={Ratio qns-3/others},
    xlabel style={
        yshift=0.9em,
        font=\sffamily\small,
        color=black
    },
    ylabel style={
        yshift=-0.5em,
        font=\sffamily\footnotesize,
        color=black
    },
    xmin=6,
    xmax=150,
    ymin=1e0,
    ymax=4e2,
    xmode=log,
    ymode=log,
]

\addplot[
    mark=x,
    mark size=1.5pt,
    color=accentblue!15,
    mark options={draw=accentblue},
    line width=0.8pt,
    error bars/.cd,
        y dir=both,
        y explicit,
        error bar style={accentblue, line width=0.4pt}
] table[
    x=N,
    y=ratioket,
    y error=ratioket_std,
]{\swapket};

\addplot[
    mark=triangle,
    mark size=1.5pt,
    color=qnattynet-magenta!15,
    mark options={draw=qnattynet-magenta},
    line width=0.8pt,
    error bars/.cd,
        y dir=both,
        y explicit,
        error bar style={qnattynet-magenta, line width=0.4pt}
] table[
    x=N,
    y=ratiodm,
    y error=ratiodm_std,
]{\swapdm};

\addplot[
    mark=diamond,
    mark size=1.5pt,
    color=staborange!15,
    mark options={draw=staborange},
    line width=0.8pt,
    error bars/.cd,
        y dir=both,
        y explicit,
        error bar style={staborange, line width=0.4pt}
] table[
    x=N,
    y=ratiostab,
    y error=ratiostab_std,
]{\swapstab};

\addplot[
    mark=square,
    mark size=1.5pt,
    color=qnslocteal!15,
    mark options={draw=qnslocteal},
    line width=0.8pt,
    error bars/.cd,
        y dir=both,
        y explicit,
        error bar style={qnslocteal, line width=0.4pt}
] table[
    x=N,
    y=ratioqnsloc,
    y error=ratioqnsloc_std,
]{\swapqnsloc};

\end{groupplot}

\end{tikzpicture}

%% file: Tikz/Fig5a.tex
\begin{tikzpicture}
\input{Tikz/Fig5_setup}

\begin{axis}[
  width=7.75cm,
  height=3.5cm,
  scale only axis,
  ymin=0, ymax=310,
  xmode=log, log basis x=2,
  xmin=6, xmax=144,
  xtick={8,16,32,64,128},
  xticklabels={$8$,$16$,$32$,$64$,$128$},
  xlabel={Nodes ($N$)},
  xmajorgrids, ymajorgrids,
  x grid style={very thin, gray!10},
  y grid style={very thin, gray!10},
  axis line style={line width=0.65pt, color=black},
  tick style={line width=0.65pt, color=black},
  xlabel style={yshift=0.25em, font=\sffamily\small},
  ylabel style={yshift=-0.5em, font=\sffamily\small},
  ticklabel style={font=\sffamily\small},
  ylabel={Time (ms)},
  ybar stacked,
  bar width=6pt,
  enlarge x limits=0.08,
]

\addplot+[ybar, bar shift=-16pt, fill=qnattynet-magenta!40, draw=qnattynet-magenta,
  postaction={pattern=diaglines, pattern color=qnattynet-magenta!70}, forget plot]
  table[x expr=\thisrow{N}, y=mean_setup_ms]{\swapdm};
\addplot+[ybar, bar shift=-16pt, fill=qnattynet-magenta!85, draw=qnattynet-magenta!80!black, line width=0.8pt]
  table[x expr=\thisrow{N}, y=mean_run_ms]{\swapdm};
\resetstackedplots

\addplot+[ybar, bar shift=-9.5pt, fill=accentblue!40, draw=accentblue,
  postaction={pattern=diaglines, pattern color=accentblue!70}, forget plot]
  table[x expr=\thisrow{N}, y=mean_setup_ms]{\swapket};
\addplot+[ybar, bar shift=-9.5pt, fill=accentblue!85, draw=accentblue!80!black, line width=0.8pt]
  table[x expr=\thisrow{N}, y=mean_run_ms]{\swapket};
\resetstackedplots

\addplot+[ybar, bar shift=-3pt, fill=staborange!40, draw=staborange,
  postaction={pattern=diaglines, pattern color=staborange!70}, forget plot]
  table[x expr=\thisrow{N}, y=mean_setup_ms]{\swapstab};
\addplot+[ybar, bar shift=-3pt, fill=staborange!85, draw=staborange!80!black, line width=0.8pt]
  table[x expr=\thisrow{N}, y=mean_run_ms]{\swapstab};
\resetstackedplots

\addplot+[ybar, bar shift=3.5pt, fill=qnscolor!40!white, draw=qnscolor,
  postaction={pattern=diaglines, pattern color=qnscolor!70}, forget plot]
  table[x expr=\thisrow{N}, y=mean_setup_ms]{\swapqns};
\resetstackedplots

\addplot+[ybar, bar shift=10pt, fill=qnslocteal!40, draw=qnslocteal,
  postaction={pattern=diaglines, pattern color=qnslocteal!70}, forget plot]
  table[x expr=\thisrow{N}, y=mean_setup_ms]{\swapqnsloc};

\end{axis}
\end{tikzpicture}

%% file: Tikz/Fig5b.tex
\begin{tikzpicture}
\input{Tikz/Fig5_setup}

\begin{axis}[
    width=7.75cm,
    height=3.5cm,
    scale only axis,
    ymin=0,
    ymax=510,
    xmode=log,
    log basis x=2,
    xmin=7,
    xmax=129,
    xtick={8, 16, 32, 64, 128},
    xticklabels={$8$, $16$, $32$, $64$, $128$},
    xlabel={Nodes ($N$)},
    xmajorgrids,
    ymajorgrids,
    x grid style={very thin, gray!10},
    y grid style={very thin, gray!10},
    axis line style={line width=0.8pt, color=black},
    tick style={line width=0.8pt, color=black},
    xlabel style={
        font=\sffamily\small,
        yshift=0.25em
    },
    ylabel style={
        font=\sffamily\small,
        yshift=-0.5em
    },
    ticklabel style={font=\sffamily\small},
    ylabel={Time (ms)},
    ybar stacked,
    bar width=6pt,
    enlarge x limits=0.08,
]

\addplot+[ybar, bar shift=-16pt,
    fill=qnattynet-magenta!40,
    draw=qnattynet-magenta,
    postaction={pattern=diaglines, pattern color=qnattynet-magenta!70},
    forget plot
]
table[
    x expr=\thisrow{N},
    y=mean_setup_ms
]{\swapdm};

\addplot+[ybar, bar shift=-16pt,
    fill=qnattynet-magenta!85,
    draw=qnattynet-magenta!80!black,
    line width=0.8pt
]
table[
    x expr=\thisrow{N},
    y=mean_run_ms
]{\swapdm};
\resetstackedplots

\addplot+[ybar, bar shift=-9.5pt,
    fill=accentblue!40,
    draw=accentblue,
    postaction={pattern=diaglines, pattern color=accentblue!70},
    forget plot
]
table[
    x expr=\thisrow{N},
    y=mean_setup_ms
]{\swapket};

\addplot+[ybar, bar shift=-9.5pt,
    fill=accentblue!85,
    draw=accentblue!80!black,
    line width=0.8pt
]
table[
    x expr=\thisrow{N},
    y=mean_run_ms
]{\swapket};
\resetstackedplots

\addplot+[ybar, bar shift=-3pt,
    fill=staborange!40,
    draw=staborange,
    postaction={pattern=diaglines, pattern color=staborange!70},
    forget plot
]
table[
    x expr=\thisrow{N},
    y=mean_setup_ms
]{\swapstab};

\addplot+[ybar, bar shift=-3pt,
    fill=staborange!85,
    draw=staborange!80!black,
    line width=0.8pt
]
table[
    x expr=\thisrow{N},
    y=mean_run_ms
]{\swapstab};
\resetstackedplots

\addplot+[ybar, bar shift=3.5pt,
    fill=qnslocteal!85,
    draw=qnslocteal!80!black,
    line width=0.8pt
]
table[
    x expr=\thisrow{N},
    y=mean_run_ms
]{\swapqnsloc};

\end{axis}

\end{tikzpicture}

%% file: Tikz/Fig6.tex
\begin{tikzpicture}

\definecolor{cryred}{RGB}{165,30,30}       
\definecolor{crycyan}{RGB}{0,120,165}      
\definecolor{crymagenta}{RGB}{115,40,85} 
\definecolor{cryyellow}{RGB}{200,140,0}

\begin{groupplot}[
    group style={
        group size=2 by 1,
        horizontal sep=0.5cm,
        group name=plots
    },
    width=3.25cm,
    height=3.35cm,
    scale only axis,
    xmajorgrids,
    ymajorgrids,
    x grid style={very thin, gray!10},
    y grid style={very thin, gray!10},
    axis line style={line width=0.65pt},
    tick style={line width=0.65pt},
    legend cell align={left},
    legend style={
        at={(1.08,-0.3)},
        anchor=north,
        draw=black,
        line width=0.6pt,
        fill=white,
        fill opacity=0.95,
        font=\sffamily\footnotesize
    },
    label style={font=\sffamily\small},
    legend columns=2,
    ticklabel style={font=\sffamily\small},
    xlabel={Distance ($km$)},
    title style={font=\sffamily\normalsize, yshift=-1.25ex},
    xmin=-0.05, xmax=310,
    ymin=0.40, ymax=1.05
]

\nextgroupplot[
    ylabel={Fidelity of Final State},
    ylabel shift=-0.15cm,
    title={(\texttt{TCP})}
]

\addplot[black, solid, mark=o, error bars/.cd, y dir=both, y explicit] coordinates {
    (10.0, 1.0) +- (0, 0.0)
    (100.0, 1.0) +- (0, 0.0)
    (200.0, 1.0) +- (0, 0.0)
    (300.0, 1.0) +- (0, 0.0)
};
\addlegendentry{$T_{dep}=\infty$ ms, idle}

\addplot[cryyellow, dashdotted, mark=triangle*, error bars/.cd, y dir=both, y explicit] coordinates {
    (10.0, 1.0) +- (0, 0.0)
    (100.0, 1.0) +- (0, 0.0)
    (200.0, 1.0) +- (0, 0.0)
    (300.0, 1.0) +- (0, 0.0)
};
\addlegendentry{$T_{dep}=\infty$ ms, congested}

\addplot[crymagenta, solid, mark=square*, error bars/.cd, y dir=both, y explicit] coordinates {
    (10.0, 0.95) +- (0, 0.0219)
    (100.0, 0.76) +- (0, 0.0429)
    (200.0, 0.59) +- (0, 0.0494)
    (300.0, 0.51) +- (0, 0.0502)
};
\addlegendentry{$T_{dep}=5$ ms, idle}

\addplot[crycyan, dashdotted, mark=diamond*, error bars/.cd, y dir=both, y explicit] coordinates {
    (10.0, 0.87) +- (0, 0.0338)
    (100.0, 0.73) +- (0, 0.0446)
    (200.0, 0.59) +- (0, 0.0494)
    (300.0, 0.53) +- (0, 0.0502)
};
\addlegendentry{$T_{dep}=5$ ms, congested}

\addplot[dotted, thick, black!30] coordinates {(0,1/3) (310,1/3)};

\nextgroupplot[
    title={(\texttt{UDP})},
    ylabel={},
    yticklabels={},
    ytick=\empty
]

\addplot[black, solid, mark=o, error bars/.cd, y dir=both, y explicit] coordinates {
    (10.0, 1.0) +- (0, 0.0)
    (100.0, 1.0) +- (0, 0.0)
    (200.0, 1.0) +- (0, 0.0)
    (300.0, 1.0) +- (0, 0.0)
};

\addplot[cryyellow, dashdotted, mark=triangle*, error bars/.cd, y dir=both, y explicit] coordinates {
    (10.0, 1.0) +- (0, 0.0)
    (100.0, 1.0) +- (0, 0.0)
    (200.0, 1.0) +- (0, 0.0)
    (300.0, 1.0) +- (0, 0.0)
};

\addplot[crymagenta, solid, mark=square*, error bars/.cd, y dir=both, y explicit] coordinates {
    (10.0, 0.95) +- (0, 0.0219)
    (100.0, 0.87) +- (0, 0.0338)
    (200.0, 0.77) +- (0, 0.0423)
    (300.0, 0.71) +- (0, 0.0456)
};

\addplot[crycyan, dashdotted, mark=diamond*, error bars/.cd, y dir=both, y explicit] coordinates {
    (10.0, 0.95) +- (0, 0.0219)
    (100.0, 0.75) +- (0, 0.0435)
    (200.0, 0.58) +- (0, 0.0496)
    (300.0, 0.51) +- (0, 0.0502)
};

\addplot[dotted, thick, black!30] coordinates {(0,1/3) (310,1/3)};

\end{groupplot}
\end{tikzpicture}

%% file: Tikz/Fig8a.tex
\begin{tikzpicture}

\definecolor{staborange}{RGB}{170,120,0}
\definecolor{softgray}{RGB}{220,220,220}
\definecolor{graygrid}{RGB}{230,230,230}

\pgfplotstableread[col sep=comma]{Data/qlan/qlan_scaling_aggregated.csv}\qlansdata

\begin{axis}[
    width=6.5cm,
    height=3cm,
    scale only axis,
    xmin=10,
    xmax=100,
    ymin=0,
    ymax=260,
    xmajorgrids,
    ymajorgrids,
    x grid style={very thin, gray!10},
    y grid style={very thin, gray!10},
    axis line style={line width=0.65pt},
    tick style={line width=0.65pt},
    xtick={10,20,30,40,50,60,70,80,90,100},
    xlabel={Number of Clients (M)},
    ylabel={Runtime (ms)},
    ylabel style={
        yshift=-0.35em,
        font=\sffamily\small
        },
    label style={font=\sffamily\small},
    ticklabel style={font=\sffamily\small},
    ybar stacked,
    bar width=4.5pt,  
    enlarge x limits=0.08,
    legend style={
        at={(0.44,0.92)},
        anchor=north,
        legend columns=2,
        /tikz/every even column/.append style={column sep=0.45cm},
        draw=black,
        line width=0.65pt,
        fill=white,
        fill opacity=0.95,
        font=\sffamily\scriptsize,
    },
    legend cell align={left},
]

\addplot+[
    ybar,
    fill=staborange!30,  
    draw=staborange!80!black,
    line width=0.8pt,
    area legend
]
table[
    x=N,
    y=mean_config_ms,
]{\qlansdata};
\addlegendentry{Configuration}

\addplot+[
    ybar,
    fill=staborange!70,  
    draw=staborange!80!black,
    line width=0.8pt,
    area legend
]
table[
    x=N,
    y=mean_sim_ms,
]{\qlansdata};
\addlegendentry{Simulation}

\end{axis}

\end{tikzpicture}

%% file: Tikz/Fig8b.tex
\begin{tikzpicture}

\definecolor{accentblue}{RGB}{0,100,145}      
\definecolor{dmgreen}{RGB}{140,25,25}        
\definecolor{stimorange}{RGB}{170,120,0}      

\begin{axis}[
    width=6.5cm,
    height=3cm,    
    scale only axis,
    xmajorgrids,
    ymajorgrids,
    x grid style={very thin, gray!10},
    y grid style={very thin, gray!10},
    axis line style={line width=0.65pt},
    tick style={line width=0.65pt},
    xmin=-0.035, xmax=0.635,
    ymin=-0.47, ymax=17.65,
    xlabel={Loss Probability $p$},
    ylabel={Completion Time (ms)},
    ylabel style={
        yshift=-0.35em,
        font=\sffamily\small
        },
    label style={font=\sffamily\small},
    ticklabel style={font=\sffamily\small},
    title style={font=\sffamily\normalsize, yshift=-1.25ex},
    legend cell align={left},
    legend style={
        at={(0.055,0.92)},
        anchor=north west,
        draw=black,
        line width=0.6pt,
        fill=white,
        fill opacity=0.95,
        font=\sffamily\footnotesize,
        legend columns=3
    },
]

\addplot[
    color=accentblue!25, 
    mark=square,
    mark size=1.5,
    line width=0.8pt,
    mark options={thick, solid, draw=accentblue},
    error bars/.cd, 
        y dir=both, 
        y explicit,
        error bar style={accentblue, line width=0.7pt}
]
coordinates {
    (0, 0.521) +- (0, 0.0)
    (0.1, 0.785) +- (0, 0.44119214073973)
    (0.2, 1.169) +- (0, 0.77700278849778)
    (0.3, 1.665) +- (0, 1.09805337398562)
    (0.4, 2.185) +- (0, 1.59870654788883)
    (0.5, 3.185) +- (0, 2.10298104911321)
    (0.6, 3.665) +- (0, 2.35857858238663)
};
\addlegendentry{1 km}

\addplot[
    color=dmgreen!25,
    line width=0.8pt,
    mark=square,
    mark size=1.5,
    mark options={thick, solid, draw=dmgreen},
    error bars/.cd, 
        y dir=both,
        y explicit,
        error bar style={dmgreen, line width=0.7pt}]
coordinates {
    (0, 2.736) +- (0, 0.0)
    (0.1, 2.96) +- (0, 0.46992799792343)
    (0.2, 3.344) +- (0, 0.77220398205331)
    (0.3, 3.768) +- (0, 1.03151159063588)
    (0.4, 4.36) +- (0, 1.39008901684298)
    (0.5, 5.136) +- (0, 2.07184102974269)
    (0.6, 6.576) +- (0, 3.03834085505809)
};
\addlegendentry{50 km}

\addplot[
    color=stimorange!25, 
    line width=0.8pt,
    mark=square,
    mark size=1.5,
    mark options={thick, solid, draw=stimorange},
    error bars/.cd, 
        y dir=both,
        y explicit,
        error bar style={stimorange, line width=0.7pt}]
coordinates {
    (0, 7.236) +- (0, 0.0)
    (0.1, 7.58) +- (0, 0.52422525808294)
    (0.2, 7.804) +- (0, 0.70365280694905)
    (0.3, 8.164) +- (0, 0.93571989247409)
    (0.4, 8.66) +- (0, 1.29436385061457)
    (0.5, 9.54) +- (0, 1.8659046063505)
    (0.6, 11.26) +- (0, 2.93763871368166)
};
\addlegendentry{150 km}

\end{axis}

\end{tikzpicture}

%% file: Tikz/Fig9_legend.tex
\begin{tikzpicture}
\definecolor{accentblue}{RGB}{0,160,230}
\definecolor{dmgreen}{RGB}{220,20,20}
\definecolor{staborange}{RGB}{255,165,0}
\definecolor{qnslocteal}{RGB}{100,200,190}
\definecolor{qnattynet-magenta}{RGB}{158, 57, 106}
\definecolor{qnscolor}{RGB}{100,35,150} 

\begin{axis}[
  hide axis,
  xmin=0, xmax=1, ymin=0, ymax=1,
  legend style={
    draw=black,
    fill=white,
    line width=0.6pt,
    fill opacity=0.95,
    font=\sffamily\small,
    at={(0.5,0.5)},
    anchor=center,
    legend columns=6,
    /tikz/every even column/.append style={column sep=0.8em},
  },
]

\addlegendimage{only marks, mark=x, mark options={draw=accentblue}, color=accentblue!15}
\addlegendentry{Q2NS (ket)}

\addlegendimage{only marks, mark=triangle, mark options={draw=qnattynet-magenta}, color=qnattynet-magenta!15}
\addlegendentry{Q2NS (dm)}

\addlegendimage{only marks, mark=diamond, mark options={draw=staborange}, color=staborange!15}
\addlegendentry{Q2NS (stab)}

\addlegendimage{only marks, mark=square, mark options={draw=qnscolor!90!black}, color=qnscolor!10}
\addlegendentry{qns-3 GR}

\addlegendimage{only marks, mark=square, mark options={draw=qnslocteal}, color=qnslocteal!15}
\addlegendentry{qns-3 GR-ID}

\addlegendimage{no marks, dashed, black, line width=0.35pt, dash pattern=on 2pt off 2pt}
\addlegendentry{Fits}

\end{axis}
\end{tikzpicture}

%% file: Tikz/Fig9a.tex
\begin{tikzpicture}
\definecolor{accentblue}{RGB}{0,160,230}
\definecolor{axisgray}{RGB}{60,60,60}
\definecolor{staborange}{RGB}{255,165,0}
\definecolor{qnslocteal}{RGB}{100,200,190}
\definecolor{qnattynet-magenta}{RGB}{158, 57, 106}
\definecolor{qnscolor}{RGB}{100,35,150} 

\def\SwapKetFile{Data/swap/swap_ket_lowconfig.csv}
\def\SwapDmFile{Data/swap/swap_dm_lowconfig.csv}
\def\SwapStabFile{Data/swap/swap_stab_lowconfig.csv} 
\def\SwapQnsFile{Data/swap/swap_qns3.csv}
\def\SwapQnsLocFile{Data/swap/swap_qns3_local.csv}

\def\FitAllAlphaKetT{-2.306323635} \def\FitAllBetaKetT{1.524916066}
\def\FitAllAlphaDmT {-2.278765818} \def\FitAllBetaDmT {1.520862369}
\def\FitAllAlphaStabT{-2.015350000} \def\FitAllBetaStabT{1.557770000}
\def\FitAllAlphaQnsT{-0.492802166} \def\FitAllBetaQnsT{3.113349052}
\def\FitAllAlphaQnsLocT{-0.190411399} \def\FitAllBetaQnsLocT{1.512880230}

\newcommand{\AddFitCurve}[5]{
  \addplot[
    dashed,
    line width=0.75pt,
    opacity=0.5,
    dash pattern=on 2pt off 2pt,
    color=#1,
    samples=120,
    domain=#4:#5
  ] {exp(#2) * x^(#3)};
}

\pgfplotstableread[col sep=comma]{\SwapKetFile}\swapket
\pgfplotstableread[col sep=comma]{\SwapDmFile}\swapdm
\pgfplotstableread[col sep=comma]{\SwapStabFile}\swapstab
\pgfplotstableread[col sep=comma]{\SwapQnsFile}\swapqns
\pgfplotstableread[col sep=comma]{\SwapQnsLocFile}\swapqnsloc

\begin{axis}[
    width=0.65\linewidth, height=0.5\linewidth,
    scale only axis,
    xmode=log, ymode=log,
    xmin=6, xmax=1500,
    ymin=1, ymax=1e7,
    xmajorgrids, ymajorgrids,
    x grid style={very thin, gray!10},
    y grid style={very thin, gray!10},
    axis line style={line width=0.85pt, color=black},
    tick style={line width=0.85pt, color=black},
    tick label style={font=\sffamily\small, color=black},
    label style={font=\sffamily\small, color=black},
    xlabel={Nodes (N)},
    ylabel={Total Time (ms)},
]

\addplot+[
    only marks,
    color=accentblue!15, mark=x, mark options={draw=accentblue},
    mark size=2.0pt,
    error bars/.cd, y dir=both, y explicit, error bar style={accentblue, line width=0.4pt}
] table[x=N, y=mean_total_ms, y error=std_total_ms]{\swapket};

\addplot+[
    only marks,
    color=qnattynet-magenta!15, mark=triangle, mark options={draw=qnattynet-magenta},
    mark size=1.9pt,
    error bars/.cd, y dir=both, y explicit, error bar style={qnattynet-magenta, line width=0.4pt}
] table[x=N, y=mean_total_ms, y error=std_total_ms]{\swapdm};

\addplot+[
    only marks,
    color=staborange!15, mark=diamond, mark options={draw=staborange},
    mark size=2.0pt,
    error bars/.cd, y dir=both, y explicit, error bar style={staborange, line width=0.4pt}
] table[x=N, y=mean_total_ms, y error=std_total_ms]{\swapstab};

\addplot+[
    only marks,
    color=qnscolor!10, mark=square, mark options={draw=qnscolor!90!black},
    mark size=1.8pt,
    error bars/.cd, y dir=both, y explicit, error bar style={qnscolor!90!black, line width=0.4pt}
] table[x=N, y=mean_total_ms, y error=std_total_ms]{\swapqns};

\addplot+[
    only marks,
    color=qnslocteal!15, mark=square, mark options={draw=qnslocteal},
    mark size=1.8pt,
    error bars/.cd, y dir=both, y explicit, error bar style={qnslocteal, line width=0.4pt}
] table[x=N, y=mean_total_ms, y error=std_total_ms]{\swapqnsloc};

\AddFitCurve{accentblue}{\FitAllAlphaKetT}{\FitAllBetaKetT}{6}{1500}
\AddFitCurve{qnattynet-magenta}{\FitAllAlphaDmT}{\FitAllBetaDmT}{6}{1500}
\AddFitCurve{staborange}{\FitAllAlphaStabT}{\FitAllBetaStabT}{6}{1500}
\AddFitCurve{qnscolor!90!black}{\FitAllAlphaQnsT}{\FitAllBetaQnsT}{6}{1500}
\AddFitCurve{qnslocteal}{\FitAllAlphaQnsLocT}{\FitAllBetaQnsLocT}{6}{1500}

\end{axis}
\end{tikzpicture}

%% file: Tikz/Fig9b.tex
\begin{tikzpicture}
\definecolor{accentblue}{RGB}{0,160,230}
\definecolor{dmgreen}{RGB}{220,20,20}
\definecolor{staborange}{RGB}{255,165,0}
\definecolor{qnslocteal}{RGB}{100,200,190}

\def\SwapKetFile{Data/swap/swap_ket_lowconfig.csv}
\def\SwapDmFile{Data/swap/swap_dm_lowconfig.csv}
\def\SwapStabFile{Data/swap/swap_stab_lowconfig.csv}
\def\SwapQnsFile{Data/swap/swap_qns3.csv}
\def\SwapQnsLocFile{Data/swap/swap_qns3_local.csv}


\def\FitLowAlphaKetT{0.122546}    \def\FitLowBetaKetT{0.698982}
\def\FitMidAlphaKetT{-2.61593}   \def\FitMidBetaKetT{1.49197}
\def\FitHighAlphaKetT{-5.14432}  \def\FitHighBetaKetT{2.01156}

\def\FitLowAlphaDmT{0.0443909}   \def\FitLowBetaDmT{0.734074}
\def\FitMidAlphaDmT{-2.48565}    \def\FitMidBetaDmT{1.46632}
\def\FitHighAlphaDmT{-5.08385}   \def\FitHighBetaDmT{2.00163}

\def\FitLowAlphaStabT{0.383431}  \def\FitLowBetaStabT{0.741473}
\def\FitMidAlphaStabT{-2.37129}  \def\FitMidBetaStabT{1.53759}
\def\FitHighAlphaStabT{-4.85840} \def\FitHighBetaStabT{2.04469}

\def\FitLowAlphaQnsT{-0.510757}  \def\FitLowBetaQnsT{3.11983}
\def\FitMidAlphaQnsT{-0.414346}  \def\FitMidBetaQnsT{3.09573}

\def\FitLowAlphaQnsLocT{2.48661}   \def\FitLowBetaQnsLocT{0.606638}
\def\FitMidAlphaQnsLocT{-0.527758} \def\FitMidBetaQnsLocT{1.46851}
\def\FitHighAlphaQnsLocT{-3.30258} \def\FitHighBetaQnsLocT{2.04771}

\newcommand{\AddFitCurve}[5]{
  \addplot[
    no marks,
    dashed,
    line width=0.75pt,
    opacity=0.5,
    dash pattern=on 2pt off 2pt,
    color=#1,
    samples=120,
    domain=#4:#5
  ] {exp(#2) * x^(#3)};
}

\pgfplotstableread[col sep=comma]{\SwapKetFile}\swapket
\pgfplotstableread[col sep=comma]{\SwapDmFile}\swapdm
\pgfplotstableread[col sep=comma]{\SwapStabFile}\swapstab
\pgfplotstableread[col sep=comma]{\SwapQnsFile}\swapqns
\pgfplotstableread[col sep=comma]{\SwapQnsLocFile}\swapqnsloc

\begin{axis}[
    width=0.65\linewidth, height=0.5\linewidth,
    scale only axis,
    xmode=log, ymode=log,
    xmin=6, xmax=1500,
    ymin=1, ymax=1e7,
    xmajorgrids, ymajorgrids,
    x grid style={very thin, gray!10},
    y grid style={very thin, gray!10},
    axis line style={line width=0.85pt, color=black},
    tick style={line width=0.85pt, color=black},
    tick label style={font=\sffamily\small, color=black},
    label style={font=\sffamily\small, color=black},
    xlabel={Nodes (N)},
    ylabel={Total Time (ms)},
]

\addplot+[
    only marks,
    color=accentblue!15, mark=x, mark options={draw=accentblue},
    mark size=2.0pt,
    error bars/.cd, y dir=both, y explicit, error bar style={accentblue, line width=0.4pt}
] table[x=N, y=mean_total_ms, y error=std_total_ms]{\swapket};

\addplot+[
    only marks,
    color=dmgreen!15, mark=triangle, mark options={draw=dmgreen},
    mark size=1.9pt,
    error bars/.cd, y dir=both, y explicit, error bar style={dmgreen, line width=0.4pt}
] table[x=N, y=mean_total_ms, y error=std_total_ms]{\swapdm};

\addplot+[
    only marks,
    color=staborange!15, mark=diamond, mark options={draw=staborange},
    mark size=2.0pt,
    error bars/.cd, y dir=both, y explicit, error bar style={staborange, line width=0.4pt}
] table[x=N, y=mean_total_ms, y error=std_total_ms]{\swapstab};

\addplot+[
    only marks,
    color=violet!10, mark=square, mark options={draw=violet!90!black},
    mark size=1.8pt,
    error bars/.cd, y dir=both, y explicit, error bar style={violet!90!black, line width=0.4pt}
] table[x=N, y=mean_total_ms, y error=std_total_ms]{\swapqns};

\addplot+[
    only marks,
    color=qnslocteal!15, mark=square, mark options={draw=qnslocteal},
    mark size=1.8pt,
    error bars/.cd, y dir=both, y explicit, error bar style={qnslocteal, line width=0.4pt}
] table[x=N, y=mean_total_ms, y error=std_total_ms]{\swapqnsloc};

\AddFitCurve{accentblue}{\FitLowAlphaKetT}{\FitLowBetaKetT}{8}{32}
\AddFitCurve{accentblue}{\FitMidAlphaKetT}{\FitMidBetaKetT}{32}{128}
\AddFitCurve{accentblue}{\FitHighAlphaKetT}{\FitHighBetaKetT}{128}{1500}

\AddFitCurve{dmgreen}{\FitLowAlphaDmT}{\FitLowBetaDmT}{8}{32}
\AddFitCurve{dmgreen}{\FitMidAlphaDmT}{\FitMidBetaDmT}{32}{128}
\AddFitCurve{dmgreen}{\FitHighAlphaDmT}{\FitHighBetaDmT}{128}{1500}

\AddFitCurve{staborange}{\FitLowAlphaStabT}{\FitLowBetaStabT}{8}{32}
\AddFitCurve{staborange}{\FitMidAlphaStabT}{\FitMidBetaStabT}{32}{128}
\AddFitCurve{staborange}{\FitHighAlphaStabT}{\FitHighBetaStabT}{128}{1500}

\AddFitCurve{violet!90!black}{\FitLowAlphaQnsT}{\FitLowBetaQnsT}{8}{32}
\AddFitCurve{violet!90!black}{\FitMidAlphaQnsT}{\FitMidBetaQnsT}{32}{128}

\AddFitCurve{qnslocteal}{\FitLowAlphaQnsLocT}{\FitLowBetaQnsLocT}{8}{32}
\AddFitCurve{qnslocteal}{\FitMidAlphaQnsLocT}{\FitMidBetaQnsLocT}{32}{128}
\AddFitCurve{qnslocteal}{\FitHighAlphaQnsLocT}{\FitHighBetaQnsLocT}{128}{1500}

\end{axis}
\end{tikzpicture}

%% file: Tikz/Fig9c.tex
\begin{tikzpicture}
\definecolor{accentblue}{RGB}{0,160,230}
\definecolor{dmgreen}{RGB}{220,20,20}
\definecolor{staborange}{RGB}{255,165,0}
\definecolor{qnslocteal}{RGB}{100,200,190}

\def\SwapKetFile{Data/swap/swap_ket_lowconfig.csv}
\def\SwapDmFile{Data/swap/swap_dm_lowconfig.csv}
\def\SwapStabFile{Data/swap/swap_stab_lowconfig.csv}
\def\SwapQnsFile{Data/swap/swap_qns3.csv}
\def\SwapQnsLocFile{Data/swap/swap_qns3_local.csv}

\def\FitAllAlphaKetM{2.222376936} \def\FitAllBetaKetM{0.231314919}
\def\FitAllAlphaDmM {2.223232442} \def\FitAllBetaDmM {0.231574413}
\def\FitAllAlphaStabM{2.067158194} \def\FitAllBetaStabM{0.296162000}
\def\FitAllAlphaQnsM{1.230795346} \def\FitAllBetaQnsM{1.052515973}
\def\FitAllAlphaQnsLocM{0.291696583} \def\FitAllBetaQnsLocM{1.082004979}

\newcommand{\AddFitCurve}[5]{
  \addplot[
    no marks,
    dashed,
    line width=0.75pt,
    opacity=0.85,
    dash pattern=on 2pt off 2pt,
    color=#1,
    samples=120,
    domain=#4:#5
  ] {exp(#2) * x^(#3)};
}

\pgfplotstableread[col sep=comma]{\SwapKetFile}\swapket
\pgfplotstableread[col sep=comma]{\SwapDmFile}\swapdm
\pgfplotstableread[col sep=comma]{\SwapStabFile}\swapstab
\pgfplotstableread[col sep=comma]{\SwapQnsFile}\swapqns
\pgfplotstableread[col sep=comma]{\SwapQnsLocFile}\swapqnsloc

\begin{axis}[
    width=0.65\linewidth, height=0.5\linewidth,
    scale only axis,
    xmode=log, ymode=log,
    xmin=6, xmax=1500,
    ymin=5, ymax=1e4,
    xmajorgrids, ymajorgrids,
    x grid style={very thin, gray!10},
    y grid style={very thin, gray!10},
    axis line style={line width=0.85pt, color=black},
    tick style={line width=0.85pt, color=black},
    tick label style={font=\sffamily\small, color=black},
    label style={font=\sffamily\small, color=black},
    xlabel={Nodes (N)},
    ylabel={Peak Memory (MB)},
]

\addplot+[
    only marks,
    color=accentblue!15, mark=x, mark options={draw=accentblue},
    mark size=2.0pt,
    error bars/.cd, y dir=both, y explicit, error bar style={accentblue, line width=0.4pt}
] table[
    x=N,
    y expr=\thisrow{mean_peak_kb}/1024,
    y error expr=\thisrow{std_peak_kb}/1024
]{\swapket};

\addplot+[
    only marks,
    color=dmgreen!15, mark=triangle, mark options={draw=dmgreen},
    mark size=1.9pt,
    error bars/.cd, y dir=both, y explicit, error bar style={dmgreen, line width=0.4pt}
] table[
    x=N,
    y expr=\thisrow{mean_peak_kb}/1024,
    y error expr=\thisrow{std_peak_kb}/1024
]{\swapdm};

\addplot+[
    only marks,
    color=staborange!15, mark=diamond, mark options={draw=staborange},
    mark size=2.0pt,
    error bars/.cd, y dir=both, y explicit, error bar style={staborange, line width=0.4pt}
] table[
    x=N,
    y expr=\thisrow{mean_peak_kb}/1024,
    y error expr=\thisrow{std_peak_kb}/1024
]{\swapstab};

\addplot+[
    only marks,
    color=violet!10, mark=square, mark options={draw=violet!90!black},
    mark size=1.8pt,
    error bars/.cd, y dir=both, y explicit, error bar style={violet!90!black, line width=0.4pt}
] table[
    x=N,
    y expr=\thisrow{mean_peak_kb}/1024,
    y error expr=\thisrow{std_peak_kb}/1024
]{\swapqns};

\addplot+[
    only marks,
    color=qnslocteal!15, mark=square, mark options={draw=qnslocteal},
    mark size=1.8pt,
    error bars/.cd, y dir=both, y explicit, error bar style={qnslocteal, line width=0.4pt}
] table[
    x=N,
    y expr=\thisrow{mean_peak_kb}/1024,
    y error expr=\thisrow{std_peak_kb}/1024
]{\swapqnsloc};

\AddFitCurve{accentblue}{\FitAllAlphaKetM}{\FitAllBetaKetM}{6}{1500}
\AddFitCurve{dmgreen}{\FitAllAlphaDmM}{\FitAllBetaDmM}{6}{1500}
\AddFitCurve{staborange}{\FitAllAlphaStabM}{\FitAllBetaStabM}{6}{1500}
\AddFitCurve{violet!90!black}{\FitAllAlphaQnsM}{\FitAllBetaQnsM}{6}{1500}
\AddFitCurve{qnslocteal}{\FitAllAlphaQnsLocM}{\FitAllBetaQnsLocM}{6}{1500}

\end{axis}
\end{tikzpicture}

%% file: Tikz/Fig9d.tex
\begin{tikzpicture}
\definecolor{accentblue}{RGB}{0,160,230}
\definecolor{dmgreen}{RGB}{220,20,20}
\definecolor{staborange}{RGB}{255,165,0}
\definecolor{qnslocteal}{RGB}{100,200,190}

\def\SwapKetFile{Data/swap/swap_ket_lowconfig.csv}
\def\SwapDmFile{Data/swap/swap_dm_lowconfig.csv}
\def\SwapStabFile{Data/swap/swap_stab_lowconfig.csv}
\def\SwapQnsFile{Data/swap/swap_qns3.csv}
\def\SwapQnsLocFile{Data/swap/swap_qns3_local.csv}


\def\FitLowAlphaKetM{2.823528194}   \def\FitLowBetaKetM{0.0389032}   
\def\FitMidAlphaKetM{2.468188194}   \def\FitMidBetaKetM{0.139341}    
\def\FitHighAlphaKetM{0.865298194}  \def\FitHighBetaKetM{0.459639}   

\def\FitLowAlphaDmM{2.816398194}    \def\FitLowBetaDmM{0.0423207}    
\def\FitMidAlphaDmM{2.487798194}    \def\FitMidBetaDmM{0.135088}     
\def\FitHighAlphaDmM{0.864168194}   \def\FitHighBetaDmM{0.460220}    

\def\FitLowAlphaStabM{2.824638194}  \def\FitLowBetaStabM{0.0507678}  
\def\FitMidAlphaStabM{2.308218194}  \def\FitMidBetaStabM{0.197714}   
\def\FitHighAlphaStabM{0.530758194} \def\FitHighBetaStabM{0.555316}  

\def\FitLowAlphaQnsM{2.702468194}   \def\FitLowBetaQnsM{0.511048}    
\def\FitMidAlphaQnsM{-1.588691806}  \def\FitMidBetaQnsM{1.71041}     

\def\FitLowAlphaQnsLocM{2.910078194}  \def\FitLowBetaQnsLocM{0.21438}   
\def\FitMidAlphaQnsLocM{0.566028194}  \def\FitMidBetaQnsLocM{0.882105}  
\def\FitHighAlphaQnsLocM{-3.717691806} \def\FitHighBetaQnsLocM{1.76446} 

\newcommand{\AddFitCurve}[5]{
  \addplot[
    no marks,
    dashed,
    line width=0.75pt,
    opacity=0.5,
    dash pattern=on 2pt off 2pt,
    color=#1,
    samples=120,
    domain=#4:#5
  ] {exp(#2) * x^(#3)};
}

\pgfplotstableread[col sep=comma]{\SwapKetFile}\swapket
\pgfplotstableread[col sep=comma]{\SwapDmFile}\swapdm
\pgfplotstableread[col sep=comma]{\SwapStabFile}\swapstab
\pgfplotstableread[col sep=comma]{\SwapQnsFile}\swapqns
\pgfplotstableread[col sep=comma]{\SwapQnsLocFile}\swapqnsloc

\begin{axis}[
    width=0.65\linewidth, height=0.5\linewidth,
    scale only axis,
    xmode=log, ymode=log,
    xmin=6, xmax=1500,
    ymin=5, ymax=1e4,
    xmajorgrids, ymajorgrids,
    x grid style={very thin, gray!10},
    y grid style={very thin, gray!10},
    axis line style={line width=0.85pt, color=black},
    tick style={line width=0.85pt, color=black},
    tick label style={font=\sffamily\small, color=black},
    label style={font=\sffamily\small, color=black},
    xlabel={Nodes (N)},
    ylabel={Peak Memory (MB)},
]

\addplot+[
    only marks,
    color=accentblue!15, mark=x, mark options={draw=accentblue},
    mark size=2.0pt,
    error bars/.cd, y dir=both, y explicit, error bar style={accentblue, line width=0.4pt}
] table[
    x=N,
    y expr=\thisrow{mean_peak_kb}/1024,
    y error expr=\thisrow{std_peak_kb}/1024
]{\swapket};

\addplot+[
    only marks,
    color=dmgreen!15, mark=triangle, mark options={draw=dmgreen},
    mark size=1.9pt,
    error bars/.cd, y dir=both, y explicit, error bar style={dmgreen, line width=0.4pt}
] table[
    x=N,
    y expr=\thisrow{mean_peak_kb}/1024,
    y error expr=\thisrow{std_peak_kb}/1024
]{\swapdm};

\addplot+[
    only marks,
    color=staborange!15, mark=diamond, mark options={draw=staborange},
    mark size=2.0pt,
    error bars/.cd, y dir=both, y explicit, error bar style={staborange, line width=0.4pt}
] table[
    x=N,
    y expr=\thisrow{mean_peak_kb}/1024,
    y error expr=\thisrow{std_peak_kb}/1024
]{\swapstab};

\addplot+[
    only marks,
    color=violet!10, mark=square, mark options={draw=violet!90!black},
    mark size=1.8pt,
    error bars/.cd, y dir=both, y explicit, error bar style={violet!90!black, line width=0.4pt}
] table[
    x=N,
    y expr=\thisrow{mean_peak_kb}/1024,
    y error expr=\thisrow{std_peak_kb}/1024
]{\swapqns};

\addplot+[
    only marks,
    color=qnslocteal!15, mark=square, mark options={draw=qnslocteal},
    mark size=1.8pt,
    error bars/.cd, y dir=both, y explicit, error bar style={qnslocteal, line width=0.4pt}
] table[
    x=N,
    y expr=\thisrow{mean_peak_kb}/1024,
    y error expr=\thisrow{std_peak_kb}/1024
]{\swapqnsloc};

\AddFitCurve{accentblue}{\FitLowAlphaKetM}{\FitLowBetaKetM}{8}{32}
\AddFitCurve{accentblue}{\FitMidAlphaKetM}{\FitMidBetaKetM}{32}{128}
\AddFitCurve{accentblue}{\FitHighAlphaKetM}{\FitHighBetaKetM}{128}{1500}

\AddFitCurve{dmgreen}{\FitLowAlphaDmM}{\FitLowBetaDmM}{8}{32}
\AddFitCurve{dmgreen}{\FitMidAlphaDmM}{\FitMidBetaDmM}{32}{128}
\AddFitCurve{dmgreen}{\FitHighAlphaDmM}{\FitHighBetaDmM}{128}{1500}

\AddFitCurve{staborange}{\FitLowAlphaStabM}{\FitLowBetaStabM}{8}{32}
\AddFitCurve{staborange}{\FitMidAlphaStabM}{\FitMidBetaStabM}{32}{128}
\AddFitCurve{staborange}{\FitHighAlphaStabM}{\FitHighBetaStabM}{128}{1500}

\AddFitCurve{violet!90!black}{\FitLowAlphaQnsM}{\FitLowBetaQnsM}{8}{32}
\AddFitCurve{violet!90!black}{\FitMidAlphaQnsM}{\FitMidBetaQnsM}{32}{128}

\AddFitCurve{qnslocteal}{\FitLowAlphaQnsLocM}{\FitLowBetaQnsLocM}{8}{32}
\AddFitCurve{qnslocteal}{\FitMidAlphaQnsLocM}{\FitMidBetaQnsLocM}{32}{128}
\AddFitCurve{qnslocteal}{\FitHighAlphaQnsLocM}{\FitHighBetaQnsLocM}{128}{1500}

\end{axis}
\end{tikzpicture}

%% file: biblio.bib
@misc{ns3quantum2025,
  title        = {Quantum Network Simulation Module for ns-3},
  author       = {{NITK Surathkal Quantum Research Group}},
  year         = {2025},
  howpublished = {\url{https://apps.nsnam.org/app/quantum/}}
}

@article{coopmans2021netsquid,
  title={NetSquid: A discrete-event simulator for quantum networks},
  author={Coopmans, Tim and Knegjens, Robert and Dahlberg, Axel and Maier, David and Nijsten, Loek and de Oliveira Filho, Julio and Papendrecht, Martijn and Rabbie, Julian and Rozpędek, Filip and Skrzypczyk, Matthew and Wubben, Leon and de Jong, Walter and Podareanu, Damian and Torres-Knoop, Ariana and Elkouss, David and Wehner, Stephanie},
  journal={Quantum Science and Technology},
  year={2021}
}

@article{wu2021sequence,
  title = {{{SeQUeNCe}}: {{A Customizable Discrete-Event Simulator}} of {{Quantum Networks}}},
  author = {Wu, Xiaoliang and Kolar, Alexander and Chung, Joaquin and Jin, Dong and Zhong, Tian and Kettimuthu, Rajkumar and Suchara, Martin},
  journal = {Quantum Science and Technology}, 
  year = {2021}
}

@article{dervisevic2024qkdnetsim,
    author = {Emir, Dervisevic and Voznak, Miroslav and Mehic, Miralem},
    title = {Large-scale quantum key distribution network simulator},
    journal = {Journal of Optical Communications and Networking},
    year = {2024}
}

@article{soler2024qkdnetsimplus,
  title = {{{QKDNetSim}}+: {{Improvement}} of the {{Quantum Network Simulator}} for {{NS-3}}},
  author = {Soler, David and Cillero, Iv{\'a}n and Dafonte, Carlos and {Fern{\'a}ndez-Veiga}, Manuel and {Fern{\'a}ndez-Vilas}, Ana and N{\'o}voa, Francisco J.},
  year = {2024},
  journal = {SoftwareX}
}

@inproceedings{Lin2025CFA,
  author       = {Huiping Lin and Ruixuan Deng and Chris Z. Yao and Zhengfeng Ji and Mingsheng Ying},
  title        = {Control Flow Adaption: An Efficient Simulation Method for Noisy Quantum Networks},
  booktitle    = {Proceedings of the IEEE Conference on Computer Communications (INFOCOM 2025)},
  year         = {2025},
  pages        = {1--10},
  doi          = {10.1109/INFOCOM55648.2025.11044741},
  url          = {https://ieeexplore.ieee.org/document/11044741},
}

@article{bartlett2018squanch,
  title        = {SQUANCH: A Distributed Simulation Framework for Quantum Networks and Channels},
  author       = {Bartlett, Stephen D. and Co-authors},
  journal      = {arXiv preprint arXiv:1808.07047},
  year         = {2018},
  url          = {https://arxiv.org/abs/1808.07047}
}

@inproceedings{satoh2021quisp,
  title        = {QuISP: A Quantum Internet Simulation Package},
  author       = {Satoh, Tatsuo and Matsuo, Tomohiro and Van Meter, Rodney},
  booktitle    = {Proceedings of the IEEE International Conference on Quantum Computing and Engineering (QCE)},
  pages        = {197--206},
  year         = {2021},
  publisher    = {IEEE},
  doi          = {10.1109/QCE52317.2021.00033}
}

@article{diadamo2021qunetsim,
  title        = {QuNetSim: A Software Framework for Quantum Networks},
  author       = {DiAdamo, Samuel and Humphreys, Peter and Elkouss, David},
  journal      = {IEEE Transactions on Quantum Engineering},
  volume       = {2},
  pages        = {1--12},
  year         = {2021},
  publisher    = {IEEE},
  doi          = {10.1109/TQE.2021.3076440},
  url          = {https://arxiv.org/abs/2003.06397}
}

@article{dahlberg2022netqasm,
  title        = {NetQASM: A Low-level Instruction Set Architecture for Hybrid Quantum–Classical Programs in a Quantum Internet},
  author       = {Dahlberg, Axel and Skrzypczyk, Matthew and Coopmans, Thomas and Fu, Xiao and Kozlowski, Wojciech and Van Meter, Rodney and Wehner, Stephanie},
  journal      = {Quantum Science and Technology},
  volume       = {7},
  number       = {3},
  pages        = {035023},
  year         = {2022},
  publisher    = {IOP Publishing},
  doi          = {10.1088/2058-9565/ac7338}
}

@misc{qneadk2022,
  title        = {QNE-ADK: Quantum Network Explorer Application Development Kit},
  author       = {{QuTech}},
  howpublished = {\url{https://github.com/QuTech-Delft/qne-adk}},
  year         = {2022},
  note         = {Accessed: 2025-10-02}
}

@article{Gheorghiu2018Quantumpp,
  author  = {V. Gheorghiu},
  title   = {Quantum++: A modern C++ quantum computing library},
  journal = {PLOS ONE},
  volume  = {13},
  number  = {12},
  pages   = {e0208073},
  year    = {2018},
  doi     = {10.1371/journal.pone.0208073},
  url     = {https://journals.plos.org/plosone/article?id=10.1371/journal.pone.0208073}
}

@article{Beaudrap2022faststabiliser,
  title        = {Fast Stabiliser Simulation with Quadratic Form Expansions},
  author       = {Beaudrap, Niel de and Herbert, Steven},
  journal      = {Quantum},
  volume       = {6},
  pages        = {803},
  year         = {2022},
  month        = {sep},
  doi          = {10.22331/q-2022-09-15-803},
  url          = {https://doi.org/10.22331/q-2022-09-15-803},
  issn         = {2521-327X},
  publisher    = {Verein zur F{\"o}rderung des Open Access Publizierens in den Quantenwissenschaften}
}

@article{10.3389/fams.2022.838601,
  title = {{{ExaTN}}: {{Scalable GPU-accelerated}} High-Performance Processing of General Tensor Networks at Exascale},
  author = {Lyakh, Dmitry I. and Nguyen, Thien and Claudino, Daniel and Dumitrescu, Eugene and McCaskey, Alexander J.},
  year = 2022,
  journal = {Frontiers in Applied Mathematics and Statistics},
  volume = {Volume 8 - 2022},
  issn = {2297-4687},
  doi = {10.3389/fams.2022.838601}
}

@misc{ns3,
  author       = "{ns-3 Consortium}",
  title        = "{ns-3: Discrete-Event Network Simulator}",
  year         = 2025,
  howpublished = "\url{https://www.nsnam.org/}",
  note         = "Accessed: October 2025"
}

@article{CirEkeHue-99,
    author = {Cirac, J. I. and Ekert, A. K. and Huelga, S. F. and Macchiavello, C.},
    doi = {10.1103/PhysRevA.59.4249},
    issue = {6},
    journal = {Phys. Rev. A},
    month = {Jun},
    numpages = {0},
    pages = {4249--4254},
    publisher = {American Physical Society},
    title = {Distributed quantum computation over noisy channels},
    url = {http://link.aps.org/doi/10.1103/PhysRevA.59.4249},
    volume = {59},
    year = {1999}
}

@article{DegReiFri-17,
  title={Quantum sensing},
  author={Degen, Christian L and Reinhard, Friedemann and Cappellaro, Paola},
  journal={Reviews of modern physics},
  volume={89},
  number={3},
  pages={035002},
  year={2017},
  publisher={APS}
}

@article{ZhaZhu-21,
  title={Distributed quantum sensing},
  author={Zhang, Zheshen and Zhuang, Quntao},
  journal={Quantum Science and Technology},
  volume={6},
  number={4},
  pages={043001},
  year={2021},
  publisher={IOP Publishing}
}

@ARTICLE{GiaWinCon-25,
  author={Giani, Andrea and Win, Moe Z. and Conti, Andrea},
  journal={IEEE Journal on Selected Areas in Information Theory}, 
  title={Quantum Sensing and Communication via Non-Gaussian States}, 
  year={2025},
  volume={6},
  number={},
  pages={18-33},
  doi={10.1109/JSAIT.2024.3491692}
}

@article{CalCac-25,
  title={{Quantum Internet Architecture: unlocking Quantum-Native Routing via Quantum Addressing}},
  author={Caleffi, Marcello and Cacciapuoti, Angela Sara},
  journal={IEEE Transactions on Communications},
  doi = {10.1109/TCOMM.2025.3650397},
  year={2026}
}

@article{CacCal-26,
  title={{A Quantum Internet Protocol Suite Beyond Layering}},
  author={Cacciapuoti, Angela Sara and Caleffi, Marcello},
  journal={arxiv},
  year={2025}
}

@article{PeaMazCal-26,
  title={{Q2NS: A Modular Framework for Quantum Network Simulation in ns-3}},
  author={Pearson, Adam and Mazza, Francesco and Caleffi, Marcello and Cacciapuoti, Angela Sara},
  journal={arxiv},
  year={2026}
}

@article{viola1998dd,
  title   = {Dynamical suppression of decoherence in two-state quantum systems},
  author  = {Viola, Lorenza and Lloyd, Seth},
  journal = {Physical Review A},
  volume  = {58},
  number  = {4},
  pages   = {2733--2744},
  year    = {1998},
  doi     = {10.1103/PhysRevA.58.2733}
}

@article{viola1999dd,
  title   = {Dynamical decoupling of open quantum systems},
  author  = {Viola, Lorenza and Knill, Emanuel and Lloyd, Seth},
  journal = {Physical Review Letters},
  volume  = {82},
  number  = {12},
  pages   = {2417--2421},
  year    = {1999},
  doi     = {10.1103/PhysRevLett.82.2417}
}

@article{souza2011dd,
  title   = {Robust dynamical decoupling},
  author  = {Souza, A. M. and {\'A}lvarez, G. A. and Suter, D.},
  journal = {Philosophical Transactions of the Royal Society A: Mathematical, Physical and Engineering Sciences},
  volume  = {370},
  number  = {1976},
  pages   = {4748--4769},
  year    = {2012},
  doi     = {10.1098/rsta.2011.0355}
}

@article{CirZolKim-97,
	title={Quantum state transfer and entanglement distribution among distant nodes in a quantum network},
	author={Cirac, Juan Ignacio and Zoller, Peter and Kimble, H Jeff and Mabuchi, Hideo},
	journal = {Phys. Rev. Lett.},
	volume={78},
	number={16},
	pages={3221},
	year={1997},
	publisher={APS}
}

@article{Kim-08,
	title={The quantum internet},
	author={Kimble, H Jeff},
	journal={Nature},
	volume={453},
	number={7198},
	pages={1023--1030},
	year={2008},
	publisher={Nature Publishing Group}
}

@book{NieChu-11,
	author = {Nielsen, Michael A. and Chuang, Isaac L.},
	title = {Quantum Computation and Quantum Information},
	year = {2011},
	publisher = {Cambridge University Press},
}

@article{BenBra-14,
	author = {Charles H. Bennett and Gilles Brassard},
	title = {Quantum cryptography: Public key distribution and coin tossing},
	journal = {Theoretical Computer Science},
	volume = {560},
	pages = {7-11},
	year = {2014},
}

@article{DurLamHeu-17,
	title={Towards a quantum internet},
	author={D{\"u}r, Wolfgang and Lamprecht, Raphael and Heusler, Stefan},
	journal={European Journal of Physics},
	volume={38},
	number={4},
	pages={043001},
	year={2017},
	publisher={IOP Publishing}
}

@article{PirDur-19,
	year = 2019,
	month = {mar},
	publisher = {{IOP} Publishing},
	volume = {21},
	number = {3},
	pages = {033003},
	author = {A Pirker and W Dür},
	title = {A quantum network stack and protocols for reliable entanglement-based networks},
	journal = {New Journal of Physics},
}

@article{CacCalVan-20,
 title={When entanglement meets classical communications: Quantum teleportation for the quantum internet},
	author={Cacciapuoti, Angela Sara and Caleffi, Marcello and Van Meter, Rodney and Hanzo, Lajos},
	journal={IEEE TCOM},
	volume={68},
	number={6},
	pages={3808--3833},
	year={2020},
	publisher={IEEE},
	note = {invited paper},
}

@article {VanSatBen-21,
author = {R. Van Meter and R. Satoh and N. Benchasattabuse and K. Teramoto and T. Matsuo and M. Hajdusek and T. Satoh and S. Nagayama and S. Suzuki},
journal = {2022 IEEE International Conference on Quantum Computing and Engineering (QCE)},
title = {A Quantum Internet Architecture},
year = {2022},
pages = {341-352},
doi = {10.1109/QCE53715.2022.00055},
publisher = {IEEE Computer Society},
month = {sep}
}

@article{IllCalMan-22,
    title={Quantum Internet Protocol Stack: a Comprehensive Survey},
    author={Illiano, Jessica and Caleffi, Marcello and Manzalini, Antonio and Cacciapuoti, Angela Sara},
    journal={Computer Networks},
    volume={213},
    year={2022}
}

@ARTICLE{CaoZhaWan-22,
	author={Cao, Yuan and Zhao, Yongli and Wang, Qin and Zhang, Jie and Ng, Soon Xin and Hanzo, Lajos},
	journal={IEEE Communications Surveys Tutorials}, 
	title={{The Evolution of Quantum Key Distribution Networks: On the Road to the Qinternet}}, 
	year={2022},
	volume={},
	number={},
	pages={1-1},
}

@article{CheIllCac-25,
  title={Entanglement-based artificial topology: Neighboring remote network nodes},
  author={Chen, Si-Yi and Illiano, Jessica and Cacciapuoti, Angela Sara and Caleffi, Marcello},
  journal={IEEE Open Journal of the Communications Society},
  year={2025},
  publisher={IEEE}
}

@article{CacPelIll-25,
  title={Quantum Data Centers: Why Entanglement Changes Everything},
  author={Cacciapuoti, Angela Sara and Pellitteri, Claudio and Illiano, Jessica and d'Avossa, Laura and Mazza, Francesco and Chen, Siyi and Caleffi, Marcello},
  journal={arXiv preprint arXiv:2506.02920},
  year={2025}
}

@article{de2022fast,
  title={Fast stabiliser simulation with quadratic form expansions},
  author={De Beaudrap, Niel and Herbert, Steven},
  journal={Quantum},
  volume={6},
  pages={803},
  year={2022},
  publisher={Verein zur F{\"o}rderung des Open Access Publizierens in den Quantenwissenschaften}
}

@article{CaldAvHan-25,
    author={Caleffi, Marcello and d’Avossa, Laura and Han, Xu and Sara Cacciapuoti, Angela},
  journal={IEEE Communications Surveys \& Tutorials}, 
  title={Quantum Transduction: Enabling Quantum Networking}, 
  year={2026},
  volume={28},
  number={},
  pages={4195-4214},
  doi={10.1109/COMST.2025.3631150}
}

@article{CalAmoFer-22,
  title = {Distributed quantum computing: A survey},
journal = {Computer Networks},
volume = {254},
pages = {110672},
year = {2024},
issn = {1389-1286},
doi = {https://doi.org/10.1016/j.comnet.2024.110672},
url = {https://www.sciencedirect.com/science/article/pii/S1389128624005048},
author = {Marcello Caleffi and Michele Amoretti and Davide Ferrari and Jessica Illiano and Antonio Manzalini and Angela Sara Cacciapuoti},

}

@ARTICLE{MazCalCac-25,
  author={Mazza, Francesco and Caleffi, Marcello and Cacciapuoti, Angela Sara},
  journal={IEEE Transactions on Network Science and Engineering}, 
  title={{Intra-QLAN Connectivity via Graph States: Beyond the Physical Topology}}, 
  year={2025},
  volume={12},
  number={2},
  pages={870-887},
  doi={10.1109/TNSE.2024.3520856}
}

@article{MinPitRob-19,
  title={Experimental quantum key distribution beyond the repeaterless secret key capacity},
  author={Minder, Mariella and Pittaluga, Mirko and Roberts, George Lloyd and Lucamarini, Marco and Dynes, James F and Yuan, ZL and Shields, Andrew J},
  journal={Nature Photonics},
  volume={13},
  number={5},
  pages={334--338},
  year={2019},
  publisher={Nature Publishing Group UK London}
}

@book{MontgomeryRunger2018,
  author    = {D. C. Montgomery and G. C. Runger},
  title     = {Applied Statistics and Probability for Engineers},
  edition   = {7th},
  publisher = {Wiley},
  address   = {Hoboken, NJ, USA},
  year      = {2018}
}

@book{BurnhamAnderson2002,
  author    = {Kenneth P. Burnham and David R. Anderson},
  title     = {Model Selection and Multimodel Inference: A Practical Information-Theoretic Approach},
  publisher = {Springer},
  year      = {2002},
  edition   = {2nd},
  isbn      = {978-0387953649}
}

@article{RaussendorfBriegel2001,
  title={A One-Way Quantum Computer},
  author={Raussendorf, Robert and Briegel, Hans J.},
  journal={Physical Review Letters},
  volume={86},
  number={22},
  pages={5188--5191},
  year={2001},
  doi={10.1103/PhysRevLett.86.5188}
}

@article{RaussendorfBrowneBriegel2003,
  title={Measurement-based quantum computation on cluster states},
  author={Raussendorf, Robert and Browne, Daniel E. and Briegel, Hans J.},
  journal={Physical Review A},
  volume={68},
  number={2},
  pages={022312},
  year={2003},
  doi={10.1103/PhysRevA.68.022312}
}

@article{PirkerDuer2019,
  title={A quantum network architecture},
  author={Pirker, Andreas and D{\"u}r, Wolfgang},
  journal={New Journal of Physics},
  volume={21},
  number={3},
  pages={033003},
  year={2019},
  doi={10.1088/1367-2630/ab0648}
}

@article{HahnPappaEisert2019,
  title={Quantum network routing and local complementation},
  author={Hahn, Fabian and Pappa, Anna and Eisert, Jens},
  journal={npj Quantum Information},
  volume={5},
  number={1},
  pages={76},
  year={2019},
  doi={10.1038/s41534-019-0183-8}
}

@article{FreundPirkerVandreDuer2025,
   title={Graph state extraction from two-dimensional cluster states},
   volume={27},
   ISSN={1367-2630},
   url={http://dx.doi.org/10.1088/1367-2630/ae02bd},
   DOI={10.1088/1367-2630/ae02bd},
   number={9},
   journal={New Journal of Physics},
   publisher={IOP Publishing},
   author={Freund, Julia and Pirker, Alexander and Vandré, Lina and Dür, Wolfgang},
   year={2025},
   month=sep, pages={094505} }

@article{Vidal2003SlightlyEntangled,
  author  = {Guifr{\'e} Vidal},
  title   = {Efficient classical simulation of slightly entangled quantum computations},
  journal = {Phys. Rev. Lett.},
  volume  = {91},
  number  = {14},
  pages   = {147902},
  year    = {2003},
  doi     = {10.1103/PhysRevLett.91.147902},
  eprint  = {quant-ph/0301063}
}

@article{MarkovShi2008TNContraction,
  author  = {Igor L. Markov and Yaoyun Shi},
  title   = {Simulating Quantum Computation by Contracting Tensor Networks},
  journal = {SIAM J. Comput.},
  volume  = {38},
  number  = {3},
  pages   = {963--981},
  year    = {2008},
  doi     = {10.1137/050644756},
  eprint  = {quant-ph/0511069}
}

@article{Hein2006GraphStatesReview,
  author  = {Matthias Hein and Wolfgang D{\"u}r and Jens Eisert and 
             Robert Raussendorf and Maarten Van den Nest and Hans J. Briegel},
  title   = {Entanglement in Graph States and its Applications},
  journal = {Proc. Int. School Phys. Enrico Fermi},
  volume  = {162},
  pages   = {115--218},
  year    = {2006},
  eprint  = {quant-ph/0602096}
}

@article{VanDenNest2007ClassicalSimVsUniversality,
  author  = {Maarten Van den Nest and Wolfgang D{\"u}r and Guifr{\'e} Vidal and 
             Hans J. Briegel},
  title   = {Classical simulation versus universality in measurement-based quantum computation},
  journal = {Phys. Rev. A},
  volume  = {75},
  pages   = {012337},
  year    = {2007},
  doi     = {10.1103/PhysRevA.75.012337},
  eprint  = {quant-ph/0608060}
}

@article{VerstraeteCirac2004PEPS,
  author  = {Frank Verstraete and J. Ignacio Cirac},
  title   = {Renormalization algorithms for quantum-many body systems in two and higher dimensions},
  journal = {arXiv preprint},
  year    = {2004},
  eprint  = {cond-mat/0407066}
}

@article{Schuch2007ComplexityOfPEPS,
  author  = {Norbert Schuch and Michael M. Wolf and Frank Verstraete and 
             J. Ignacio Cirac},
  title   = {Computational Complexity of Projected Entangled Pair States},
  journal = {Phys. Rev. Lett.},
  volume  = {98},
  number  = {14},
  pages   = {140506},
  year    = {2007},
  doi     = {10.1103/PhysRevLett.98.140506},
  eprint  = {quant-ph/0611050}
}

@article{O_Gorman2019TNShape,
  author  = {Bryan O'Gorman and Earl Campbell},
  title   = {Quantum computation and tensor networks: the shape of efficient simulation},
  journal = {arXiv preprint},
  year    = {2019},
  eprint  = {1904.11044}
}

@article{Kourtis2019PractitionersGuide,
  author  = {Stefanos Kourtis and Laurent E. C. Rosenthal and Jelmer J. Renema 
             and Aleksander Cvitanovic and others},
  title   = {A practitioner's guide to tensor network contraction},
  journal = {arXiv preprint},
  year    = {2019},
  eprint  = {1908.08833}
}

@phdthesis{Gottesman1997StabilizerFormalism,
  author       = {Daniel Gottesman},
  title        = {Stabilizer Codes and Quantum Error Correction},
  school       = {Caltech},
  year         = {1997},
  url          = {https://arxiv.org/abs/quant-ph/9705052},
  eprint       = {quant-ph/9705052}
}

@article{Gottesman1998HeisenbergPicture,
  author  = {Daniel Gottesman},
  title   = {The Heisenberg Representation of Quantum Computers},
  journal = {Proceedings of the XXII International Colloquium on Group-Theoretical Methods},
  year    = {1998},
  url     = {https://arxiv.org/abs/quant-ph/9807006},
  eprint  = {quant-ph/9807006}
}

@inproceedings{AaronsonGottesman2004ImprovedSimulation,
  author    = {Scott Aaronson and Daniel Gottesman},
  title     = {Improved Simulation of Stabilizer Circuits},
  booktitle = {Proceedings of the 37th STOC},
  year      = {2004},
  pages     = {417--426},
  doi       = {10.1145/1007352.1007422}
}

@book{NielsenChuang2010QuantumComputation,
  author    = {Michael A. Nielsen and Isaac L. Chuang},
  title     = {Quantum Computation and Quantum Information},
  edition   = {10th Anniversary Edition},
  publisher = {Cambridge University Press},
  year      = {2010}
}

@article{DeRaedt2007QCSimulation,
  author  = {Hans De Raedt and Kristel Michielsen and others},
  title   = {Quantum computing: A computer science perspective},
  journal = {IEEE Computer},
  year    = {2007},
  pages   = {40--50},
  url     = {https://arxiv.org/abs/0707.1536},
  eprint  = {0707.1536}
}

@article{PerezGarcia2007MPS,
  author  = {David Perez-Garcia and Frank Verstraete and Michal M. Wolf and J. Ignacio Cirac},
  title   = {Matrix Product State Representations},
  journal = {Quantum Information and Computation},
  volume  = {7},
  number  = {5--6},
  pages   = {401--430},
  year    = {2007},
  eprint  = {quant-ph/0608197}
}

@article{Schollwoeck2011DMRG,
  author  = {Ulrich Schollw{\"o}ck},
  title   = {The density-matrix renormalization group in the age of matrix product states},
  journal = {Annals of Physics},
  volume  = {326},
  number  = {1},
  pages   = {96--192},
  year    = {2011},
  doi     = {10.1016/j.aop.2010.09.012}
}

@article{Orus2014TNReview,
  author  = {Rom{\'a}n Or{\'u}s},
  title   = {A practical introduction to tensor networks: Matrix product states and projected entangled pair states},
  journal = {Annals of Physics},
  volume  = {349},
  pages   = {117--158},
  year    = {2014},
  doi     = {10.1016/j.aop.2014.06.013}
}

@article{Eisert2010,
  title = {Colloquium: Area laws for the entanglement entropy},
  author = {Eisert, J. and Cramer, M. and Plenio, M. B.},
  journal = {Rev. Mod. Phys.},
  volume = {82},
  issue = {1},
  pages = {277--306},
  numpages = {0},
  year = {2010},
  month = {Feb},
  publisher = {American Physical Society},
  doi = {10.1103/RevModPhys.82.277},
  url = {https://link.aps.org/doi/10.1103/RevModPhys.82.277}
}

@article{Haferkamp2023,
  title={Contracting projected entangled pair states is average-case hard},
  author={Haferkamp, Jonas and Brydges, Thomas and Eisert, Jens and Huang, Yunchao and Kueng, Richard and Yang, Yifan},
  journal={Nature Physics},
  volume={19},
  pages={1424--1430},
  year={2023},
  doi={10.1038/s41567-023-02180-7}
}

@INPROCEEDINGS{Mazza2024QCNCTwocolorable,
  author={Mazza, Francesco and Caleffi, Marcello and Cacciapuoti, Angela Sara},
  booktitle={2024 International Conference on Quantum Communications, Networking, and Computing (QCNC)}, 
  title={Quantum LAN: On-Demand Network Topology via Two-Colorable Graph States}, 
  year={2024},
  pages={127--134},
  doi={10.1109/QCNC62729.2024.00029}
}

@misc{Mazza2024SeQUeNCe,
  title     = {{Simulation of Entanglement-Enabled Connectivity in QLANs using SeQUeNCe}},
  author    = {Francesco Mazza and Caitao Zhan and Joaquin Chung and Rajkumar Kettimuthu and Marcello Caleffi and Angela Sara Cacciapuoti},
  year      = {2024},
  eprint    = {2411.11031},
  archivePrefix = {arXiv},
  primaryClass  = {quant-ph},
  url       = {https://arxiv.org/abs/2411.11031}
}

@article{Hein2004GraphStates,
  author    = {M. Hein and J. Eisert and H. J. Briegel},
  title     = {Multiparty entanglement in graph states},
  journal   = {Physical Review A},
  volume    = {69},
  number    = {6},
  pages     = {062311},
  year      = {2004},
  doi       = {10.1103/PhysRevA.69.062311}
}

@article{VanDenNest2004LC,
  author    = {M. Van den Nest and J. Dehaene and B. De Moor},
  title     = {Graphical description of the action of local Clifford transformations on graph states},
  journal   = {Physical Review A},
  volume    = {69},
  number    = {2},
  pages     = {022316},
  year      = {2004},
  doi       = {10.1103/PhysRevA.69.022316}
}
